\documentclass{cernrep} 
\usepackage{texnames}
\usepackage[T1]{fontenc}
\begin{document}

\title{LHC Results Highlights}

\author{Oscar Gonz\'alez}

\institute{CIEMAT, Spain}

\maketitle 


\begin{abstract}
The good performance of the LHC provided enough data at 7~TeV and 8~TeV to
allow the experiments to perform very competitive measurements and to expand the
knowledge about the fundamental interaction far beyond that from previous
colliders.
This report summarizes the highlights of the results obtained with these data samples
by the four large experiments, covering all the topics of the physics program and
focusing on those exploiting the possibilities of the LHC.
\end{abstract}


\section{Introduction}

The standard model (SM)~\cite{bib:sm1,bib:sm2,bib:sm3} of particles and interactions is currently the 
most successful theory describing the Universe at the smallest distances,
or equivalently, highest energies. Such task is performed with the use
of three families of fermions and a number of bosons associated to the interactions
as given by the $SU(3)_C\times SU(2)_L\times U(1)_Y$ symmetry group. Since
in Nature the $SU(2)_L\times U(1)_Y$ is not an exact symmetry, we require an
additional field, the so-called Higgs field, which sponteneously breaks the
symmetry according to the BEH mechanism~\cite{bib:higgs},
giving rise to the weak and electromagnetic interactions as they
are observed at lower energies. In addition this field is responsible to
give mass to the fermions.

Although successful, the SM does not appear to be complete since several esperimental
evidences are not included in the model. In this group, it should be remarked
that gravitational effects are not described, neither are all the related effects, such
as Dark Matter or Dark Energy. In addition the current structure of the SM does not include
enough CP violation to justify the observed matter-antimatter imbalance in the Universe. Finally
the neutrinos in the model are assumed to be massless, something that currently is experimentally
discarded after the measurements of neutrino mixing.

In addition to the missing parts in the SM there are several points in which the model
is not completely satisfactory, concretely related to theoretical aspects of it. Several
issues are always mentioned in this context, but they are summarized in three main
issues: the need of fine-tuning to understand the low scale of the electroweak symmetry breaking
and other parameters~(the
hierarchy problem),
the lack of understanding on why there are three families with double-nature
sets (i.e. quark and leptons) and the lack of apparent relation between the different interactions
(i.e. the origin of the observed values for couplings, including fermionic masses). In practice, the
SM has clear limitations since it misses too many explanations about why things are as they are and
it requires too many parameters to actually describe things as they are.

The proposed solution to both the experimentally-motivated limitations and the theoretical dissatisfaction
is to add more interactions or particles which complete the model. In such scenario, the SM would become
a low-energy approximation, or visible part, of a larger theory. By increasing the energy in our studies
we gain access to the additional particles and effects, which are usually referred to as ``new physics''
or ``physics beyond the SM''~(BSM). These effects that are not explained by the SM will provide additional
information about the limitations of the SM, opening the correct doors towards a more accurate description
of our Universe.

With this motivation we are led to the design of a powerful hadronic collider which maximizes the
reach in sensitivity to the possible BSM physics. This is achieved by maximizing the available energy, which
would provide the possibility to produce more massive particles, and the number of collisions per
time unit~(luminosity), which increase the yield for the produced particles and effects. This is exactly
the motivation for the Large Hadron Collider~(LHC)~\cite{bib:lhc} located
at CERN, near Geneva~(Switzerland), which is recognized as ``the discovery machine''
for physics beyond the SM providing a large amount of energy per collision and a large amount of collisions.

In the following sections we will describe the LHC and the related experiments and report on the
main results for the different part of the program, designed to take advantage of all the possibilities
given by such powerful machines.

\section{The LHC and the experiments}

The LHC
is the most energetic and most
challenging collider up to date. It is designed to collide protons or heavy ions at a maximum energy of 14~TeV
of energy and very high collision rates. Technical limitations has prevented it to reach its design parameters, and the
collected datasets contains collisions at 7 or 8~TeV of total center-of-mass energies. In any case this represents
more than 3 times more energy than the previously most energetic collider~(The Tevatron at Fermilab, USA).
This allows to reach energy scales that were not accessible before, both for particle and heavy-ion physics.

But the LHC is not just about large energy: it also provides the largest collision rate ever reached,
allowing to collect sizable data samples in record time. To quantify the amount of data, the previously
mentioned concept of luminosity is used. The integrated luminosity relates the number of a type
of events in a sample and the cross section for that type of event. Experimentally, this allows to
compute the luminosity (``calibrate'' the size of the sample) using a very well known process and
count the number of events from it, and so $L=N/\sigma$ where $L$ is the luminosity, $N$ the number of events
and $\sigma$ the cross section of the process. Once the sample luminosity is known, the value is used to measure
cross sections of processes of interest, as $\sigma=L/N$. Finally, knowing the cross section of a process,
one estimates the number of expected events from that process in the sample with $N=L\cdot\sigma$. These
are the basic tools to perform analysis of the data samples.

At the LHC during the first years of operations, samples of reasonable size were obtained at 7~TeV~(in 2010 and 2011),
accounting for 6~fb$^{-1}$ of luminosity for proton-proton collisions and 170~$\mu$b$^{-1}$ for lead-lead
collisions. Additionally, data at 8~TeV were obtained for proton-proton collisions, accounting to 23.3~fb$^{-1}$,
and proton-lead collisions with a luminosity of 32~nb$^{-1}$. The results described in this report have
been obtained by using these data samples.

\begin{figure}
\centering\includegraphics[width=.55\linewidth]{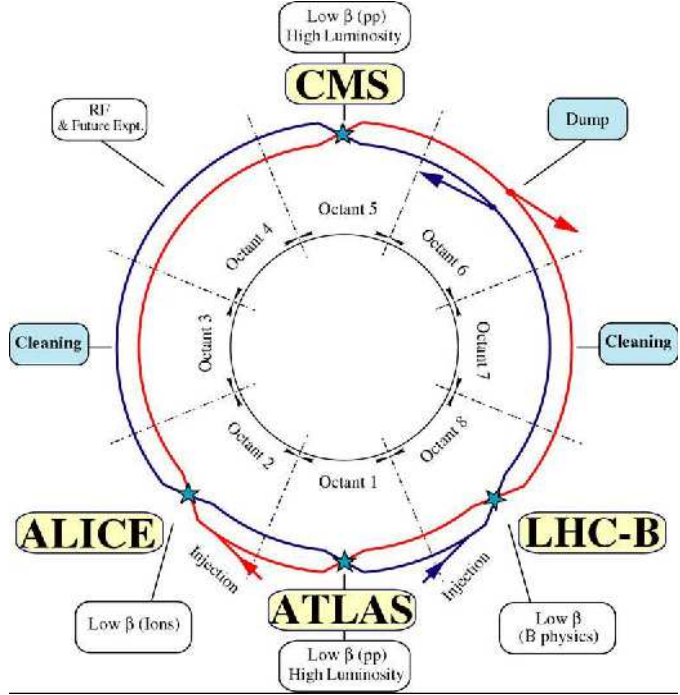}
\caption{Schematic layout of the LHC and the main experiments, identified at their location in
the accelerator ring.}
\label{fig:lhc}
\end{figure}

The collisions provided by the LHC occur at four interaction points along the 27-km ring. At those points,
several experiments are located. The main four experiments are ALICE, ATLAS, CMS and LHCb and are located
as shown in Fig.~\ref{fig:lhc}. These four experiments collect the data from the collisions and provide
the results of the physics analyses, as described in the following sections.

In addition to the main experiments, other three \emph{minor} experiments are intended for more
dedicated studies: TOTEM~\cite{bib:totem:webpage}, LHCf~\cite{bib:lhcf:webpage}
and MoEDAL~\cite{bib:moedal:webpage}.
Neither their results nor plans will be covered here since their
scientific output is very specific and beyond the aim of this report. However, this should not minimize
their importance in order to understand forward production~(as it is the case of the first two) or dedicated
search for magnetic monoples (as it is the aim of MoEDAL).

Each of the main experiments deserved some specific description to put into context the physics
output they provide.

\subsection{The ATLAS experiment}

ATLAS~\cite{bib:atlas} is the largest experiment at the LHC. It is intended to study all possible physics topic
by analysing the full final state of the LHC collisions. It is characterized by its great
capabilities in tracking and calorimetry surrounded by huge muon-detection chambers
in a toroidal field.

The detector has almost full solid-angle coverage with a forward-backward 
symmetric distribution. It is also azimuthally symmetric, as expected for the physics
in the collisions. The hermetic design allows to infer the presence of undetected particles
via the transverse momentum embalanace, the so-called ``missing $E_T$'' ($E_T^{\text{miss}}$),
which can be computed as:

\begin{equation*}
    E_T^{\text{miss}} = \sqrt{\biggl[\sum p_x\biggr]^2 + \biggl[\sum p_y\biggr]^2}\,\,\,;
\end{equation*}

\noindent
where the sum runs over the observed particles (regardless on the way they are detected and
reconstructed).

This quantity is expected to be small due to the conservation of the momentum and therefore a significantly
large value is interpreted as the presence of particle(s) that escape detection, as if the case
of neutrinos and other weakly-interacting particles which do not interact with matter by mean of the nuclear
or electromagnetic forces.

In order to quantify the coverage of the detector, another interesting variable is the
pseudorapidity, an alternative to the polar angle $\theta$ defined as:

\begin{equation*}
     \eta = -\ln \bigl[ \tan \bigl(\theta/2\bigr)\bigr] 
          = \frac{1}{2}\ln \bigl[ \frac{|p|+p_z}{|p|-p_z} \bigr]
\end{equation*}

\noindent
which is well suited for cylindrical
description of events, as it is the case of collisions involving hadrons in the initial state.

The structure of ATLAS allows to reconstruct jets up to $|\eta|\sim 4.5$, muons up to $|\eta|\sim 3$ and
electrons and photons up to $|\eta|\sim 2.47$, providing a very large coverage for the main pieces
to study the final states in the LHC collisions.

\subsection{The CMS experiment}

CMS~\cite{bib:cms} is the other multipurpose detector of the LHC. Similar to ATLAS in aim and 
capabilities, it present a more compact
structure for a similar performance due to its stronger magnetic field.
It is also hermetic and provides an impressive energy resolution for electrons
and photons, for a coverage of $|\eta|<2.5$. Muons are detected up to $|\eta|\sim 2.4$
with a more traditional approach that takes advantage of the redundancy with the inner
tracking. Finally jets are reconstructed up to $|\eta|\sim 4.5$.

When comparing both detectors, the strong point of CMS is the great resolution in the inner
tracking, which becomes the core of the detector, specifically when used as redundancy for
reconstruction of muons and other particles. On the other hand, ATLAS has better global
calorimetry and more precise and sophisticated muon detection.

However, these differences are in practice more technical than real, since the treatment
of the data in the reconstruction of objects allows both collaborations to obtain very comparable
results. The idea is to compensate the limitations of the detectors with the information coming
from the stronger parts or redundant informations from other components.

One good example of this is provided by the concept of \emph{particle flow} that has been 
extensively used in the last years, specially in the CMS analyses. The idea is that instead 
of reconstructing the event quantities from the detector information~(calorimeter cells, tracks),
an intermediate step is taken and that detector information is combined to identify ``objects''
that are associated to particles. From the detector information, the kinematic recontruction
of each ``object'' is performed in an optimal way, since each class of object (lepton, photon,
neutral or charge hadron and so on) is treated differently. It is then from these ``objects''
that the event quantities are then reconstructed.

These idea represent a big gain since each object is treated as close as possible to its expected
behaviour with the detector components. Additionally, the combination of the detector parts allows
to get the most of the detector information as a whole, leading to the final goal of having
a global event description. The case of CMS is extremely clear since the particle-flow approach
allows to use as much tracking information as possible, reducing the impact of the lower quality
hadronic reconstruction in the calorimeters.

By the use of this kind of ideas and even more sophisticated techniques, the LHC experiments
have been able to extract the most of the data samples, going beyond the most optimistic
expectations, as we will describe in future sections.

\subsection{The LHCb experiment}

The LHCb detector~\cite{bib:lhcb} has been designed to perform studies on flavour physics, specifically
of hadrons containing bottom quarks.
Since their production is specially large in the forward region, the detector design
is mostly oriented to maximize rate and provide very accurate reconstruction instead of maximazing
the coverage. It therefore detects particles in the forward region and it reaches an impresive
track and vertex reconstruction due to dedicated sophisticated components.

The main limitation of the measurements in the forward region is the high sensitivity to processes
in which multiplicities are large. For this reason, the LHCb did not collect lead-lead data
and required \emph{luminosity leveling} to keep the number of collisions in the same event at
reasonable levels. This leveling is the reason why the integrated luminosity of the
data samples is smaller for this experiment.

On the other hand, its great coverage in the forward region allows this detector to
perform measurements beyond the coverage of ATLAS and CMS, providing a nice complementarity
at the LHC that is not limited to the topics for which the LHCb was intended.
As we will see below, the LHCb experiment is
providing nice and competetive results in areas where CMS and ATLAS were expected to be
dominant.

\subsection{The ALICE experiment}

The ALICE detector~\cite{bib:alice} has been designed to maximize the physics output from heavy-ion collisions.
The aim of the experiment is not the detection of exotic or striking signatures but to maximize the
particle identification in order to retrieve as much information as possible about the properties
of the medium created in the collision and how it affects the behaviour of the produced particles.
Therefore, the detector components mostly focus on measurements that allows to study 
the dependence of statistical properties of the final states with respect of variables that
correlates with the production of new matter states, i.e. the production of high energy density,
high temperature and high pression states.

Due to this, the strong point of ALICE with respect to the other LHC experiments is the
impressive particle identification, in order to identify relevant particles immersed in
high multiplicity events. The limitations that this impose is the reduced coverage for
each type of particle and the lack of symmetry in the detector: more types of different
subdetectors covering different solid angle regions.
This makes that the muon coverage is limited to the forward region ($2.5<\eta<4$) while electrons
and photons are detected centrally ($\eta<0.9$). 

The specific design of the ALICE detector makes the results from ATLAS and CMS also very
atractive for heavy-ion physics, due to its complementarity to ALICE, although they are
not in competition when the particle identification is a key part of the study, as
we will discuss later in this report.

\subsection{Data adquisition and event reconstruction at the LHC experiments}

The data-acquisition~(DAQ) systems of the experiments have been designed to collect the information
of the collisions happening at the LHC. They are very sophisticated in order to efficiently collect
the information from all the detector components and store it to tape for future analysis.

On the other hand, the DAQ need to deal with the problem that having collisions
every 50~ns~(or 25~ns in the future) it is impossible to store all or even part of the information
for every single event. For that it is needed to have an automated decision system which selects
the events as soon as they are produced in order to reduce the amount of data that is physically
stored to a manageable level. This system, called \emph{trigger}, has therefore the goal of reducing the
rate from tens of MHz to hundreds of Hz, providing data of 100 MB/s, which is a storable quantity.

Although the concept is simple, it should be noticed that events that are not accepted
by the trigger are lost forever, implying a big responsability. Additionally,
the trigger conditions at the LHC are very challenging and represent a new frontier in data acquisition
due to high rates and event sizes. However, there is the need for those required rates ans event sizes
since the aim of the experiments is to study rare processes with high precision, even at the cost of
suffering at the DAQ level.

In addition to the DAQ challenges, other difficulty arises from the high rate: since the collision
cross section is so large, it is very likely that several proton-proton pairs collide in the same
event (i.e. crossing). Most of the collisions are soft uninteresting collisions that would appear
at the same time as interesting ones.
This situation is usually referred as \emph{pile-up} collisions and it complicates the reconstruction
of interesting events since it becomes harder to distinct them from usual background, something
that is specially dramatic at the trigger level. The reason underneath being that reconstructed quantities,
specially the global ones like the $E_T^\text{miss}$, are modified and led to misleading values.

This problem with the \emph{pile-up} is what motivated the luminosity leveling at the LHCb interaction point: to
avoid the deterioration of the performance due to the overlap of collisions. Since statistics is not
really the issue due to the large cross section, it is more practical to reduce the collision rate to
collect higher purity events than just reject good events due to trigger limitations. It should be noticed
that a similar idea may be required for the other experiments in the future when running at the highest rates.

After the data has been collected and stored in tape, it is analyzed to investigate the
characterization of the physics producing it. The analysis consists on the identification 
and quantification of the
objects contained in the event. 

We have already described how to reconstruct the $E_T^\text{miss}$
quantity that allows to associate undetected particles to the event. Additionally we also
described how the reconstruction of the final state may be simplified with the use
of the concept of \emph{particle
flow}. 

As a specific case of the later, the presence of leptons in the final state is a fundamental
tool in a hadron collider to recognize important physic events. Electrons are identified using the
properties of its interaction with the calorimeters. Muons are identified using the chambers
specifically designed for its detection, using the property that they are charged and highly penetrating.

Photons are also identified using the deposits in the calorimeters, where they look similar to electrons, but
are distinguishable from them due to the absence of electric charge, and therefore the lack of hits in the
tracking system.

The $\tau$ leptons are the hardest objects to identify in a detector, but their use is strongly motivated
by their common presence in final states for BSM physics, or for Higgs searches, as we will see later.
Their leptonic decays are hard to distinguish from electrons and muons, but their hadronic decays, the
dominant ones, are separated from other hadron production due to their low multiplicity
and the kinematical properties. The main issue is that is commonly hard to separate them from the
large background of hadron production, and specially at the trigger, where the usable resources
are more limited. On the other hand the experiments at LHC has used experience at previous
colliders to really exploit all the possibilities of analysis with $\tau$ leptons, as it is described below.

Finally, apart from leptons and photons, it is very common the production of hadrons. They are originated
from quarks and gluons that are not observed because the strong force confines them within colourless hadrons.
The mechanisms transforming those coloured particles into hadrons cannot be understood in the pertubartion
approach used to perform estimations from the theory, but fortunately they can be treated in such a way
that their effects do not affect too much the predictions. The simpler technique to reduce this
effect is by using \emph{jets} of hadrons to reconstruct and characterize the final states.

The idea is that the processes that are not perturbately calculable occur at energy scales that are
much lower than the usual hard processes taking part in the LHC collisions. Therefore they do not modify
sustantially the global topology of the event and hadrons appear as collimated bunches of particles
that are kinematically close to that of the hard partons produced in the event.

This qualitative description, only valid for studies of hard parton production, should be quantified
with the use of a specific and well-suited algorithm that reconstruct the jets. The results are
usually dependent on the algorithm, but when the same algorithm is used for comparing measurements
and theory, the conclusions are independent of the algorithm, if the application is sounded.

Data analyses at the LHC experiments are performed with all these objects:
 leptons, photons, $E_T^\text{miss}$, hadrons and jets,
with very
satisfactory results, mostly due to the high quality of the data acquisition and reconstruction.

\section{Measurements to rediscover the SM} 

As mentioned above, the aim of the LHC is to produce unknown particles
and increase sensitivity to new possible interactions by colliding protons at high
energies. However, on top of the possible interesting processes there are other SM-related
processes that tend to hide the most interesting ones. For a hadron collider, QCD jet production
has a so large cross section that is the basic process happening in the collisions.

In fact, this makes the LHC a QCD machine aiming for discovery. Independently of what is
actually done, everything depends on QCD-related effects: parton radiaton,
parton distribution functions~(PDFs) of the initial-state protons, hadronization processes
for the final-state partons and so. Unfortunately most of these cannot be calculated
due to our limited knowledge on how to deal with the QCD theory and therefore, in
order to understand them requires the realization of measurements which allow to refine
the existing phenomenological models used to obtain predictions on what to expect in the
proton-proton collisions at the LHC.

For this reason it is
impossible to simply ignore the ``less interesting'' events which are considered as background
of the events containing effects and particles beyond the SM. In fact, at the LHC, as in
any other hadron collider, the understanding of QCD is not just something needed nor a priority: it is the only
possibility.

As a good example, it is needed to realize that the first measurements performed at the LHC
are the total cross section and the differential cross sections for producing charged particles.
They are not calculable in the pertubative approach of QCD, but they are required to perform realistic
predictions~(via the \emph{tunings} of the model generators). They were performed
at the beginning of the collisions by all 
the experiments~(see e.g. ~\cite{bib:alice:particleproduction,bib:atlas:particleproduction}) and
from the beginning have become important tools to understand the collisions at the several energies
the LHC has been operating.

In addition, even in these preliminary studies the LHC experiments proved that the LHC is crossing
the lines to a new regime: an interesting effect observed looking at the correlations
between charged particles: CMS observed~\cite{bib:cms:theridge} that in addition to the
usual \emph{large $\Delta\phi$} correlations (i.e. opposite hemispheres), there are
additional \emph{near-side} (i.e. small $\Delta\phi$ and large $\Delta\eta$) correlations
in events with very high multiplicities, specifically with more than 100 produced charged particles.

Figure~\ref{fig:theridge} shows the mentioned observation of the so-called ``rigde''. Similar effects
were observed previosly in heavy-ion collisions, although it is not completely clear the source
of them is the same. Currently there is not a clear explanation of the source, but the LHC
data has confirmed its presence in lead-lead and proton-lead collisions,
see e.g.~\cite{bib:cms:ridgeplead}.

\subsection{Studies of jet production at the LHC}

Apart from these soft-QCD measurements that are a fundamental piece to adjust the phenomenological
models, measurements related to hard QCD are also performed at the LHC experiments in order to
validate the QCD expectations on the perturbative regime, and to learn about the interactions
between partons at the shortest distances and also about the partonic content of the proton.

Measurements are done for inclusive jet production, as those by ATLAS
in~\cite{bib:atlas:jetproduction}, and compared
to the NLO predictions, which are able to reproduce the data after soft-physics corrections (that are
not large). Some kinematic regions are sometimes off, but they are correlated to problematic
areas, in which proton PDFs are not well known or the effects from higher orders or soft physics
are large. Similar conclusions are drawn from studies of multijet production, in which the
sensitivity to QCD is enhanced using ratios, as the three-to-two jet
ratio by CMS~\cite{bib:cms:multijetproduction}, in which many uncertainties cancel and the senstitivity
to QCD shows up via the emission of hard partons.
In fact the direct sensitivity to the strong coupling constant, $\alpha_S(Q)$, 
allows a measurement of this
value for the first time beyond 400~GeV, confirming the expectation from the running of that coupling.

With a different aim, instead of measuring quantities that are more accurately known, there
is interest in measuring in regions where uncertainties may be larger, but sensitive to
unknown quantities, as it is the case of the PDFs. Measurements at the
LHC experiments~\cite{bib:atlas:pdfjets,bib:cms:pdfjets} are sensitive to PDFs in regions
where they are not well constrained and able to distinguish between prediction of
different sets. Specially useful for the high-x gluon and sea quark PDF which is loosely constrained
by the HERA data.
It is worth to remark that even if the LHC aims for discovering of BSM physics, it is a very useful
machine to increase the knowledge about the internal structure of the proton, via the 
measurements sensitive to the PDFs. In incoming sections this will be mentioned a few times.

When studying the production of jets, an important topic by its own is the measurement of
production of heavy-flavour (charm and bottom) jets. Since they are not present in the
proton in a sizable way, its study provides important information about QCD, specially for
specific flavour production, something which is not possible for the light quarks and gluons.
The fact that it is possible to perform separated studies for charm and bottom jets is
due to the possibility of tagging the jets as originating/containing a heavy-flavour quark.

This has been a recent possibility due to the improvement in tracking, specifically at the closest
distances to the collision. After surpassing the challenges involved in the LEP and
Tevatron experiments, the detectors have reached to possibility to reconstruct vertices so
precisely, that resolving secondary vertices coming from ``long lived'' hadrons containing
a bottom and a charm quark has become a standard tool in accelerator physics.

The fact behind this
\emph{heavy-flavour tagging} is 
the presence of hadrons that live long enough so their decay products appear
in the detector as displaced tracks and vertices within jets that are incompatible with
originating at the so-called primary vertex, in which the interaction took place. These displaced
tracks and vertices are resolved and conveniently used to tag jets containing these heavy-flavour
hadrons and therefore likely to originate from a charm or bottom quark. 
The information provided by them is used either on a simple and straightforward way (that is safer
and more traditional) or on multivariate techniques that allow to increase performance of the
tagger. The later has become more popular as expertise with this kind of tool is well established.

\begin{figure}
\begin{minipage}{0.48\linewidth}
\centering\includegraphics[width=1\linewidth]{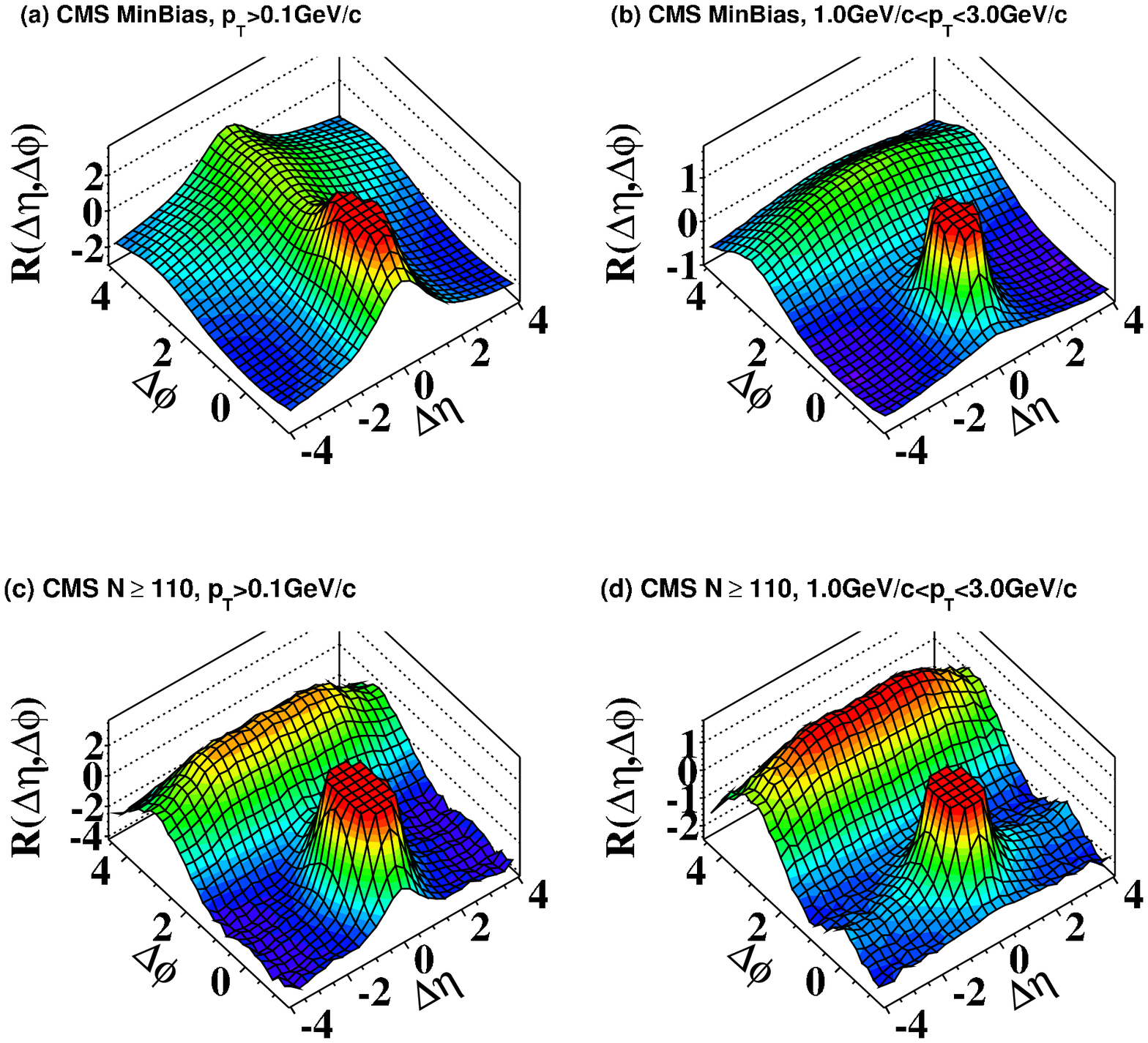}
\caption{Relative distribution of the charged particles in proton-proton at 7~TeV as measured by CMS
in several selections 
in the $\Delta\eta-\Delta\phi$ plane. Apart from the expected back-to-back correlation, a near-side
correlation is observed even at large $\Delta\eta$ for high-multiplicity events~(plots below).}
\label{fig:theridge}
\end{minipage}
\hfill
\begin{minipage}{0.48\linewidth}
\centering\includegraphics[width=.94\linewidth]{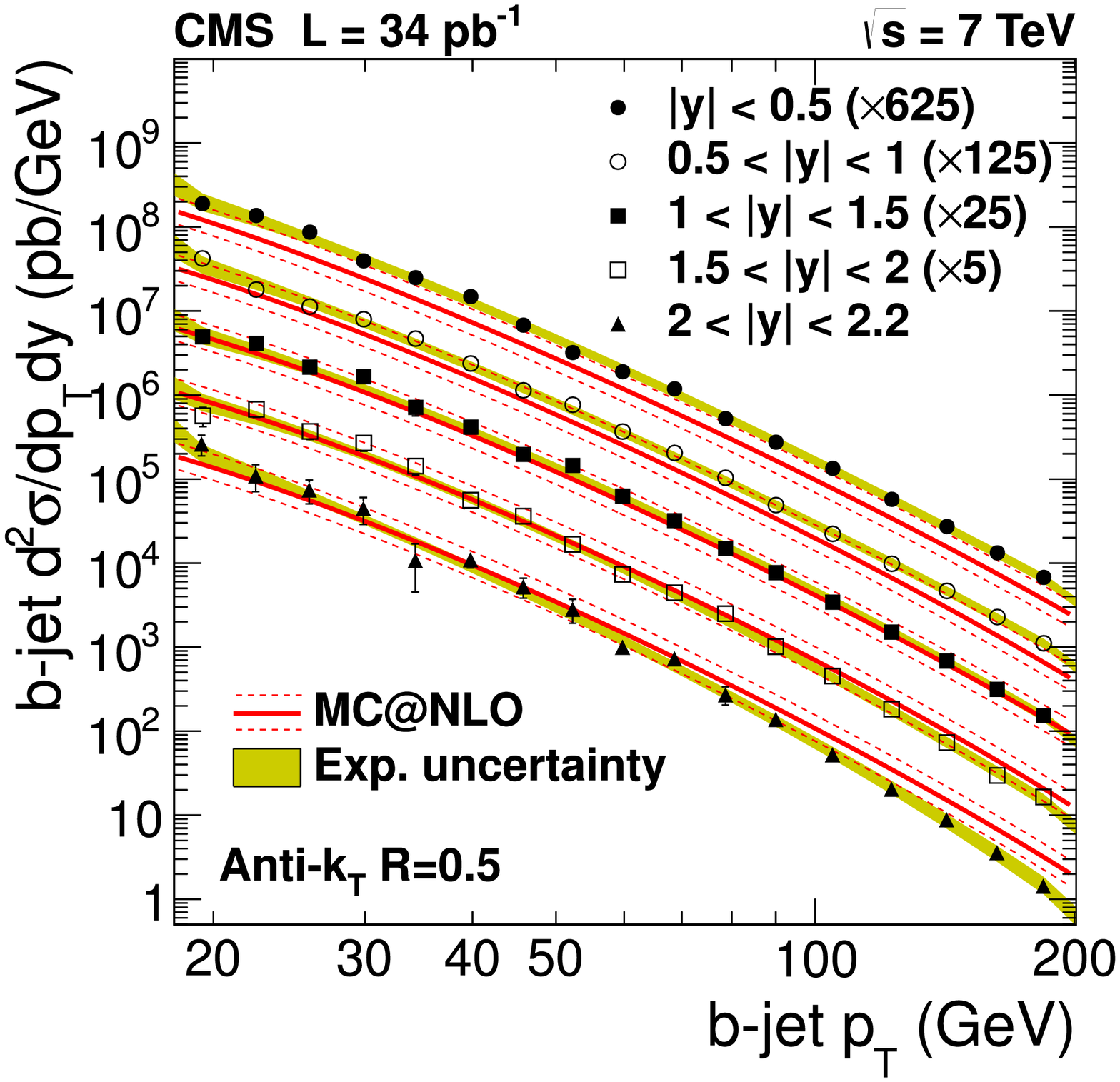}
\caption{Differential cross section of bottom jet production at the LHC with a
center of mass energy of 7~TeV. Measurements in different rapidity regions~(dots) are plotted
as a function of the transverse momentum of the jet and compared to the MC@NLO predictions~(lines).}
\label{fig:bjets}
\end{minipage}
\end{figure}

Making use of the tagging tools it is possible to study the production of jets originating from
a bottom quark, or b-jets. Measurements by the two collaborations has been
made~\cite{bib:atlas:bjets,bib:cms:bjets} and compared to QCD precitions for heavy-flavour
production computed with the MC@NLO~\cite{bib:mcatnlo} program. As shown in Fig.~\ref{fig:bjets}, a
good agreement is observed overall although there are some small discrepancies in specific
kinematic regions, similarly at what was observed in inclusive jet production. It should be noticed
that the level of agreement is good due to the improvements in the theoretical calculations
during the last decade. Predictions are difficult for the kind of process under study, so the
level of discrepancy observed is considered a complete success of the QCD calculations. Of course
further work is still needed, emphasizing the importance of the precise measurements at the LHC.

In a similar topic, one important measurement at the LHC experiments will be to try to
disentangle the production of jets containing two heavy-flavour quarks. In the past the
quality of the heavy-flavour tagging only allowed the separation of jets with at least
a heavy-flavour quark. However, at the LHC, the improved detection techniques and the
experience with tagging tools will also allow to investigate the production of multi-b jets,
which are of importance in topologies with merged jets or to reject the presence of
gluon jets containing a gluon-splitting process into heavy-flavour quarks.

Exploiting the subtle differences in the displacement of tracks, studies are performed
on this issue~\cite{bib:atlas:bbjet}, and good rejection power 
of gluon jets has been observed while keeping a big fraction of the single b jets. More dedicated
studies will be needed to improve the related tools for rejecting this background, but
current results has confirmed its feasibility and also that the heavy-flavour taggers at
the LHC experiments are taking advantage of the improved detector capabilities.

Regarding the LHC in a new kinematic regime, it should be remarked the development
during the last years of tools to investigate the production of boosted objects. Since
available energies at the LHC are much larger than the masses of the SM particles, it
is likely to observe their production with very large transverse momenta, giving rise
to the merging of objects.
This is specially worrisome in the case of jets since they are hard to separate after
their constituents have been merged together. For that reason, several dedicated
studies and the development of new techniques has been done at the LHC
experiments~\cite{bib:atlas:boostedjets,bib:cms:boostedjets} in order to deal with
the topology of boosted jets. The idea is to exploit the properties of the internal
structure to recover information of the original partons whose jets have
been merged, and separate them from single parton jets that are boosted in the
transverse direction, i.e. produced with large transverse momentum.

Many techniques have been developed and tested in the identification of merged jet
and check how the simulation reproduce the characteristics of the jets allowing the
distinction of the jets containing one or more ``hard'' partons. Currently its performance
has been proven to identify merged jets coming from boosted W bosons and top quarks, 
and used for searches. However its principal motivation is still the need of this
kind of tools for the future
running at higher energy.

\subsection{Studies of soft QCD physics at the LHC}

Apart from particle and jet production via QCD processes, the experiments are able to
perform studies related to QCD via more complicated mechanisms. Among this, one that has
become really important is the possibility of observing more than one partonic collision
from the same protons. Since a proton is a bunch of partons it is not uncommon to have
several partons colliding at the same time. And the LHC allows to have very hard collisions
since the energy of the protons is very large.

These multiparton interactions are a complicated topic since it is not clear 
up to which level 
each collision can be considered independent of the others. In addition, the probability
associated to the additional collisions to happen is not calculable and require models whose
parameters require some tuning in order to improve the modeling of the underlying event.
The validity test of the models is usually done in samples that are reasonably understood
and trying to extract the maximum possible information to get the proper parameterization. 
With this aim, ATLAS has measured the contribution from double-parton interaction
for W+dijet events~\cite{bib:atlas:mpi} to be $0.16\pm0.01\text{(stat)}\pm0.03\text{(syst)}$,
in good agreement with the expectations that were tuned to previous data.

Related to QCD in strange regions, the LHC allows studies for diffractive and forward
production of particles and jets at higher scales than previous hadron colliders. These
are relevant in order to understand hadron interaction at softer scales, and also
to adjust the models describing this kind of process.

Even the LHCb experiments has produced results for forward hadron production, which are very
competitive due to the optimization of the detector for particle ID and its very forward
coverage. Results of these studies\cite{bib:lhcb:forwardparticles} have been compared to
the predictions obtained by traditional event generator and also those used in the
simulation of cosmic-ray events, which are very sensitive to this kind of processes.

Another example of new kind of QCD measurements is the study of exclusive diboson~(WW)
production via the collision of photons performed by CMS~\cite{bib:cms:gammaww}. This makes the LHC a
photon collider at high energies, which allows dedicated studies of the electroweak
interaction.
The result with the dataset collected at 7~TeV allows to measure the cross section with
still a low significance, implying the need of more data. However, using the sample
with highest transverse momentum, it was possible to set limits on the production
via anomalous quartic couplings, showing the potential of this kind of studies.

\subsection{Electroweak boson and diboson production at the LHC}

Although the measurements described above allow to test the predictions by QCD and even of
the electroweak sector in some cases, the most sensitive studies to validate the SM predictions
are coming from the events containing photons or weak bosons. The idea is that these events
are usually simple to recognize and the perturbative calculations of the processes and the
backgrounds are usually very accurate.

The most common process of this kind is the production of photons, whose interest have been
demonstrated in the past hadron colliders, in which this was considered a ``QCD study'' since
it provided direct information on the quarks. Hard photons radiated from quarks are good probes
of the interation since they are not affected by soft processes and they are able to distinguish
mong different kind of quarks. In addition the large cross section of the $\gamma$+jet allows its
use as a fundamental calibration tool.

Additionally, studies of diphoton production yield to very stringent test of the SM predictions,
specially for a final state that is an important background in many interesting searches
of new particles, decaying in photon final states. The study by ATLAS~\cite{bib:atlas:diphoton}
performed measurements of the photon pair production as a function of several variables and
compared them to several event generators, at different orders in QCD and types of partonic
showers in order to evaluate the level of performance of the available production tools.

\begin{figure}
\begin{minipage}{0.48\linewidth}
\centering{\hspace*{-1.cm}\includegraphics[width=1.1\linewidth]{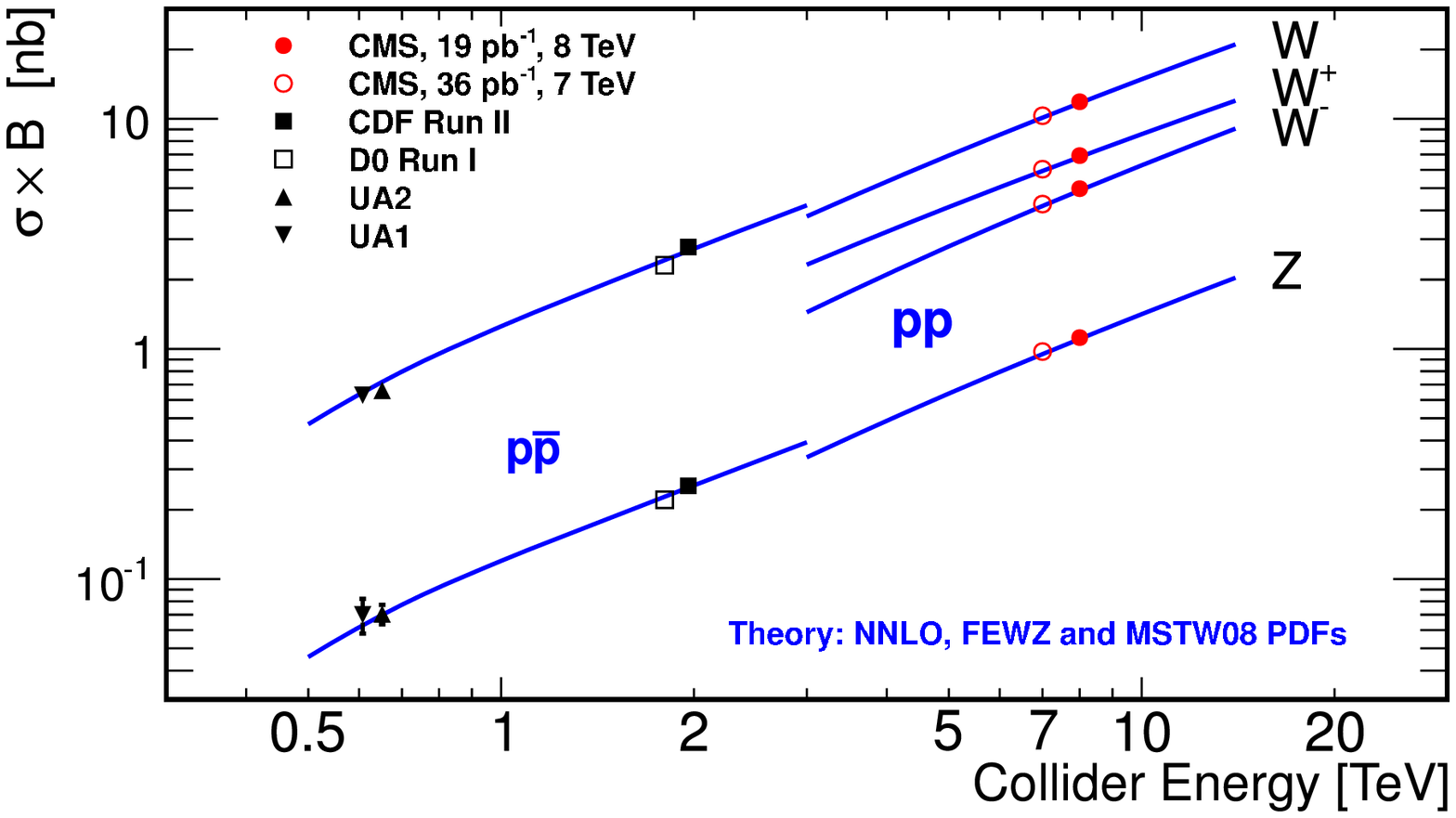}}
\vspace*{0.3cm}
\caption{Cross sections of weak boson production in hadron colliders at several center-of-mass
energies. SM predictions for proton-antiproton and proton-proton collisions are compared to
the measurements shown as different types of dots.}
\label{fig:wz}
\end{minipage}
\hfill
\begin{minipage}{0.48\linewidth}
\centering\includegraphics[width=0.95\linewidth]{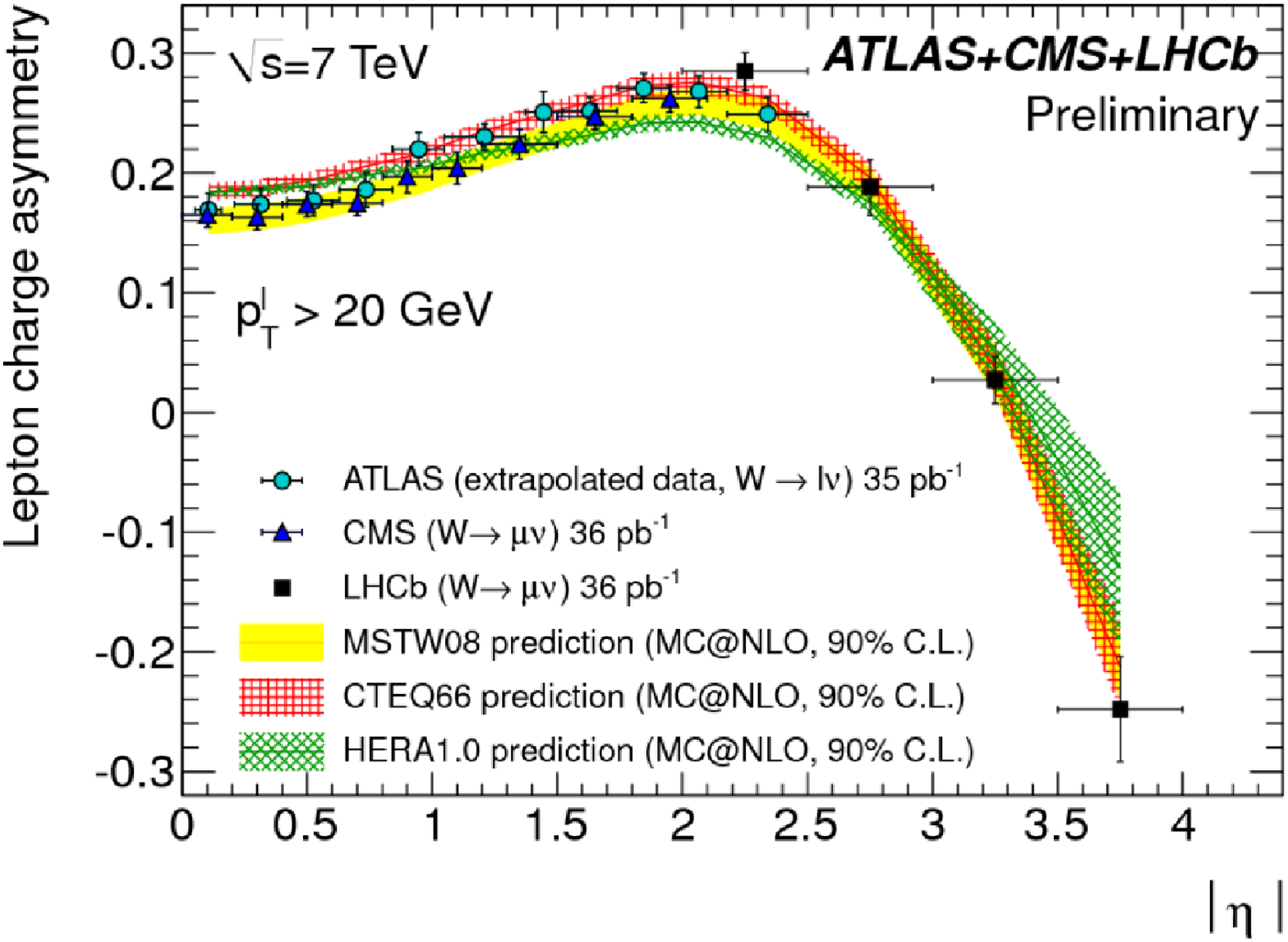}
\caption{Lepton charge asymmetry
of W production as a function of $\eta$ at the LHC with the 7-TeV data
from the ATLAS, CMS and LHCb experiment. The inclusion of the later
experiment allows to extend the measurement to very forward regions never
reached before.}
\label{fig:wasym}
\end{minipage}
\end{figure}

However, when talking on boson production, the studies related to the weak bosons become
a fundamental test of the SM predictions that were performed at the LHC in order to also check
the performance of the detectors and tools for analyses. Even after the first analyses, 
the studies of events with W and Z bosons are fundamental tools for calibration and understanding
of the object identification and reconstruction.
Measurements at several energies, as the one at 8~TeV by CMS~\cite{bib:cms:wandz}, 
have been performed and show very
good agreemeent with the expectations by the SM and also confirming the
excelent predictions of the SM at several energies for measurements
performed for W and Z production during the last three decades, as shown in Fig..~\ref{fig:wz}.

Although the basic goal for studying the production of weak bosons is to confirm
the performance of the detectors and of the basic SM prediction, dedicated measurements
related to them are also a fundamental part of the LHC program. This is the
case for measurements sensitive to the internal structure of the proton and also of the
SM details that could not be tested before at the level of precision reachable
at the LHC. This affects both kind of processes: final states that were never available
in a proton-proton collider before, like the ratio of W$^+$ to W$^-$ measured
by ATLAS~\cite{bib:atlas:wratio}, or whose yield was too small, like the measurement
of $\text{Z}\rightarrow 4l$~(as in~\cite{bib:cms:z4l}) which is a calibration piece for the Higgs
searches.

This explains the large effort at the LHC to measure the properties of the production of weak
bosons. Some of the properties are measurements for confirmation and 
validation purposes, but some are really motivated by the new possibilities opened
at the LHC experiments.
This is seen even in experiments that are not intended for boson studies,
like the results at LHCb, in which the very forward detection makes measurements of
Z and W production very competitive even with lower acceptances~\cite{bib:lhcb:zprod},
since they are measured in kinematic regions that are not available for the main detectors.
Even events compatible with forward Z bosons decaying into $\tau$ leptons have been
observed at the LHCb~\cite{bib:lhcb:ztautau}, indicating an important benchmark
for the performance of the experiment to obtain results beyond flavour physics.

In the case of W production, Fig.~\ref{fig:wasym} shows the lepton charge asymmetry
as a function of $\eta$ also confirms the complementarity of the several experiments
at the LHC, in this case how the LHCb is able to extend the region reachable by the
ATLAS and CMS, even with a reduced yield.
All these measurements of forward production will have a big impact in the fits to extract
the parton content of the proton, since most of the current uncertainty is reduced
by forward production of particles, more sensitive to the less constrained partonic
content, as gluons and sea quarks at high-x.

But not only the proton structure benefits from the large yields at the LHC for
producing weak bosons since the presence of a massive object allows studies
of QCD processes in an environment where perturbative calculations are accurate
enough to bring very stringent tests of the expectations.

The typical example is the use of bosons as ``probes'' of the underlying hard
process involving the partons, whose rules are naturally dictacted by
strong interactions. This is the case of the measurment of jet production
in association to a Z or W, as in~\cite{bib:atlas:wjets,bib:cms:zjets}, which
are sensitive to the partons interacting and also major backgrounds to most of
the new models for BSM physics. The measurements are able to constrain 
the room for the new physics, and, in other kienmatic regions, to check the validity
of the tools used to estimate these final states. It should be noted that not only the
yields are interesting, but also the kinematic distributions of the final state objects,
specially those sensitive to unexpected underlying physics, as
in~\cite{bib:atlas:wandzratio,bib:cms:wreson}, in which specific distributions of
bosons and jets are studied in order to perform accurate tests of the SM predictions,
taking advantage of the large yields.

Similarly, another topic that directly benefits from the high cross section and
luminosity at the LHC is the production of heavy flavour quarks in association
with a weak boson. Being very sensitive to the SM structure, some of the processes
have not being
accurately tested due to the limited statistics at previous colliders. In fact,
results at the Tevatron have been controversial regarding the way the event
generators reproduce the measurements. The larger statistics at the LHC 
allows the improvement in the precision
of the measurement. This is the case for the W+b-jet measurement by
ATLAS~\cite{bib:atlas:wbjet}, which clearly shows that description by event generators
could be improved, which is not a trivial case, since it is a background for many
studies for BSM physics. Understanding this discrepancy should be a clear priority
of the physics at LHC, from the theoretical and experimental point of view.

Another final state that has benefit a lot by the new frontier set at the LHC is the
production of charm in association with a W boson. Its interest is given by the fact
that since W is able to change the flavour of a quark, the production of single charm
is dominated by interactions involving down and strange quarks in the proton. Therefore
directly sensitive to the strange content of the proton. In addition, the charge of the
produced W is completely correlated to the charge of the charm and down/strange quark.
As mentioned above, the W is used as a direct probe of the structure of the
underlaying parton collision. In this case the result of the measurement by
CMS~\cite{bib:cms:wcharm}
is presented
as the fraction of charm jets in W+jet events and also of the ratio of W$^+$ to
$W^-$ in events with a charm produced in association with the W.
Both quantities are sensitive to the PDF of the strange quark and antiquark. The
measurements are
in good agreement with the expectations and they will allow to improve the accuracy of
the proton PDFs.

In the case of the Z boson, the low cross sections prevented detailed studies
of the production associated with heavy flavour quarks to be performed at the Tevatron. Again the LHC
has brought the possibility to study this in detail. The analyses studying the production
of Z+b-jet, as in~\cite{bib:cms:zbjet}, show that the event generators,
in this case MADGRAPH~\cite{bib:madgraph}, are able to describe
the distributions. However, with the explicit requirement of two b-jets the agreement
get clearly worse~\cite{bib:cms:zbjets}, implying that some theoretical work may be required:
although the processes (and calculation diagrams) are the same, the relative weight
is different due to the kinematic requirements on the second jet. 

Finally, the last topic entering the scene when talking about weak bosons and jets is
the study for electroweakly produced bosons, the so-called \emph{Vector-boson fusion}~(VBF)
production. In this case the boson is produced in association of two jets that tends
to be forward, due to the kinematics. Those forward jets are used to ``VBF-tag'' the event
and separate them from the main processes, weak radiation from partons 
or parton annihilation. Measurement by CMS~\cite{bib:cms:zvbf} allowed to measure 
a cross section in agreement with NLO calculations. In addition, this
kind of analysis also contributes to understand the production of jets in the forward
region, which is less understood due to the challenges in experimental studies and
also in theoretical calculations.

It should be remarked that the interest of all the results involving jet production in
association with weak bosons will be kept in the future, as the measurements get more precise,
implying larger challenges for the modeling of very important processes at the LHC, either for their
own interest or just as background estimations for searches of all kind.

\subsection{Diboson production at the LHC}
\label{ssec:dibosons}

As it is well known,
the production of more than one boson is one of the most sensitive test of the non-abelian
structure of the electroweak sector of the SM,
so it is very sensitive to deviation produced
by new couplings involving the SM bosons.

The main limitation is that precisely the presence of several weak couplings makes the cross section
small, and the observation of these final states has been very difficult. However, the LHC has
open a new era for this kind of studies since large samples are available to perform detailed studies,
allowing precise studies of diboson production for the first time. In fact, the LHC will allow
in the future the observation of multiboson production, which has never been observed.
In addition, the large samples available has allowed that diboson production has become a standard
reference for calibration in advanceed analyses.

The basic processes testing the SM structure and with large cross section is the production of
a weak boson and a photon (W$\gamma$ and Z$\gamma$) which are directly sensitive to the unification
of the electromagnetic and weak interactions. The results of the analysis,
like~\cite{bib:cms:wgamma}, shown that data are in good agreement with expectations, even at higher
transverse momenta, which may be sensitive to new physics affecting the unification of interactions.

In the case of two massive boson, the process with the highest cross section is the production
of two W bosons, in which the samples are large enough to allow detailed comparisons with
the predictions by the event generators, even via differential distributions~\cite{bib:cms:ww}.
The conclusion of the studies is that the SM predictions reproduce very well the shapes of the
observed distributions in data, but they underestimate the total cross section.

This discrepancy
has been observed by the two collaborations and at the two energies of the LHC. Investigation
of the origin of it is under study.
Similarly, studies of the production of two Z bosons shows a slight excess in the data with
respect to the expectations~\cite{bib:cms:ww,bib:atlas:zz}. In this case, the yields are
small and the excess is not as significant, but the clean final state, requiring four isolated
leptons, leads to very straightforward conclusions. This channel, which leads to a pure sample
of ZZ events and with fully reconstructed kinematics, 
provides the best test bed for diboson studies, specially with the amount of events expected
at the LHC.

In addition to the pure leptonic channels, that are much cleaner in a hadron collider, the
semileptonic channels are also exploited at the LHC, since it is the most precise way to
study the hadronic decays of Z and W bosons, not available in the inclusive production due to
the large dijet backgrounds.
The performed measurements in the W+dijet sample~\cite{bib:cms:wreson,bib:atlas:wreson} yield
the observation of the diboson signal. Separation of the Z and W in the hadronic channel is not
possible due to resolution, and therefore this final state is able to measure
the mixture of WW and WZ events. The result is in agreement with the observation, and the analysis
has also tested the W+dijet background, whose interest was mentioned above.
Finally it should be remarked that WZ has been also measured in the fully leptonic
channel~\cite{bib:atlas:wz} which provides the topology of three charged leptons and $E_T^{\text{miss}}$
which has a large relevance in searches for new physics, in particular supersymmetry, and therefore
the understanding of the kinematics in this diboson process is a fundamental part of the
program.

\begin{figure}
\centering\includegraphics[width=.65\linewidth]{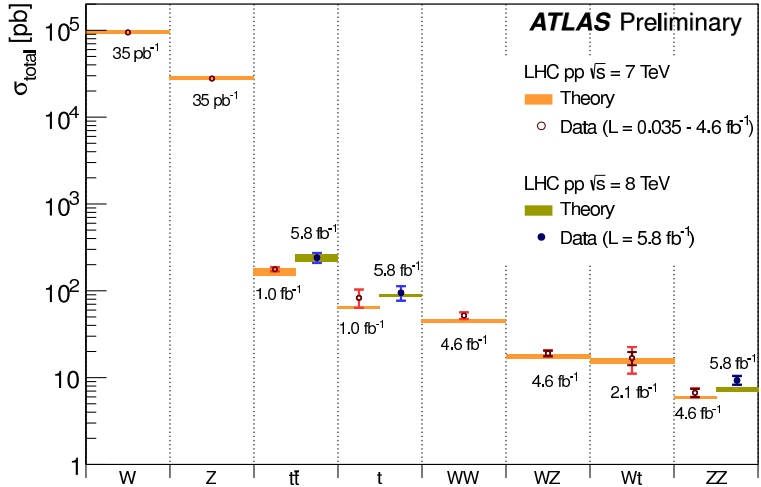}
\caption{Summary of the measurements by ATLAS for massive particles (weak bosons and
top quarks) in single and double mode at the LHC with a center-of-mass energy of 7~TeV.}
\label{fig:smsummary}
\end{figure}

In conclusion, it should be remarked that even if the LHC is intended to discover the physics
beyond the SM, measurements of the know processes has produced many interesting results, some
to confirm the observations at previous colliders, but also new results that were not previously
accessible. In this sense, and as summarized in Fig.~\ref{fig:smsummary}, the impressive
agreement of the measurements provides a solid base on which the experiments are building the
tools and confidence for the observation of unexpected results, when higher precision or
new final states are reachable in the data.

\section{Measurements on bottom and charm hadrons}

The spectroscopy of hadrons has been a fundamental source of information in particle
physics, since it has allowed to detect effects beyond the reachable energy scale 
and since it provides the only direct way to understand quarks and QCD at low energies.

The case of heavy flavour hadrons, which include at least a bottom or charm quark, is
of a broader interest due to the higher masses involved that allows to perform more
accurate theoretical calculations related to the properties of the hadrons.
With the measurements in hadron spectroscopy, it is possible to perform several classes
of studies, as the properties of bound states, production of new states, measure
branching ratios and interference effects. All of them provide information about 
possible BSM physics or improve the knowledge about partons in confinement states.

It should be remarked that in order to perform studies with hadrons, it is needed to
reconstruct them. This sets a very different approach to the ones described above in which
the hadrons are just merged together in jets that are related to the original partons. The goal
in the physics with hadrons is to explicitly identify the interesting objects.
This is achieved in several steps: The first consists in the identification of
the detected particles, as
pions, kaons and more commonly muons and electrons. Some of these objects are
(pseudo)stable and are identified as tracks or similar. Sometimes the nature of the particle
is also inferred by using specifically designed detectors, but in other cases the nature
is just assumed as part of the reconstruction process.

After the detected particles are identified, they are combined to reconstruct ``mother''
particles that may have decayed into them. The usual method is to reconstruct the invariant
mass of several identified objects and find events in which they are coming from another particle (over a
possible continuous background) as a resonant excess. Those events associated to a decaying
particle may be used to extract information about the particle, apart from the direct
identification of the particle itself in the mass distribution. Furthermore, the particles
identified this way via its decay products may be further used to reconstruct other partental
particles in a recursive reconstruction that allows the full identification of the
decay chain of the original particle.

With these tools and the goal of measuring the hadron properties in mind, the LHC
experiments have been
able to identify hadrons, some of them completely unknown. One example is the
observation by ATLAS of the new exited state, $\chi_b$(3P), belonging to
the bottomonium family decaying into $\Upsilon$(1S/2S) by the emission of a
photon~\cite{bib:atlas:chi3p}. The mass distribution showing the resonances
produced by the new state is shown in Fig.~\ref{fig:chi3p} centered at
a mass of $10.530\pm 0.005 \text{(stat)}\pm 0.009 \text{(syst)}$~GeV.
Also the CMS experiments was able to find the
$\Xi_b^{*}\rightarrow \Xi_b^\mp\pi^\pm$ state, which has been the first baryon
and fermion found at the LHC, 
and with a mass of $5945.0\pm 0.7 \text{(stat)}\pm 0.3 \text{(syst)}]\pm 2.7 \text{(PDG)}$~MeV~\cite{bib:cms:xib}.

However, and as expected, it is 
the main experiment focusing in heavy-flavour physics, LHCb with its larger
samples with higher purity who is able to measure the properties of bottom hadrons with
higher precision. 
Specially about the recently discovered
baryons, for which this experiment has already relatively large samples with high
purity selection. The measurements for $\Lambda_b$, $\Omega_b^-$ and $\Xi_b^{-}$ documented
in~\cite{bib:lhcb:baryons} required very detailed understanding of the detector
momentum scales, in order to get the most precise mass measurements in the World.

Additionally the LHCb is also leading the effort in searching for rare decays of known
hadrons. These decays are of interest for its possible sensitivity to new interactions
involving quarks because they include loop diagrams or interesting vertices that could be
affected by unknown effects.
Among the rare decays, one of the most attractive ones is $B_s/B^{0} \rightarrow \mu\mu$ since
it is associated to a well-controlled and easily identifiable final state. Additionally,
the branching ratio is very small but expected to be enhanced in several of the possible
BSM extensions. This explains the intensive search for this signal in the last decade at
the Tevatron, where exclusion limit approached the SM expectation. However, the large
sample collected by the LHCb experiment allowed to get evidence of the decay, with a significance
of $3.5\sigma$, for $B_s$ that is in good agreement with the SM
value~\cite{bib:lhcb:mumu}. The decay
for $B^{0}$, searched in the same analysis, is also
in agreement with the SM, but significance of the excess is smaller. The absence of discrepancy
has set strong limits on possible new physics affecting the decay, confirming the
negative results from direct searches at the other LHC experiments, as described in
sections~\ref{sec:exotics} and~\ref{sec:susy}.

Another interesting decay under study is $B^{0} \rightarrow K^{*}\mu\mu$, whose
branching fraction in the SM is not that small but whose kinematics is sensitive to
the presence of new physics. One is the forward-backward asymmetry as a function
of the invariant mass of the muons, measured by LHCb~\cite{bib:lhcb:mumuk} and observed
to be in agreement with the SM calculations. 

All these measurements confirm the good performance of the detectors for heavy hadron physics,
although the measurements are not bringing information about the possible BSM physics, but
setting stringent constraints on the way the new physics may modify the interaction
between quarks.

\subsection{Mixing and oscillations}

Within the properties of hadrons, one that has become of large relevance is that of the mixing
of neutral mesons, in which the flavour eigenstates differ from the mass eigenstates, leading to
a change in its nature according to the quantum mechanics rules.
These oscillations are well steblished for the $K^0$, $B^0$ and $B_s^0$ and are starting to
become accessible for the $D^0$. 

\begin{figure}
\begin{minipage}{0.48\linewidth}
\centering{\hspace*{-0.4cm}\includegraphics[width=1.08\linewidth]{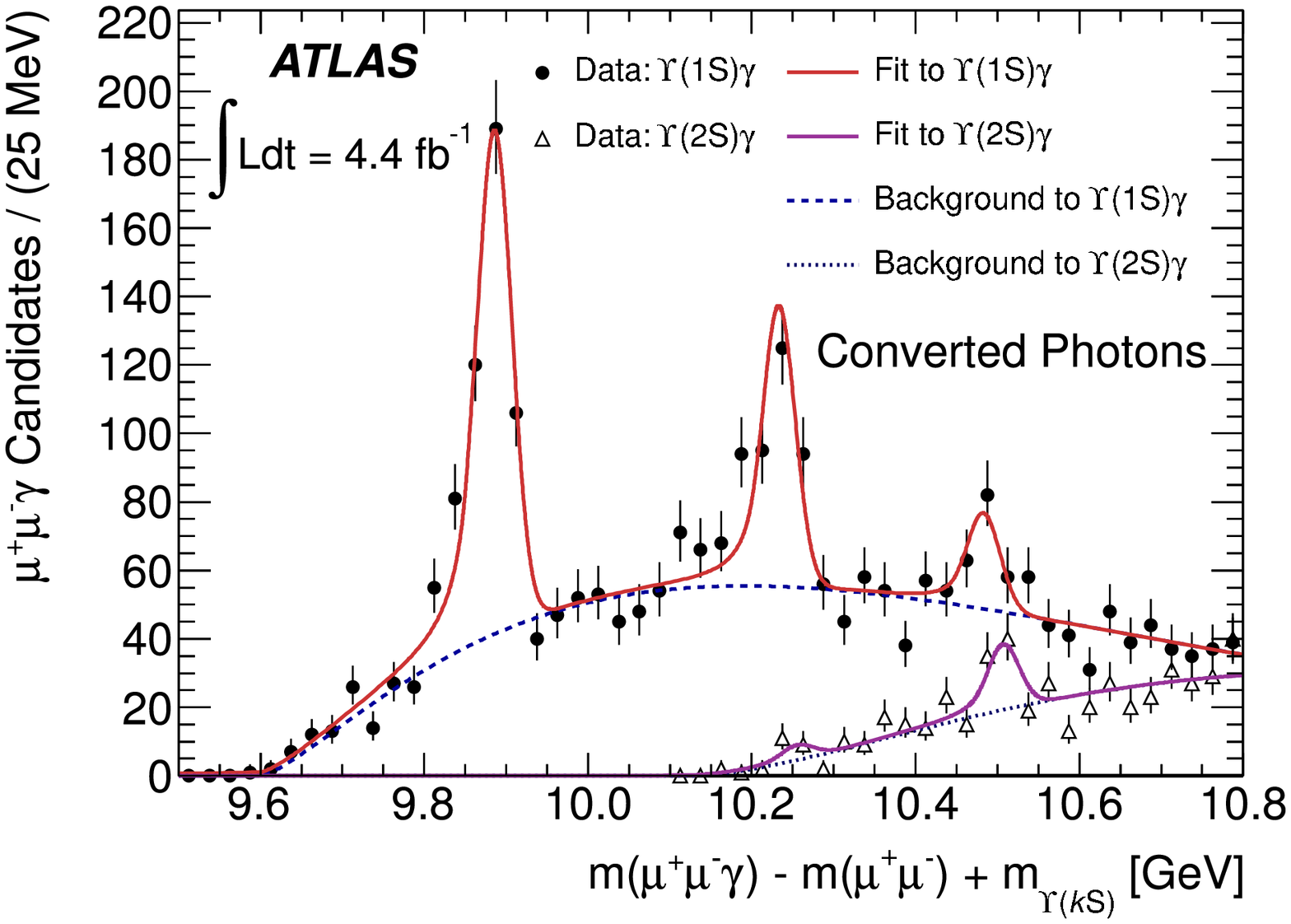}}
\caption{Invariant mass distribution of $\mu^+\mu^-\gamma$ to observe the resonances
decaying into $\Upsilon$(1S/2S) and a photon. A clear state at 10.5~GeV is observed
in both decays, compatible with being the $\chi_b$(3P) state of the bottomonium
family.}
\label{fig:chi3p}
\end{minipage}
\hfill
\begin{minipage}{0.48\linewidth}
\vspace*{-0.5cm}
\centering{\hspace*{-0.2cm}\includegraphics[width=1.08\linewidth]{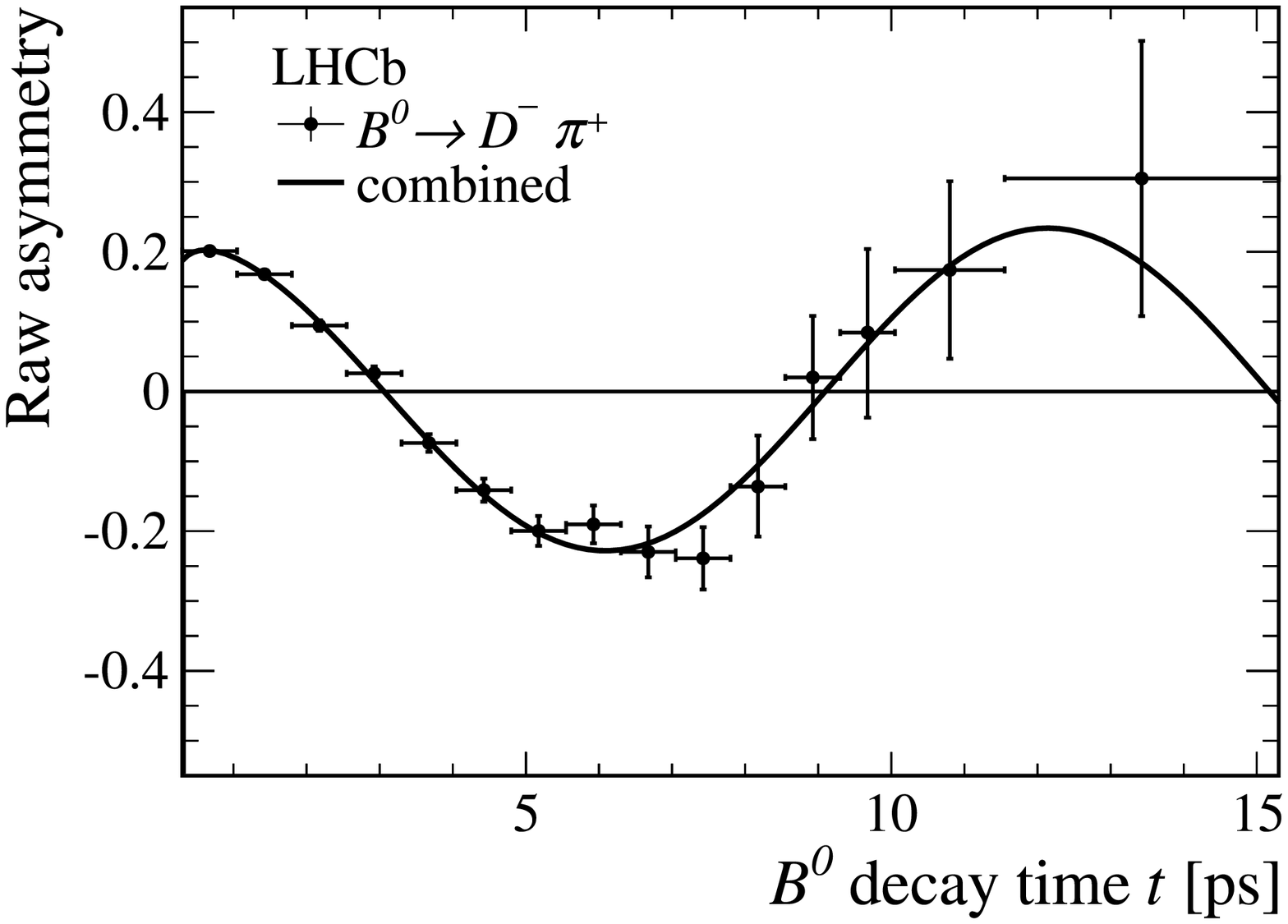}}
\caption{Raw mixing asymmetry for $B^0\rightarrow D^{-}\pi^{+}$ as a function of the decay time. 
The solid black line
is the projection of the mixing asymmetry of the combined probability function for the sample.}
\label{fig:mixing}
\end{minipage}
\end{figure}

In the case of the $B^0$, the LHCb samples are reaching
unprecedent precision and even providing new channels of observation. Figure~\ref{fig:mixing}
shows the result of the oscillations for the very pure sample of $B^0\rightarrow D^{-}\pi^{+}$
as a function of the decay time~\cite{bib:lhcb:b0mix}. As it can be observed, the
measurements are well reproduced
by the expectation obtained taking into account
the composition of the sample used to compute the raw asymmetry.

In the case of the $D^0$, the oscillations are now becoming accessible thanks to the
large samples, specially at the LHCb. Its study is strongly motivated since charm is the
only up-type quark in which mixing and CP violation are accessible. It can also provide
surprises since it is a previously unexplored region.
The study of the mixing and oscillations for the $D^0$ is done by exploiting the interference between
the mixing and the double-Cabibbo-suppressed decays. The same channel provide a right sign and
a wrong sign set of candidates that are used to perform the measurement. The first set is not sensitive to
the mixing and therefore provides a perfect reference sample.

In order to reduce uncertainties in the production, the initial $D^0$ state is tagged by using the
decay product of the $D^*\rightarrow D^0\pi_s$. Using all these events, it is possible
to measure the mixing and the LHCb has provided the first observation from a single
measurement, with a significance of 9.1$\sigma$~\cite{bib:lhcb:d0mix}. The result is
in good agreement with previous measurements, but the increased significance is another proof of
the reach available at the LHC even for studies of low-mass objects.

\subsection{Measurements of the CKM matrix and CP violation}

As remarked several times, the main goal of the studies in flavour physics is to investigate the details
of the fermion families, specially the relationship among them. In the case of the quarks, the
relation between
the flavour eigenstates (from the point
of view of the weak interaction) and the mass eigenstates.
is given by the so-called CKM matrix~\cite{bib:ckm}
which is expected to be unitary (when all families are included) and that can be parameterized
with three mixing angles and one complex phase. The unitary condition allows the representation
of combinations of elements in rows and columns of the matrix as a triangle whose area is related
to the CP violation in the family mixing. 

The goal is therefore to identify the processes that are sensitive to combinations of elements
in the matrix and extract the associated information about the matrix and the triangle. The
measurement of single elements in the matrix is associated to processes that are not observable
in hadron physics. However, that is not a complete limitation, as proben by the large set of
results in the last decade related to the CKM and CP violation parameters. Still, certain
measurements are newly coming from the LHC. As an example, the LHCb experiment has measured the
angle $\gamma$ using the tree processes $B^\pm \rightarrow D^0K^\pm$~\cite{bib:lhcb:gammangle}
which has the advantage of being very clean: as we mentioned before, processes with loops are
sensitive to new physics, so the values measured at tree level are dominated by SM-only physics.
The measured value, $\gamma=(71.1^{+16.6}_{-15.7})^{\circ}$ is in agreement with the World
average, with comparable uncertainty.

Other interesting result from the LHCb is the study of CP violation in charmless three-body decays
of B mesons~\cite{bib:lhcb:charmless}, that are sensitive to transitions between the first and third generation.
The observed asymmetry is interesting because it is opposite
in $\pi^\pm\pi^+\pi^-$ (enhacement for $B^-$)
with respect to $K^+K^-\pi^\pm$ (enhacement for $B^+$) and it seems to be enhanced locally for some
kinematics regions.

In the case of the mixing, one of the most important channel is $B_s\rightarrow J/\psi\phi$ since
it is sensitive to new physics affecting the CP violation. Measurements~\cite{bib:lhcb:bxmix}
agree with
the SM expectations, and they were also used to obtain the first measurement of the width difference
of the mass eigenstates which is not compatible with zero 
($\Delta\Gamma_s=0.116\pm0.018\text{(stat)}\pm0.006\text{(syst)}$~ps$^{-1}$).

Finally, the last open topic for CP violation is its study in charm decays, which has
been measured
by the LHCb collaboration~\cite{bib:lhcb:cpvcharm} to be significantly different from zero, an
unexpected result since most of the SM-based predictions suggest almost no violation. Although
calculations are difficult and the usual estimations may underestimate the value, the measured
value, confirmed at other experiments, seems a bit large, which may be pointing to some
BSM effects. 

As with most of the discrepancies observed, more data is needed to increase our knowledge, but
theoretical development is an additional requirement to quantify the level of disagreeement
observed and before its origin is further investigated.

\section{Results on the top quark}

In the hadron physics described in the previous section, one quark is not investigated: the
top. Being the most massive of the quarks (and of any observed
fundamental particle) it is hard to produce and also it does not hadronize
but directly decays into a W and a bottom quark. Additionally, its exceptionally
high value of the mass makes him the best candidate to be related to new physics, so
its study is mandatory and one of the big goals of the LHC program: the top quark may lead the
path to BSM physics, in the same way as neutrinos are leading the path in non-collider
results.

At the LHC the dominant process to produce a top quark is QCD pair-production that has a large
cross section. In fact the LHC is the first machine that is able to produce top quarks at
high rate, allowing detailed studies to be performed. This also applies to other production mechanisms,
as that of single-top and $t$W production, the latter being available at the LHC for the first time.
In fact the production cross sections of processes involving top are so large that it is also a 
very common background in many types of searches, which is an additional motivation for studying
its properties.

The study of the top quark at the LHC follows a similar strategy developed at Tevatron: channels are
identified with the number and type of leptons in the final state. Depending on that, events are analyzed
to extract all available information in a sample as clean as possible. Additionally all channels are
considered, in order to investigate all possible events and the presence of discrepancies with respect to the
SM expectations.

\subsection{Measurements of the top-pair production cross section}

\begin{figure}
\begin{minipage}{0.48\linewidth}
\vspace*{-0.6cm}
\centering{\hspace*{-1.cm}\includegraphics[width=1.1\linewidth]{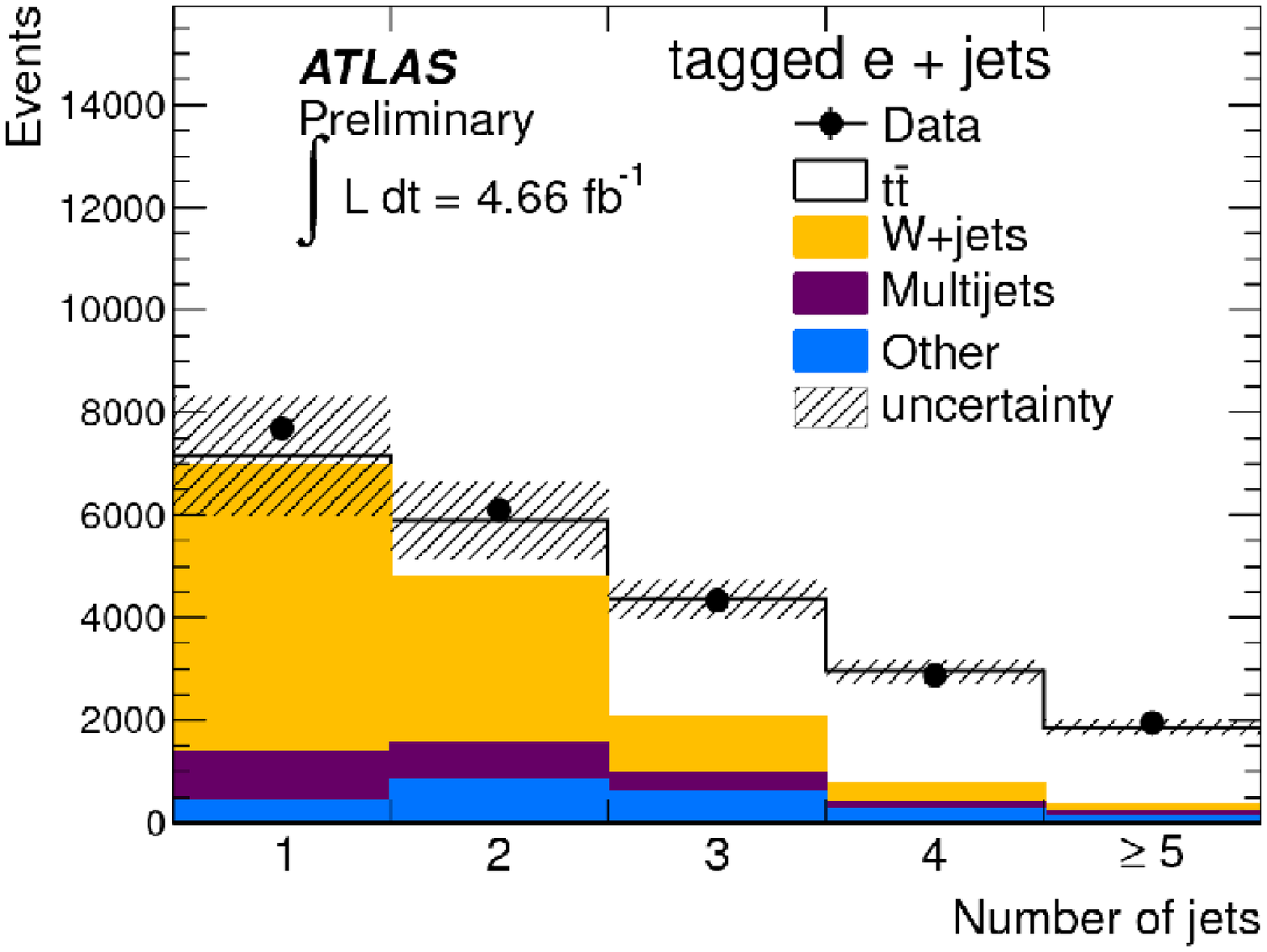}}
\vspace*{0.2cm}
\caption{Distribution of the number of jets in events with an electron or positron, a b-jet and
significant $E_T^{\text{miss}}$ as measured by ATLAS at 7~TeV.
Sample composition is split into the main
components.}
\label{fig:semileptonictop}
\end{minipage}
\hfill
\begin{minipage}{0.48\linewidth}
\centering\includegraphics[width=.9\linewidth]{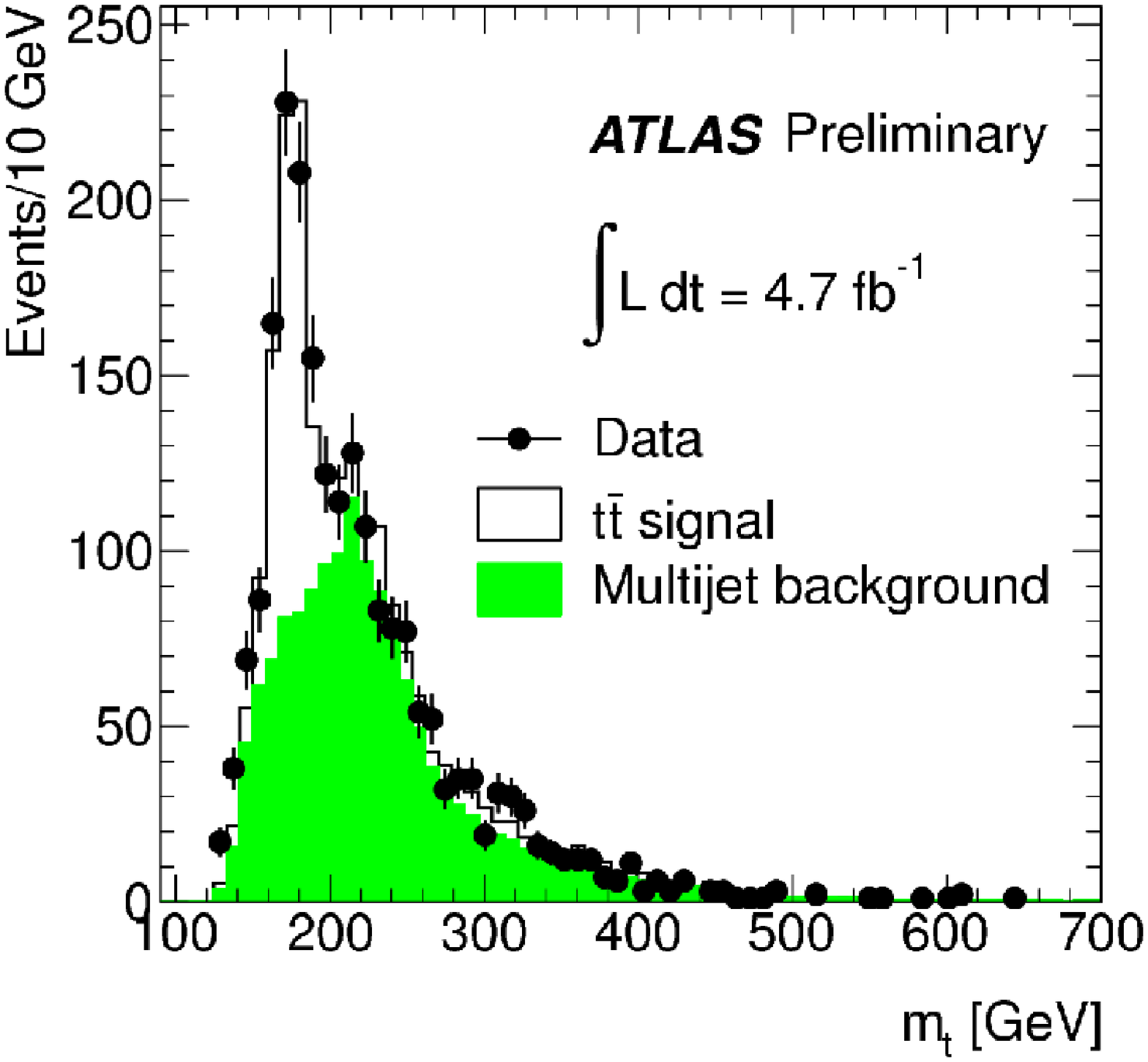}
\caption{Invariant mass distribution for three jets forming a top candidate in fully-hadronic
of top-pair production events. Measurement by ATLAS at 7~TeV. Expectations
for the top-pair signal and the
multijet background~(histograms) are shown and compared to the data~(dots).}
\label{fig:hadronictop}
\end{minipage}
\end{figure}

The first property to be measured for the top quark is the production cross section in the main
mechanism (pair production) and the simpler channel: the semileptonic events in which there is
a good identified lepton and at least one jet tagged as coming from a bottom quark. Results
were obtained for the sample collected at 7~TeV by ATLAS, giving a cross section of
$165\pm2\text{(stat)}\pm 17\text{(syst)}\pm 3\text{(lumi)}$~pb~\cite{bib:atlas:topsemi7}.
Distribution of the number
of jets is presented for events with an electron in Fig.\ref{fig:semileptonictop}, showing the clear
signal yield for high jet multiplicities.

It should be noted that the semileptonic events apply only to electrons and muons, not to the $\tau$
lepton that is considered aside. That channel has also being studied
since it is very important for the possible new physics related to the third generation and
the measurements~(like the one in~\cite{bib:cms:tautop}) are found to be in good agreement
with the expectations. Additionally the all-hadronic channel has also being
investigated~\cite{bib:atlas:allhadronictop}
in order to confirm the expectations. These two channels used the invariant mass distribution
of the top quark candidates, as shown in Fig.~\ref{fig:hadronictop}, in order to separate
the large backgrounds. It should be remarked that the lack of precision for these channels
is basically driven by the systematic uncertainties affecting the background or the acceptances.

On the other extreme, channels containing two leptons (electrons and muons) provide the cleanest
signature. At the Tevatron this channel was not precise because of the lower yield, but the LHC has proven
this is no longer an issue with the single most precise measurement of the cross section from the
dilepton channel at CMS~\cite{bib:cms:dileptontop}, 
$161.9\pm2.5\text{(stat)}^{+5.1}_{-5.0}\text{(syst)}\pm 3.6\text{(lumi)}$~pb, again at 7~TeV.

All these channels provide experimentally independent measurements of the production cross 
section that have been combined~\cite{bib:atlascms:top7tev} to give a value of
$173.3\pm2.3\text{(stat)}\pm9.8\text{(syst)}$~pb. The combination has also proven
the good consistency among the different channels and the two experiments. In addition to these results
at 7~TeV, the two collaborations are working on getting a similar picture with the data
collected at 8~TeV and measure the top-pair production cross section, whose interest is to
test the model at higher energies but also to open the possibility of performing ratios
of energies~(and even double ratios with the addition of the Z-boson production cross section)
which will enhance the sensitivity to BSM physics. The first measurements of the cross section at 8~TeV
are reported in~\cite{bib:atlas:top8tev} and~\cite{bib:cms:top8tev}.

However it should be remarked that the large samples of top events are also allowing new sets
of studies that were not available at Tevatron: measuring SM quantities using events containing
top quarks. Those provide good tests of the SM, but also a useful frame to perform precise measurements.
One example is the extraction of $\alpha_S$ from the top-pair production cross
section~\cite{bib:cms:alphastop}, which leads to a competitive value because it is determined
in an energy regime that has only been accessible to a reduced amount of measurements.

Besides of the total production cross section, the experiments are measuring differential cross
sections~\cite{bib:atlas:topdiff,bib:cms:topdiff}. These studies
provide very stringent test of the SM
predictions and of the modeling in simulation. In addition the sensitivity to possible discrepancies
is enhanced, since such discrepancies could appear in tails of distributions, as expected from
possible new physics, and not affect the bulk of them in any visible way.

The results of the measurements does not present any significant discrepancy and good agreement
is observed, which increases the confidence on the predictive power of the theoretical tools.
These are going to be fundamental when larger samples are investigated, as those collected in 2012
at 8~TeV, since precision will be much larger and the challenges and sensitivity to new physics
increases to previously unknown levels.

\subsection{Measurement of the properties of the top quark}
\label{ssec:topquarkproperties}

Until more data is available for detailed studies of the production mechanism, the
current data samples allow the measurement of the properties of the top quark to
an unprecendent precision. The first one is the determination of the mass, since
it ia a parameter that determines many other properties, and its high value is
already a motivation by itself.

The LHC experiments are exploiting the experience at the Tevatron and are already 
measuring the mass
of the top quark with very advanced techniques: template fits, jet calibration in-situ
and similar. In addition the measurements are performed in several samples that are
later combined, even to get a combined LHC result, as summarized in Fig.~\ref{fig:atlascms:topmass}
and documented by the collaborations~\cite{bib:atlascms:topmass}.
It should be remarked that the achieved precision will be very hard to improve, but still the mass
of the top quark is a relevant quantity of study at the LHC. Specfically larger samples will allow
differential measurements of the mass, d$M_{t}$/d$X$, which provides additional information
and constraints.

In addition to the direct measurement of the mass, the LHC experiments are also measuring
the mass indirectly from the measured cross section and the comparison to the theoretical
expectations. The value extracted from this~\cite{bib:atlas:masscross,bib:cms:masscross} is not as precise as the
direct measurements, but the comparison provides a new handle to find inconsistencies
in the theory predictions (and therefore opening the way to possible BSM physics). The results are
in good agreement, confirming the impressive performance of the SM predictions for top production
and properties.

Additionally to the mass there are other several quantities that have been measured for the
top at the LHC by CMS and ATLAS. As an incomplete summary, here are brief references to them:

\noindent
{\bf -- Electric charge}

\noindent
Within the SM there is a fixed expectation for the electric charge of the top quark (+2/3 of that
of the positron). However, some models would allow a charge of -4/3 (same units) which is still
fully compatible with the observed decays since the inclusive measurements do not relate the charge
of the lepton from the W boson and that of the bottom quark, specially due to the difficulties to measure the
latter.

However performing studies of the charge asasociated to the bottom quark (and the jet) and the pairing
of jet and W boson to identify the ones coming from the same top, it is possible to obtain sensitivity
to the charge of the top quark. Even with limited luminosities, analyses by the two
collaborations~\cite{bib:atlas:topq,bib:cms:topq} by testing the two models
again sensitive distributions are excluding the alternative value beyond
any reasonable doubt.

\noindent
{\bf -- Mass difference for top and antitop}

\noindent
CMS has measured the mass difference between the quark and the antiquark version of the
top~\cite{bib:cms:massantitop}, which provides a stringent test of the CPT invariance in Nature and of the possible
compositeness of the top quark state.
The result is in agreement with the SM expectation in which there is no difference.

\noindent
{\bf -- Polarization and spin correlations}

\noindent
Due to the short lifetime of the top quark, its decay happens before a change of
the spin. This allows to perform studies related to the spin that are not available
to any other quark. 

In pair production the polarization of the top quark 
is investigated by using the angle between
the quark and the lepton. Measurements by CMS in the dilepton channel~\cite{bib:cms:toppolar} and by ATLAS in the
lepton+jet sample~\cite{bib:atlas:toppolar} has confirmed that the polarization is in agreement with the SM
expectation: top quarks are produced unpolarized.

However, the SM predicts that even if the quarks are not polarized, the spins of que quark and antiquark are
correlated. The degree of correlation as measured by ATLAS in helicity basis is
$0.40^{+0.09}_{-0.08}$~\cite{bib:atlas:topcorrela}, 
in perfect agreement with NLO SM predictions, which sets additional
constraints to possible anomalous production, i.e. BSM physics.

\noindent
{\bf -- Helicity of W from top decays}

\noindent
Due to the characteristics of the coupling of the W boson to fermions, we expect that
helicity of the W decaying from top quarks to be fully determined. This property
is parameterized in different components that are accessible by studying the angular
ditributions
between the lepton from the W boson and the top quark in the W rest frame.

Measurements performed by the two collaborations~\cite{bib:atlas:whelic,bib:cms:whelic}
are in agreement with the SM expectations and the results are used to set limits
on anomalous couplings between the W boson and the top quark, basically testing the
V-A structure of the weak coupling of the only quark in which it is directly accessible.

\noindent
{\bf -- Forward-Backward asymmetry in top-pair production}

\noindent
In top-quark pair production a stricking assymetry was observed at the Tevatron regarding
the foward-backward production of the quarks, which a clear preference of the
top quark to be produced in the direction of the proton~(and the antiquark in that
of the antiproton).

Although this is somewhat expected, the observed value is much larger than the NLO
predictions. Some uncertainties involved in the calculations may be large but the
effect may be also produced by some unknown effect, specially because the effect
increases with the mass of the produced pair.

At the LHC the available energy and production yield motivates a more precise
study of the effect. However, the symmetric initial state prevents the realization
of exactly the same measurement. On the other hand, the matter-dominated
initial state introduces differences in the rapidity distributions of the quark
and antiquark that is related to the distribution studied at the Tevatron experiments.

The measurements of the asymmetry for the quantity $\Delta |y| = |y_t| - |y_{\overline{t}}|$ 
performed by the two experiments~\cite{bib:atlas:topfbasym,bib:cms:topfbasym} show good
agreement with the SM expectations. It should be remarked this does not exclude the
Tevatron result, since there are no final model explaining the asymmetry. However, the
LHC results exclude some proposed models and adds some additional information that
is very useful for this subject, that is a good candidate to be one of the hot
topics for the incoming years, specifically regarding top physics.

\noindent
{\bf -- Study of $t\overline{t}+X$ production}

\noindent
Since the pair production cross section of top quarks is so large, it has become
possible to start studying the properties of the top quark with the associated
production of additional objects, usually radiated from the top. Sizes
of the current datasamples do not allow detailed studies of the most interesting
processes, as the production of a pair of tops and electroweak bosons, but current studies
are showing the possibilities for the future running.

On the other hand, other processes that have not been studied in detail 
are already reachable for accurate comparison with the SM predictions. Two examples
are given by the production of jets in association with a top
pair~\cite{bib:atlas:ttjets} or even the production of bottom jets~\cite{bib:cms:ttbjets}.
These measurements
are in good agreement with expectations and are setting strong constraints on the
model predictions in regions that were not investigated before.

In summary, the LHC has been proven as a \emph{top factory} allowing a high rate of produced top
quarks to perform very detailed measurements of its properties. It is expected that the
precision of these will increase with the future samples, providing information and
constraints for models related to the less known of the quarks in the standard model.
Therefore it is not an exaggeration to claim that particle physics has already entered 
in the era of precision in top-quark physics.

\subsection{Single-top production}

A very important topic regarding top production is that of \emph{single top} that is dominated by
electroweak production of top quarks. The process, observed at the Tevatron, has not being
studied in detail until the arrival of the LHC, in which the available yields allow
accurate comparison to the theory.

\begin{figure}
\begin{minipage}{0.48\linewidth}
\centering{\hspace*{-0.4cm}\includegraphics[width=1.1\linewidth]{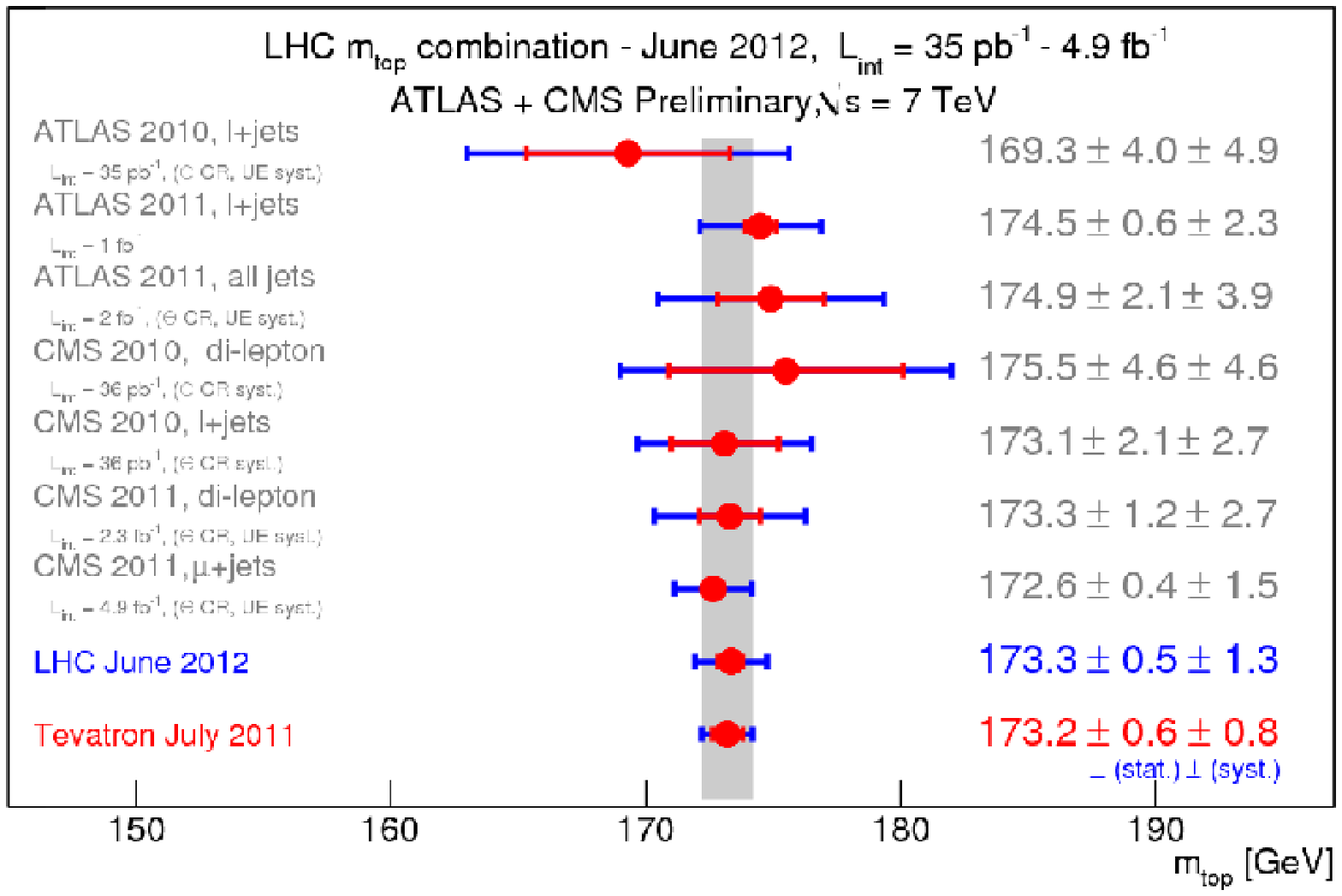}}
\vspace*{-0.65cm}
\caption{Summary of the more relevant measurements of the top-quark mass at the LHC, including the
combined from the two experiments and the comparison with the best Tevatron combination.}
\label{fig:atlascms:topmass}
\end{minipage}
\hfill
\begin{minipage}{0.48\linewidth}
\centering{\hspace*{-0.4cm}\includegraphics[width=1.1\linewidth]{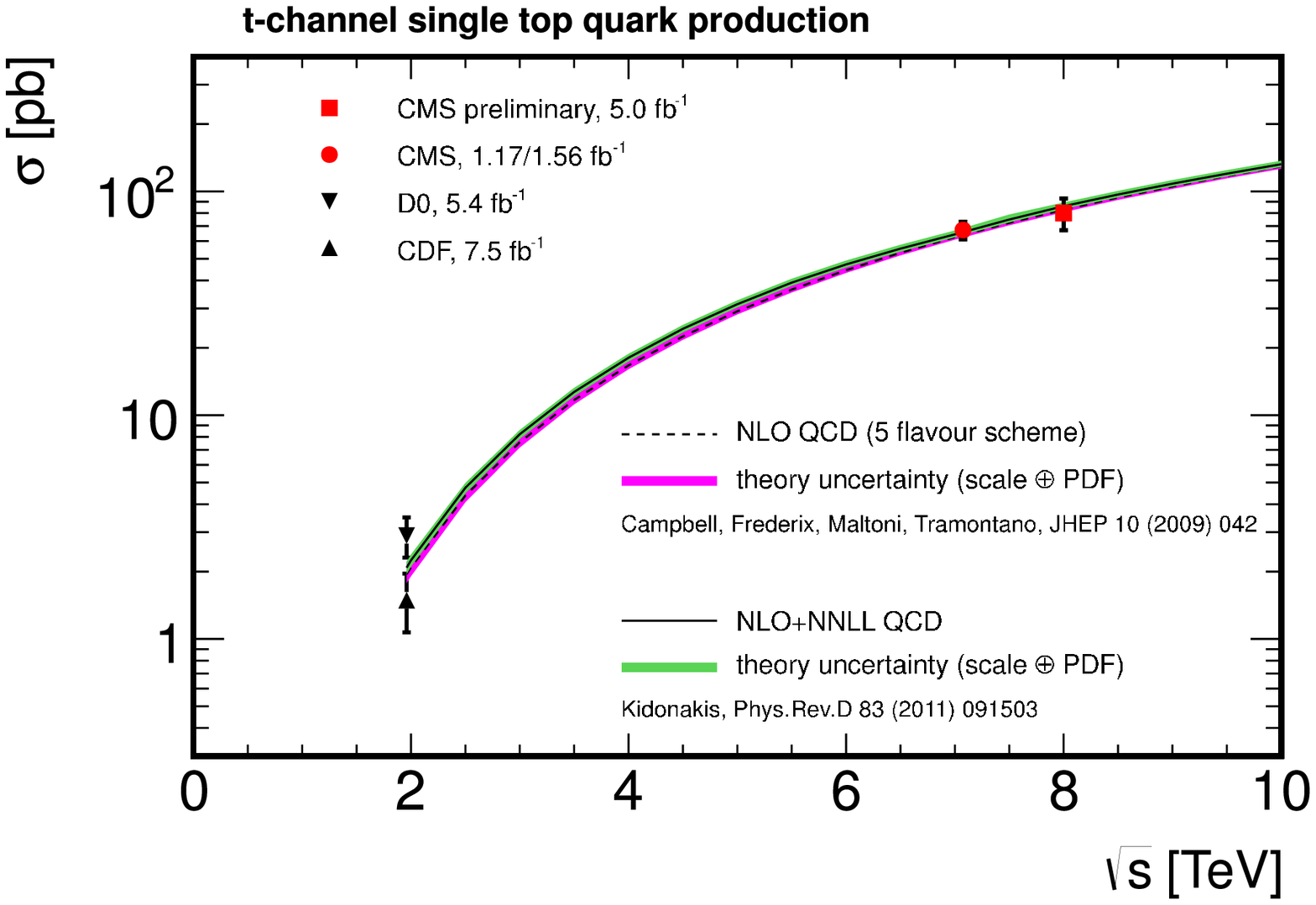}}
\vspace*{-0.8cm}
\caption{Measurements of the single-top production cross section in the t-channel by CMS at 7~TeV
and 8~TeV. For comparison, measurements at the Tevatron experiments are also shown.}
\label{fig:singletop}
\end{minipage}
\end{figure}

In the production of single top there are traditionally three channels under consideration: the t-channel
(via a W exchange) which is the one with the highest cross section and sensitive to the bottom-quark content
of the proton, the s-channel (via virtual W production) and W$t$ production, which was not observed
at the Tevatron. From them, the t-channel is relatively easy to be studied at the LHC and current
results have reached a good precision and even allowed separate sudies of the quark and antiquark
production. 
Figure~\ref{fig:singletop} show the measurements at CMS at 7~TeV and
8~TeV~\cite{bib:cms:singletop8} and comparison with Tevatron measurements.
Similar studies has been produced by ATLAS, with similar reach and conclusions~\cite{bib:atlas:singletopt}.
Additionally, results on
the s-channel were able to set limits on the process that are around 5 times the SM 
predictions~\cite{bib:atlas:singletops}. However,
the current analysis does not include the full data available. With more data the results will
become much more relevant. It should be noted that the s-channel is more sensitive to possible
anomalous production of particles.

Regarding the third channel, the associated production of a W boson and a top quark, both experiments
reached the level of evidence using the 7~TeV sample~\cite{bib:atlas:tw,bib:cms:tw}. The
observed distributions are in agreement with the SM expectations, but more data is needed to
perform accurate comparisons. The 8~TeV data should allow the observation and first precise
measurements of this process, although the analysis is a bit challenging due to the harder
conditions.

\begin{figure}[t]
\centering\includegraphics[width=0.65\linewidth]{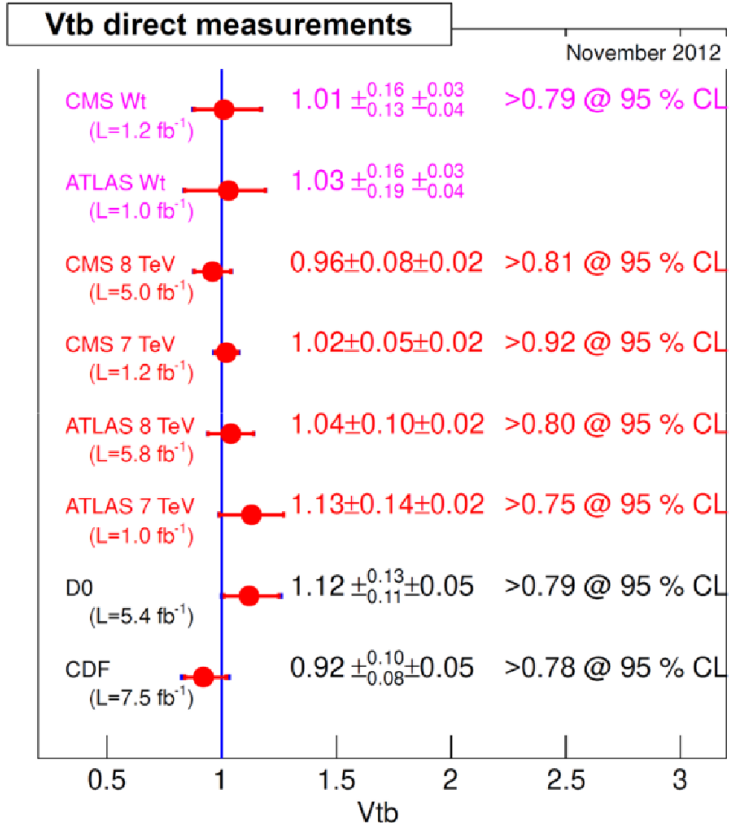}
\caption{Summary of all the direct determinations of the CKM element $V_{tb}$ at the Tevatron
and LHC experiments from single-top production.}
\label{fig:vtbsummary}
\end{figure}

Once the production of single-top events has been established, the study of them allows to 
provide information about the electroweak couplings of the top quark, specifically due to the
sensitivity of the production mechanism to the CKM element $V_{tb}$ ruling the coupling
between the top quark, the bottom quark and the W boson. Several determinations of this
quantity have been performed at Tevatron and LHC, as summarized in Fig.~\ref{fig:vtbsummary}.

In conclusion, studies of the single-top production are starting to reach a precision that
will put the SM under test in the unexplored sector of electroweak physics with top quarks.
Without doubt, this will also contribute in the next years to complete the picture we have
of this quark as a key piece of the SM and its link to its possible extensions. 

\section{Results on heavy-ion collisions}

Although the main goal of the LHC is to understand the interactions at the highest energies (or
shortest distances), this collider also allows to produce extreme conditions in terms of
energy density, pressure affecting baryonic matter. This is achieved by colliding heavy-ion
nuclei, as it is the case of lead.
The main goal is to try to study the strong interaction at lower levels, i.e. investigate concepts
as confinement, thermal phenomena, chiral symmetry and so on, more closely related to the
conditions affecting quarks and gluons in the early universe than the clean parton-parton
collisions usually studied at the LHC when colliding protons.

Also in the case of the LHC the increase in energy represents a big step forward in studies of
heavy-ion collisions: the experiments at RHIC were intended to discover the production of 
strongly-interacting perfect fluid. The LHC experiments shall characterize the details of this
new class of matter with the increased precision.
For that, one of the most useful quantities is the \emph{elliptic flow}, defined as the second momentum
of the azimuthal distribution of produced particles. It contains very important
physics information because larger values of the quantity indicates the presence of
viscosity in the medium at the early times after the collision. Such values were
observed at RHIC and by ALICE~\cite{bib:alice:ellipticflow},
confirming the expectations from hydrodynamic models. Adiditionally, ALICE has measured
the elliptic flow and production yields (and ratios) for specific particles, as
e.g. in~\cite{bib:alice:hadyields} identified
via its sophisticated detector subsystems. Some of the results are a bit unexpected, as the
reduced production of baryons with respect to pions, which may be pointing to some
presence of hadronic rescattering, an effect never observed.
Other interesting measurements have already been performed by the collaborations with the 
aim of quantifying the characteristics of the collisions, as studies of higher-order
harmonics~(as in~\cite{bib:cms:collective}), or particle correlations, and the studies
related to the measurements sensitive to the Chiral Margnetic Effect~\cite{bib:alice:cme}
which is a fundamental study in the heavy-ion program at the LHC after the first hints at
RHIC. 

However, most of the current studies in heavy-ion collisions are more pointing to the confirmation
of the results found at RHIC in order to tests new tools and fix a solid base to go beyond
in terms of energy and sizes of data samples. In fact, it is in terms of hard probes of
the created medium where the LHC experiments have clearly go beyond previous experiments.

ATLAS was the first one presented a result on jet quenching~\cite{bib:atlas:jetquenching}, in which
one expect dijet events produced from hard parton interactions in lead-lead collisions are observed
as assymetric production of jets: opposite to a produced jet with large transverse momentum it
is not straightforward to find a second jet, as in the usual proton-proton collisions. In fact a factor 2
of suppresion in central collision is observed, very independent of the jet momentum. This
is explained by the presence of a strongly interacting medium which affects more one hard parton than
its companion, and therefore giving the impression of disappearence of jets.

In addition to jets, it has been very common the use of hard photons as probes of the medium. Photons
are transparent to the medium, so they are perfect to quantify effects on jet quenching in
the production of $\gamma$+jet, as in~\cite{bib:cms:gammajetquench}. However, photons may also be
coming from the hadrons in the medium, or in the final state, so they represent as small limitation
that the LHC experiments may avoid with the use of more massive probes that were not available
at RHIC: the weak bosons.
Currently the experiments have been focusing on detecting the presence of those bosons, since available
data samples does not allow its use as actual probes, e.g. in Z+jet production. However, the detection
of leptonic Z bosons by CMS~\cite{bib:cms:zhi} and ATLAS~\cite{bib:atlas:zhi} have already allowed the first differential
measurements to characterize the production of these ideal probes, completely insensitive to
initial state or hadronization and for which the medium is transparent. Studies of the W bosons
have also been performed~\cite{bib:cms:whi} and have already provided interesting confirmation regarding
proton-neutron differences: isospin effect yields a reduced asymmetry in charge with respect to
proton-proton collisions at the same energy per nucleon. Again larger samples are needed for more detailed
studies, but the LHC is probing all its potential in heavy-ion collisions.

\begin{figure}
\begin{minipage}{0.48\linewidth}
\vspace*{-0.5cm}
\centering{\includegraphics[width=1.02\linewidth]{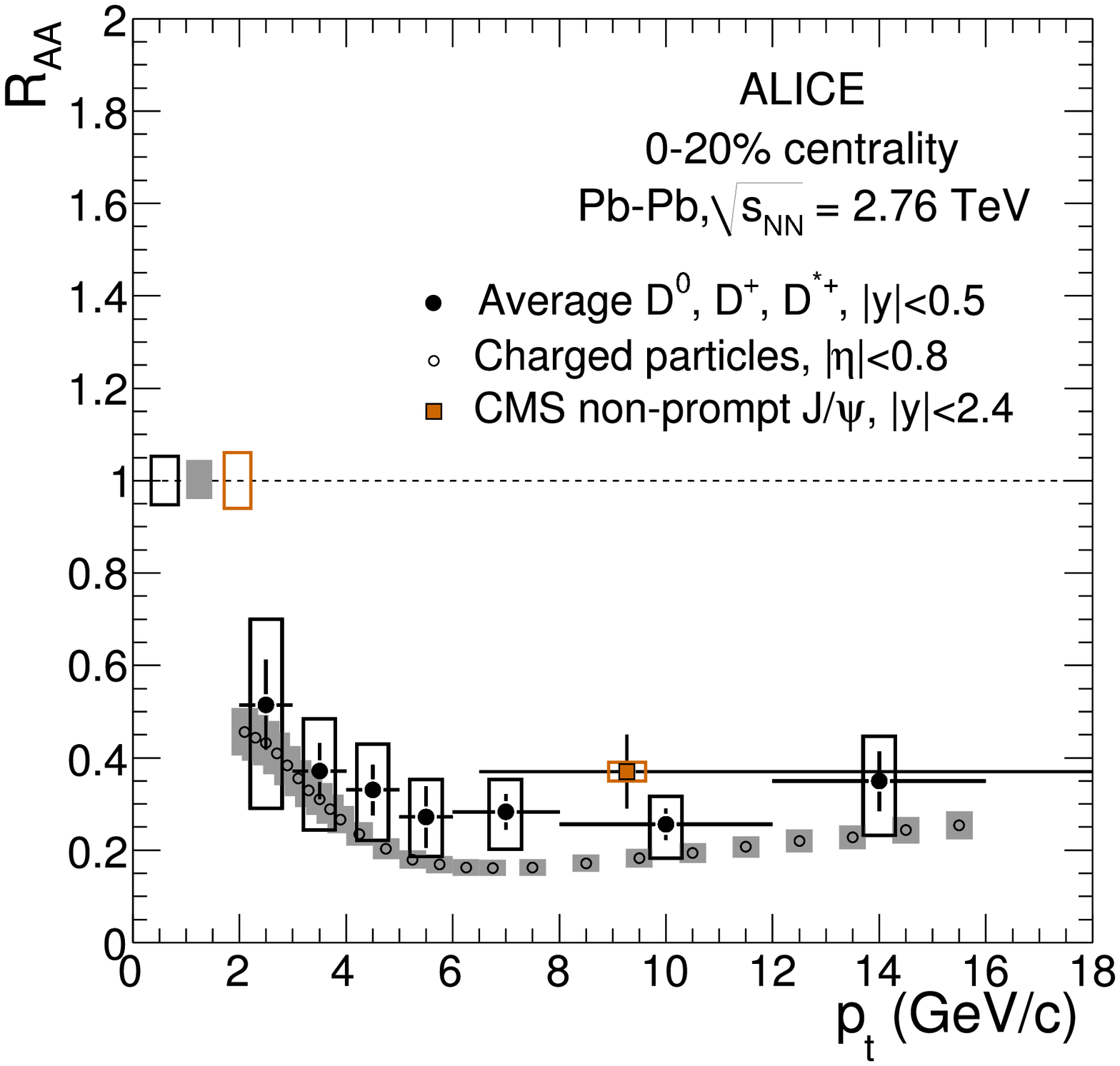}}
\caption{The nuclear modification factor with respect to proton-proton measured in
lead-lead collisions for D mesons in the most central events as measured by ALICE.
Data~(black dots) are
compared to the nuclear modification factors of charged particles~(open circles) and non-prompt $J/\psi$
from CMS~(squares). 
}
\label{fig:opencharm}
\end{minipage}
\hfill
\begin{minipage}{0.48\linewidth}
\vspace*{0.3cm}
\centering{\includegraphics[width=1.0\linewidth]{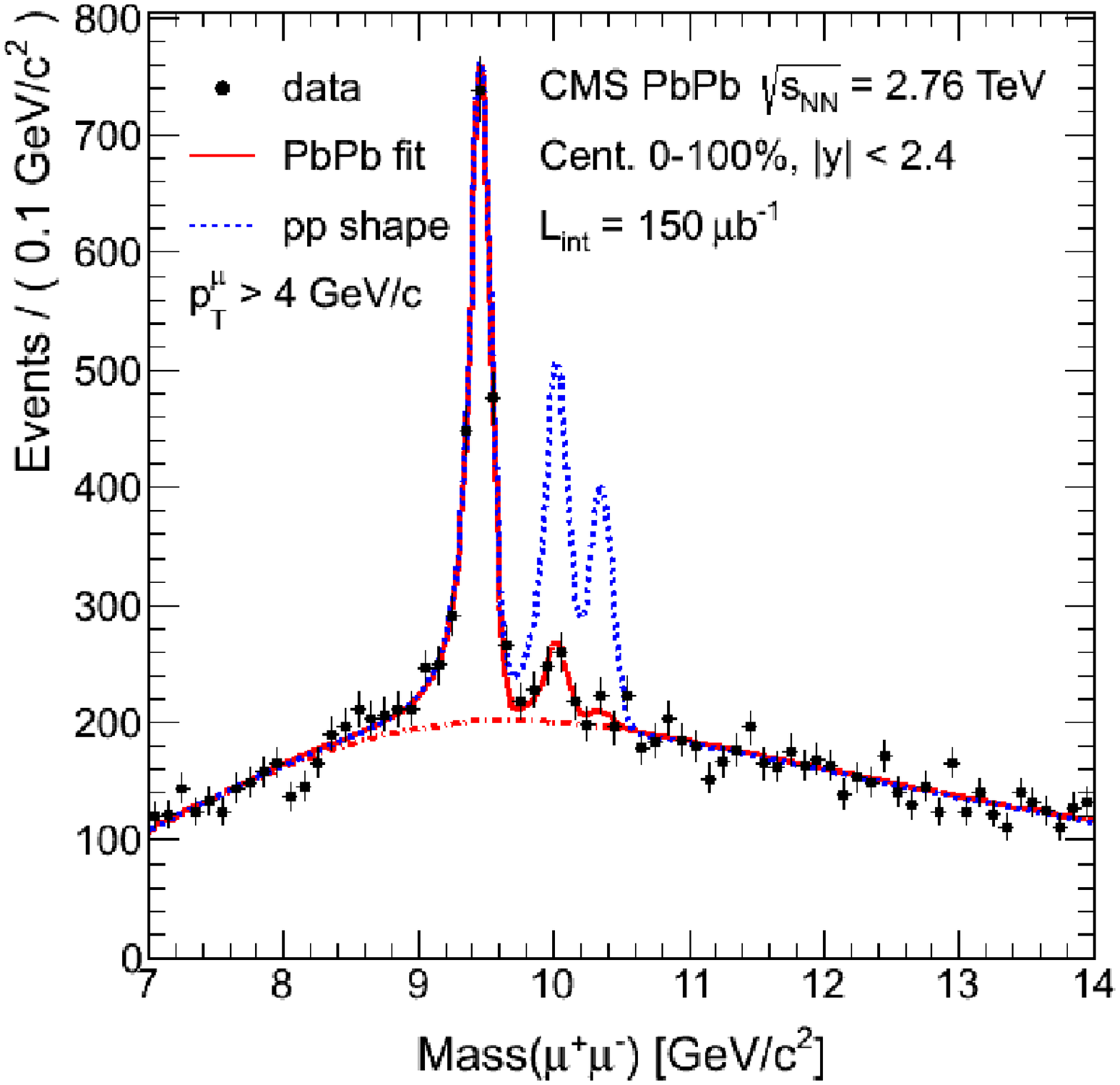}}
\caption{Invariant mass of dimuon pairs measured by CMS in the region of the $\Upsilon$ family
as produced in heavy-ion collisions~(dots and red-line fit). Comparison to the data from
proton-proton collisions normalized to the $\Upsilon$(1S) peak~(blue dashed line) shows the
sequential suppression of the family in heavy-ion collisions.}
\label{fig:upsilonsupressed}
\end{minipage}
\end{figure}

Another area in which the LHC allows to reach much further than RHIC is the sudy of heavy-flavour
production. As in the case of proton-proton collisions, the possibility of identifying secondary vertices
allows specific studies to be performed. In fact ALICE has shown its great capabilities with the
reconstruction of open-charm mesons, D mesons~\cite{bib:alice:opencharm} which are not only nicely
observed but also used to perform measurements, like the one shown in Fig.~\ref{fig:opencharm},
which probes the confirmation of suppression for open charm in central collisions, in good agreement
with more inclusive studies. The aim of using open-charm mesons (and perhaps B mesons) 
is that they bring the possibility of quantifying differences in the energy loss in the medium
between heavy or light quarks and even gluons.

But the identification of heavy-flavour states is much more powerful in the dilepton resonances, specifically
for the quarkonia states. They have a long history of being studied in heavy-ion collisions due to their
clean signature and the big theoretical/phenomenological knowledge on them. Regardless of
being colourless they are sensitive to the medium since they rely on the strong force to keep the two quarks
bounded. In fact these states are affected by screening effect and they become an actual thermometer
of the medium: the larger the radius of the system (larger for e.g. 2S states than 1S) the larger the
screening. Therefore we expect to observe a \emph{sequential suppresion} or \emph{melting}
within the quarkonia families:
less bound states are more suppressed than those that are
more bound. This has been clearly observed in
measurements by CMS~\cite{bib:cms:upsilonsup} for the $\Upsilon$ family, as shown in Fig.~\ref{fig:upsilonsupressed}.
Clearly the excited states are affected more in relative terms than the ground state when comparing
reasults from lead-lead collisions with those of proton-proton at the same energy per nucleon. This
is an additional confirmation that a strongly interacting medium is created in the relativistic
heavy-ion collisions at the LHC.

It should be noted however that even if the qualitative picture seems clean, the quantitative details
do not completely fit, so further measurements and theoretical developments will be needed in order
to fully understand the generated medium. Such kind of studies are already in place, as the measurements
of $J/\psi$ suppresion by CMS~\cite{bib:cms:jpsisup} (in central rapidities) and
ALICE~\cite{bib:alice:jpsisup} (in forward rapidities),
probing the nice complementarity between experiments. However the agreement in the suppression does not
apply to the observation by CMS that $\psi$(2S) is less suppressed than the $J/\psi$ for transverse momenta
larger than 3~GeV, something not confirmed by the ALICE measurements.

In conclusion the heavy-ion program of the LHC experiments is already providing interesting results
bringing the field to unexplored areas with a new energy regime and new possibilities, like the
use of new available tools and probes. The propects for the future, with further analyses of the data,
including the 30~nb$^{-1}$ collected for proton-lead
collisions (as the previews in~\cite{bib:cms:protonlead,bib:alice:protonlead}), will help towards the
ultimate goal of the program: detailed characterization of QCD thermal matter by means of precise
measurements from
heavy-ion collisions at LHC.

\section{Searches for the SM Higgs boson}

The SM structure and its implications in the description of the Universe is based on the
presence of a field, known as \emph{Higgs field} that is responsible for the symmetry
breaking giving rise to the electromagnetic and weak interaction and also to give masses
to the weak bosons. In this process, a single degree of freedom is translated into a 
scalar particle, \emph{the Higgs boson}, that should be observed and whose coupling to 
the fermions are introduced in such a way that these last ones acquire the masses that
are forbidden by the symmetry before it gets broken.

This particle is therefore the missing keystone of the SM and it was extensively searched for
in previous colliders without success. The good performance of the SM strongly motivated the
existence of the particle, and the measurements and fits from pre-LHC colliders pointed to
a mass of around 100~GeV~\cite{bib:prelhchiggs}.

Under this situation, the LHC started collecting the data that should provide light to the
existence of this boson and eventually find it. This was the most important search for the
first years of the LHC experiments and for this reason it deserves a full section describing
the analyses and the strategy to follow in order to observe the presence of the boson and also
the related measurements wich are aiming to confirm whether the observed resonance actually
matches the properties expected for the SM Higgs boson.

\subsection{Strategy to search for the boson at the LHC}

Before the LHC had collected enough data for being competitive in searches of the Higgs
boson, the results from LEP and the Tevatron were the richest source of information. In fact,
LEP had excluded at 95\% C.L. the SM Higgs boson below 114~GeV and its measurements had
constrained the mass of the Higgs to be around 100~GeV. 

In the case of Tevatron, the
direct searches were excluding a Higgs aroung 165~GeV, leaving the available regions
to be clearly separated into two: The low-mass region, 
for masses between 115 and 160~GeV, that was very strongly motivated. The second region, with
relatively high masses beyond 170~GeV, was less motivated, but still not discarded, specially
considering that the motivation was assuming negligible effects from possible BSM physics (or
more complex Higgs models).

The first step therefore for the LHC was to look into these two regions and during 2011 all
channels were considered to investigate all the mass ranges. For low masses, although the decay is dominated
by that to bottom quarks, the involved channels were those having the Higgs decaying into
ZZ (in 4 leptons) or $\gamma\gamma$, with some
information from the WW, $\tau^+\tau^-$ and $b\overline{b}$ decays in all accessible production
modes.
For high masses the most useful channels were those involving decays into WW and ZZ in all posible signatures.
With this approach the two experiments presented results on December 13$^\text{th}$ 2011 with the
data collected at 7~TeV. The results presented at that time led to a complete exclusion
of the Higgs boson in the high-mass region (up to more than 400-500 GeV) and most of the
low-mass one, leaving alone a small window around 125~GeV.

In that window the exclusion was not possible because both experiments saw an excess, not completely
significant but enough to prevent exclusion of the presence of a SM Higgs boson. the excess was
appearing in several of the channels 
Naturally, the presence of a resonance in the most motivated channels to detect the SM Higgs
boson was a clear suggestion that such boson was the responsible for the excess, so all the
focus from that moment was to intensively search for a possible boson with a mass around 125~GeV
whose properties were close to those expected for the SM Higgs boson.

This effort was designed to be applied to the 8~TeV data collected right after the Winter
in 2012 and the idea was to maximize sensitivity in the two most sensitive channels at that
mass~(4-lepton ZZ and $\gamma\gamma$) and also look at the complementary channels~(WW, $\tau^+\tau^-$
and $b\overline{b}$) that could provide some further sensitivity and also some additional
information regarding the nature of the boson: more couplings involved.

In parallel more analyses were still considered in order to complete the pictures, even
those that were looking (and excluding) the presence of a SM Higgs boson at higher
and higher masses. 

All these analyses are described in the following sections.

\subsection{Analyses for the discovery (ICHEP-2012 results and afterwards)}

\begin{figure}
\centerline{\includegraphics[width=0.51\linewidth]{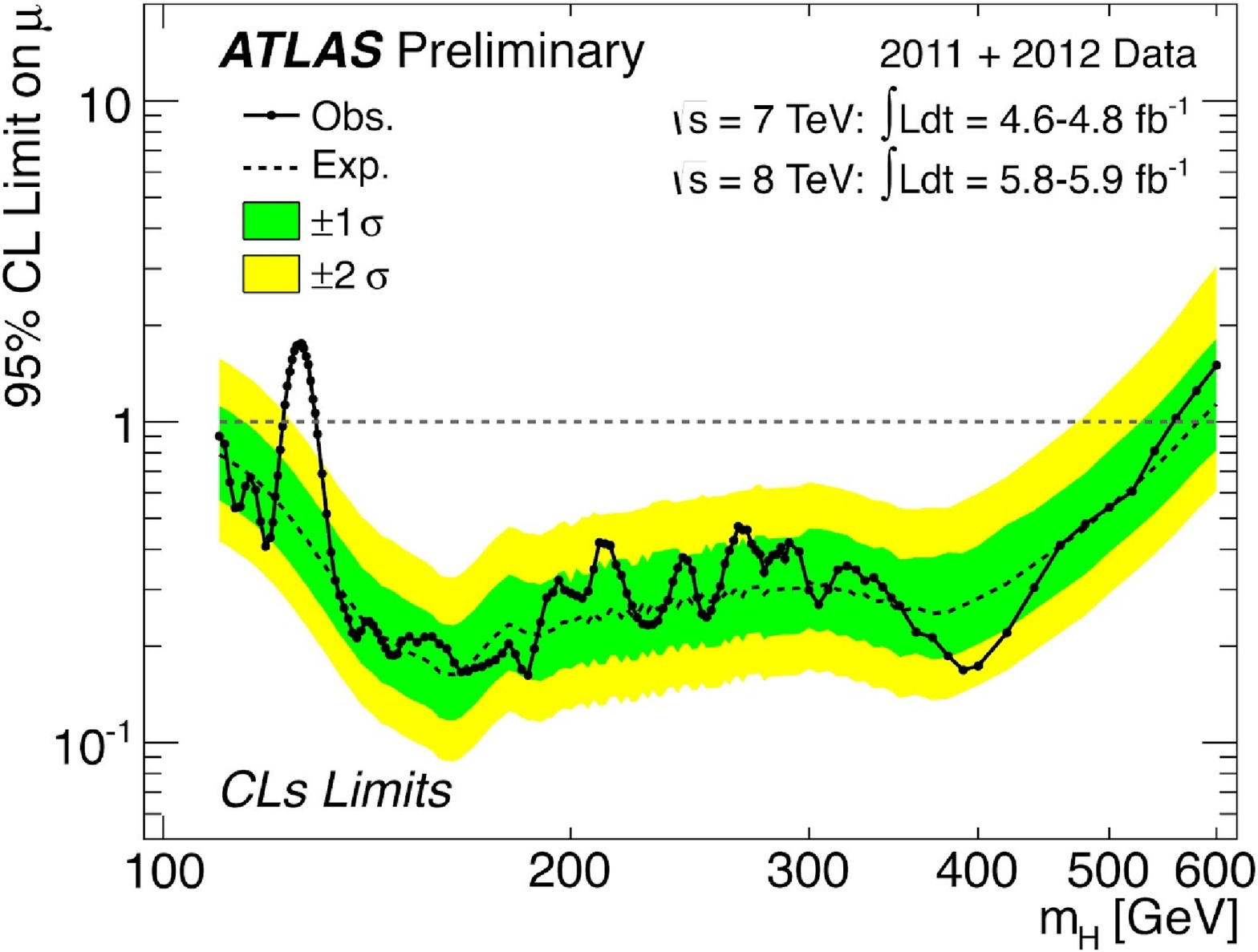}
\hfill
\raisebox{0.2cm}{\includegraphics[width=0.51\linewidth]{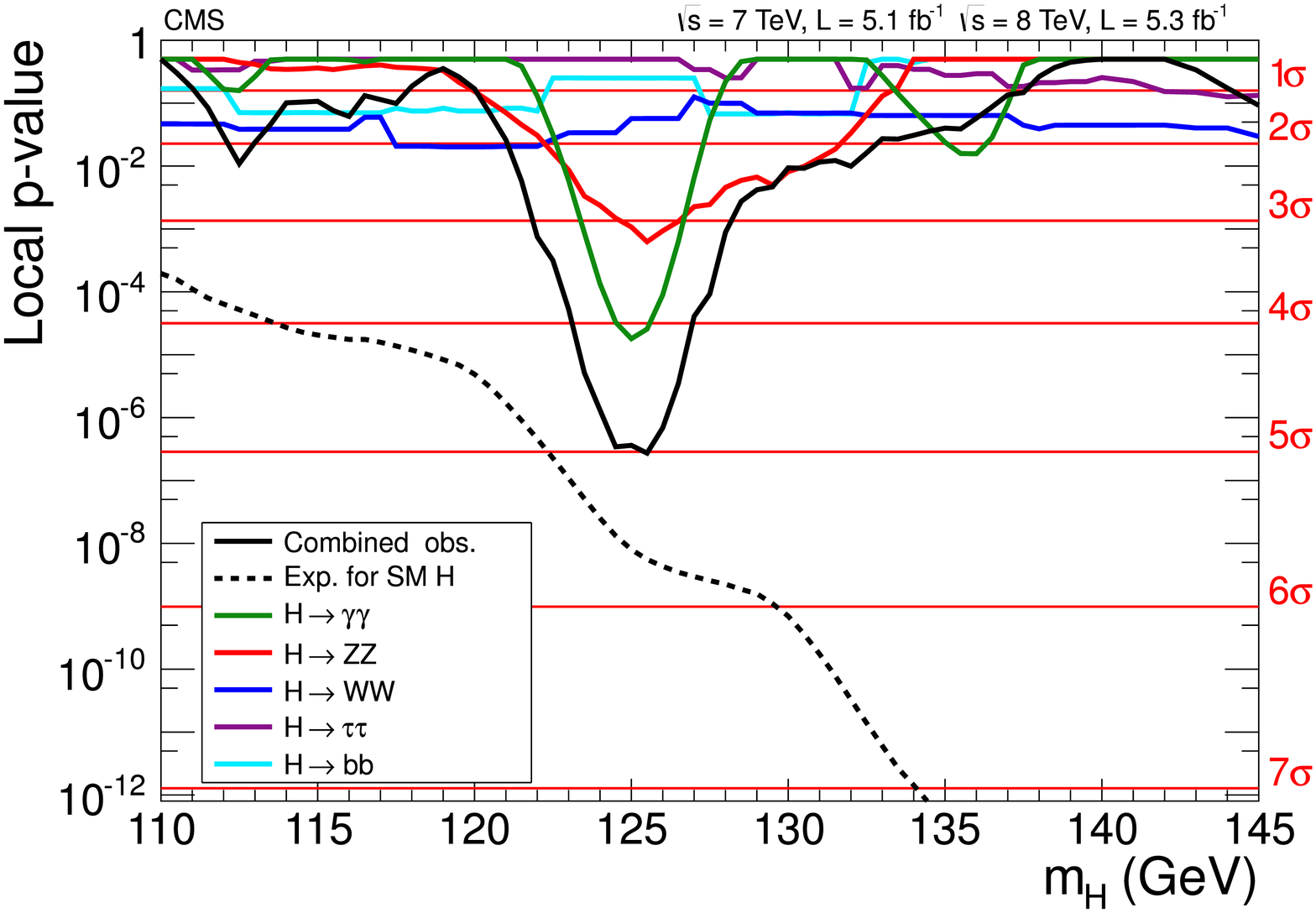}}}
\caption{On the left, 95\% C.L. limit on the ratio of the cross section over the SM expectation
for the production of a Higgs boson as a function of its mass as obtained by ATLAS at the time
of ICHEP-2012. Observed limit is compared with the expected limit (in absence of such a particle)
and the uncertainty intervals. On the right, the local p-values for a similar study
in several analyses by CMS. In both cases a very significant excess is observed around 125~GeV
that is interpreted as observation of a new particle, likely the SM Higgs boson.}
\label{fig:dicoveryichep}
\end{figure}

At the time of ICHEP-2012 the size of the available data at 8~TeV was comparable to
that collected at 7~TeV, allowing already enough sensitivity to perform statements
on the boson. Both collaborations presented results in the main channels
on July $^\text{th}$ 2012, and they
confirmed the presence of a new boson at the discovery~($5\sigma$) level. The
presented results are summarized by the plots in Fig.~\ref{fig:dicoveryichep}, where
the results from the statistical analyses of the studies are shown.

The measurements performed at 8~TeV also increased the precision on the knowledge of the boson
and in general tend to confirm its nature as that of the SM Higgs boson. Later improvements
to the analyses and the addition of the data that was provided by the LHC during 2012 have brought
additional support for this hypothesis. However, some questions are still to be investigated
and further data would allow more precise measurements in the future. Here we will discuss
some of the more relevant results bringing to the current knowledge about the boson dicovered
at a mass of 125~GeV.

In the case of CMS, the $H\rightarrow\gamma\gamma$ search~\cite{bib:cms:h:gammagamma} is performed
by using several categories of diphoton (for inclusive production mode) 
and two categories for tagging Vector-Boson Fusion~(VBF) processes. It should be noted that VBF
is very important because it is sizable~(mostly because the leading Higgs production occurs via loops)
and it involves different couplings than the dominant mechanism, e.g. it is very important for 
fermiophobic models.

With all those categories, the analysis is able to achieve a significant excess of $4.1\sigma$ with
a yield a bit higher than expectation.

\begin{figure}
\centerline{\includegraphics[width=.45\linewidth]{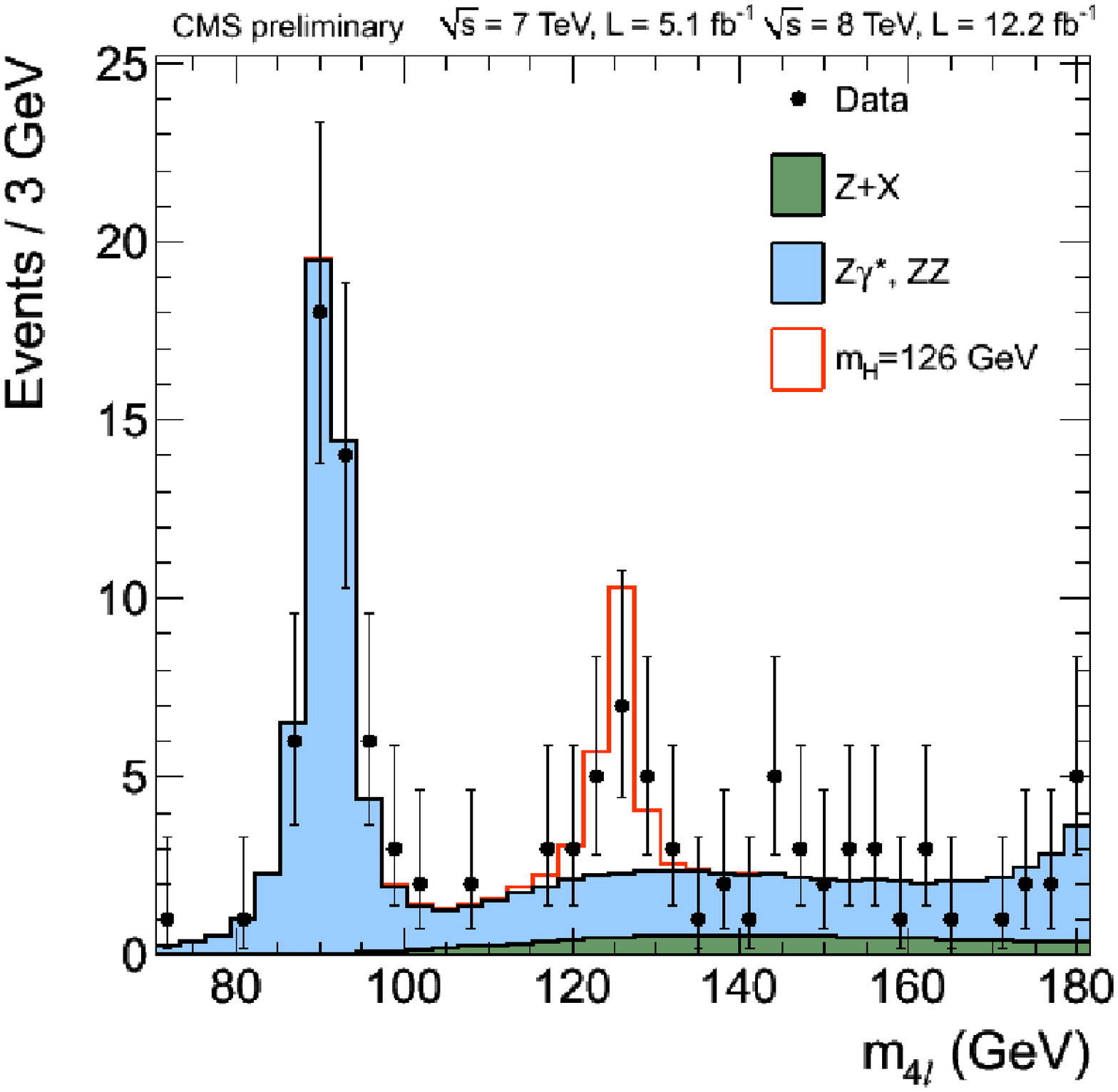}
\hfill
\includegraphics[width=.45\linewidth]{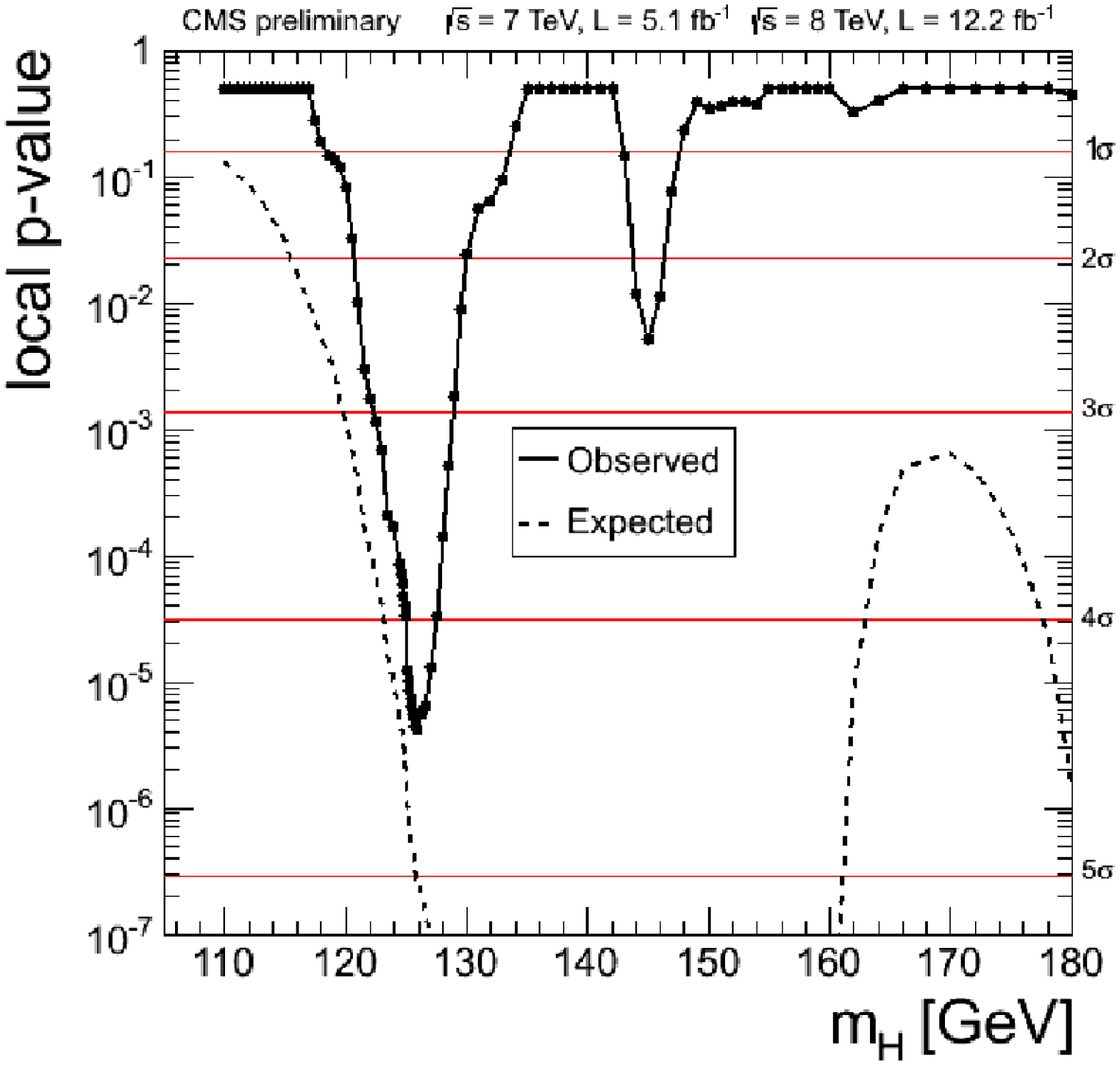}}
\caption{On the left, the mass distribution of 4 leptons in events selected for the Higgs search at CMS.
Data~(dots) are compared to the background expectation~(solid histograms)
and a Higgs signal with $m(H)$=126~GeV~(red line). On the right, local p-values associated to
the same analysis, with a clear excess for a Higgs mass around 125~GeV.}
\label{fig:cmshzz}
\end{figure}

In addition to that, the 4-lepton search was dealt in this collaboration with the use of a kinematic
discriminant that accounts for the fact that the Higgs boson is a scalar. This kind of tools
have made that this analysis~\cite{bib:cms:h:4leptons} is the central reference for measuring the
properties of the boson, as described below. As shown in Fig.~\ref{fig:cmshzz} the channel has
very little background and the signal is clearly observed in spite of the low yield. The significance
of the excess at a mass of 126~GeV is very high, although in this case the yield comes a bit
lower than the SM expectation, but still in agreement.

In addition with the most sensitive channels, CMS has put a lot of effort on the secondary channels
which are giving additional constraints about the boson, with a small sensitivity. Specifically,
the WW decay also suggests the existence of a boson, but with a yield on the lower side~\cite{bib:cms:h:ww}.
The $\tau^+\tau^-$ shows clear limitations on the size of the data sample and although the result is compatible
with a SM Higgs, it is also in agreement with the background-only hypothesis~\cite{bib:cms:h:tautau}.
A similar conclusion is extracted from the decay into bottom quarks~\cite{bib:cms:h:bb}, in which the
Higgs need to be observed in the production associated with a weak boson, in order to keep the dijet background
under reasonable limits. The studies of diboson production described in section~\ref{ssec:dibosons},
specifically in the
semileptonic channels, provide a solid support to the search of the boson in this decay channel. In any case,
more data will provide stronger constraints on the fermionic decay channels, currently compatible
with the existence of the SM Higgs boson but with small significance.

\begin{figure}
\centerline{\raisebox{0.4cm}{\includegraphics[width=0.48\linewidth]{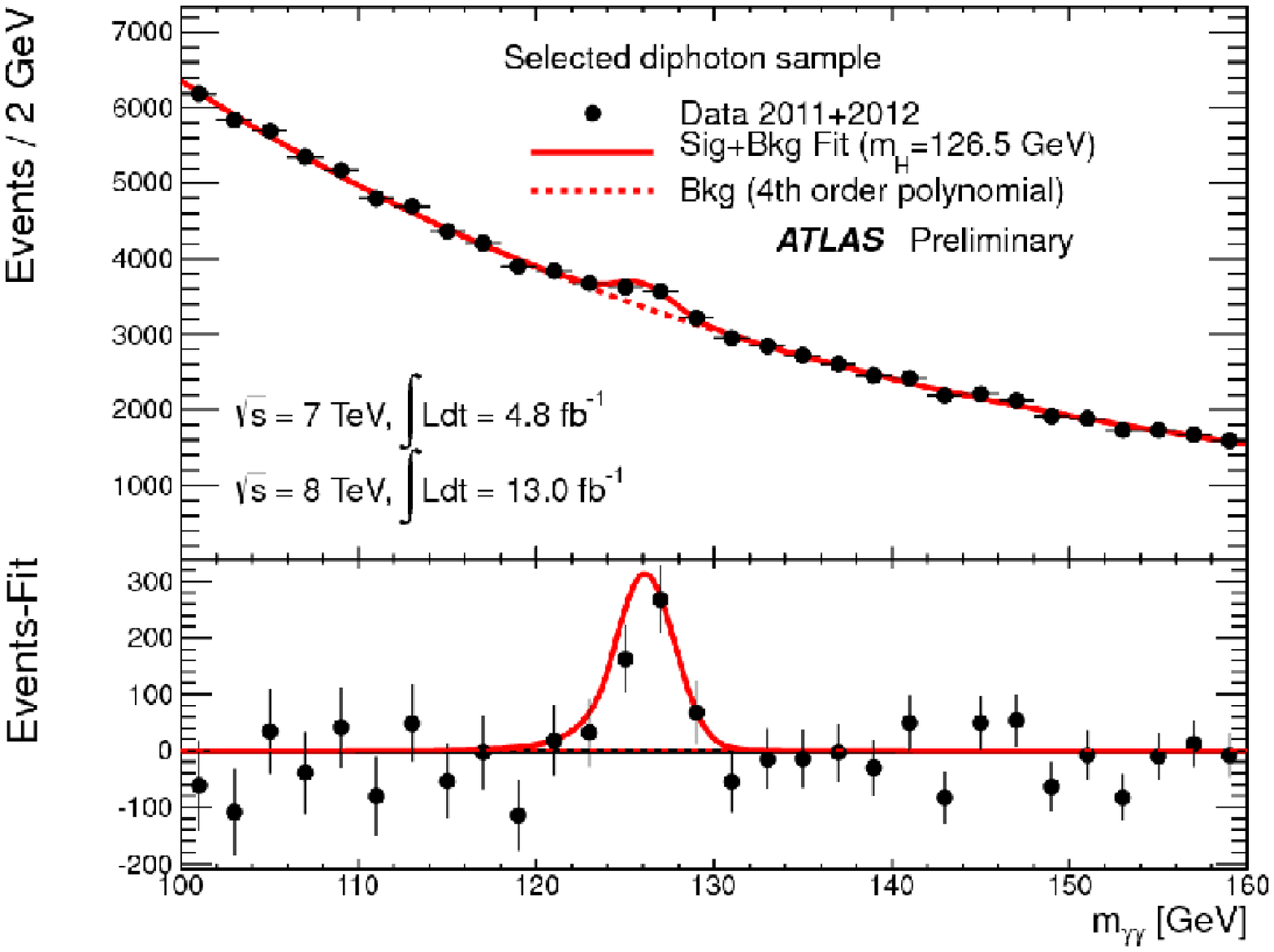}}
\hfill
\includegraphics[width=0.48\linewidth]{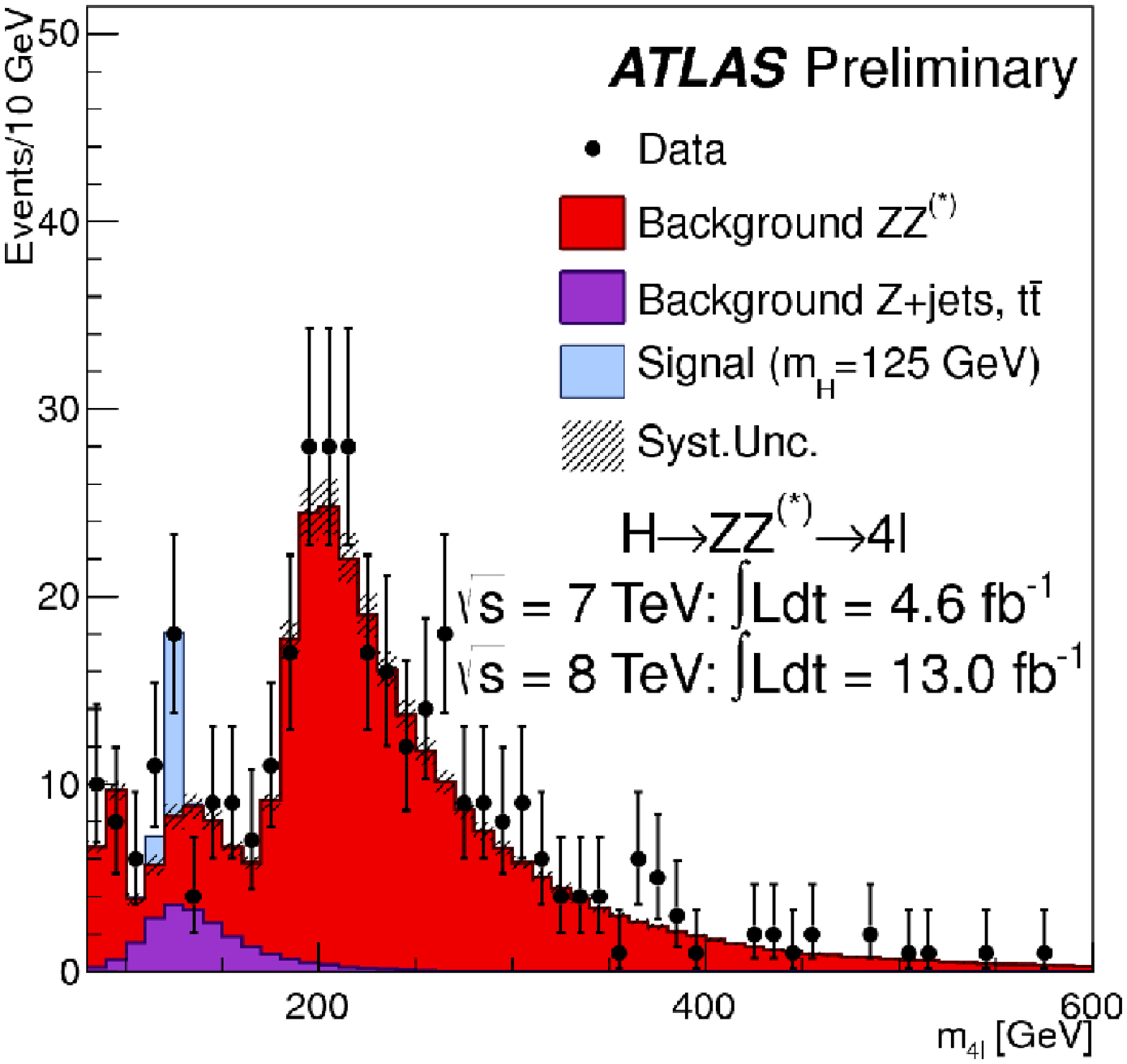}}
\vspace*{0.2cm}
\caption{On the left, invariant mass of two photons in the search for the Higgs decaying
into diphotons performed by ATLAS.
The fitted background is subtracted in the plot below in order to enhance
a resonant excess close to 125~GeV. On the right, invariant mass distribution of 4 leptons
in events selected for the Higgs search at ATLAS.Data are compared to the background and a signal
hypothesis with $m(H)$=125~GeV.}
\label{fig:atlashiggs}
\end{figure}

From the ATLAS side, also several updates came after ICHEP-2012, bringing
further confirmation to the signal and, as in the CMS case, higher precision in the results. The diphoton
search~\cite{bib:atlas:h:gammagamma}, performed with several categories, has lead to a very strong signal,
which approaches the level of being very high when compared to the SM expectation with a signal
strength value approaching a factor of 2 (being 1 the SM prediction). Dedicated studies of this
value in a per-channel basis does not indicate anything striking, but uncertainties in those cases
are large since it is the combination of them which is bringing the high significance of the signal.
Plot on the left of Fig.~\ref{fig:atlashiggs} shows the invariant mass distribution of diphotons
in which the resonance at a mass around 125~GeV is clearly observed.

As in the diphoton search, the 4-lepton channel in ATLAS gives a signal strength higher than the expectation,
although in this case in agreement with the SM value (and with the CMS result). The study of this
final state~\cite{bib:atlas:h:zz} is performed by exploiting the kinematical properties of the decay
products from a spin-0 particle. As shown in the plot on the right of 
 Fig.~\ref{fig:atlashiggs}, the signal is clearly observed with a reasonable amount of background, which
leads to this channel as the main reference to measure the properties of the boson, as in the case of CMS.

Regarding the complementary channels, ATLAS also puts a big effort on those with similar conclusions
to those obtained by CMS. In the case of the decay into bottom quarks~\cite{bib:atlas:h:bb}, sensitivity
has not yet reached the level to allow quantitative statements about the boson to be made. The other
two channels~\cite{bib:atlas:h:ww,bib:atlas:h:tautau} give higher yields than expected, but
still with large uncertainties. In the case of $\tau^+\tau^-$, the value seems to be high
in the case of the main production channel, but in VBF and in associated production with a weak boson~(VH)
the signal strength is clearly on the low side~\cite{bib:atlas:h:tautau}. It is too early to 
be considered a problem since the uncertainty is still large enough to cover the SM value within $1\sigma$.

\subsection{Post-discovery goals: measuring the properties}

As described in the previous section, a new boson has been observed and its properties
are compatible to those expected from the Higgs boson of the SM. With the additional analysis
the picture is getting more complete, but precision needs to be improved to extract further
conclusions.

One of the goals in the incoming \emph{post-discovery} years is the measurements of all the properties.
This has been already started, and some answers are already provided, as we will discuss here.

\begin{figure}
{\includegraphics[width=.48\linewidth]{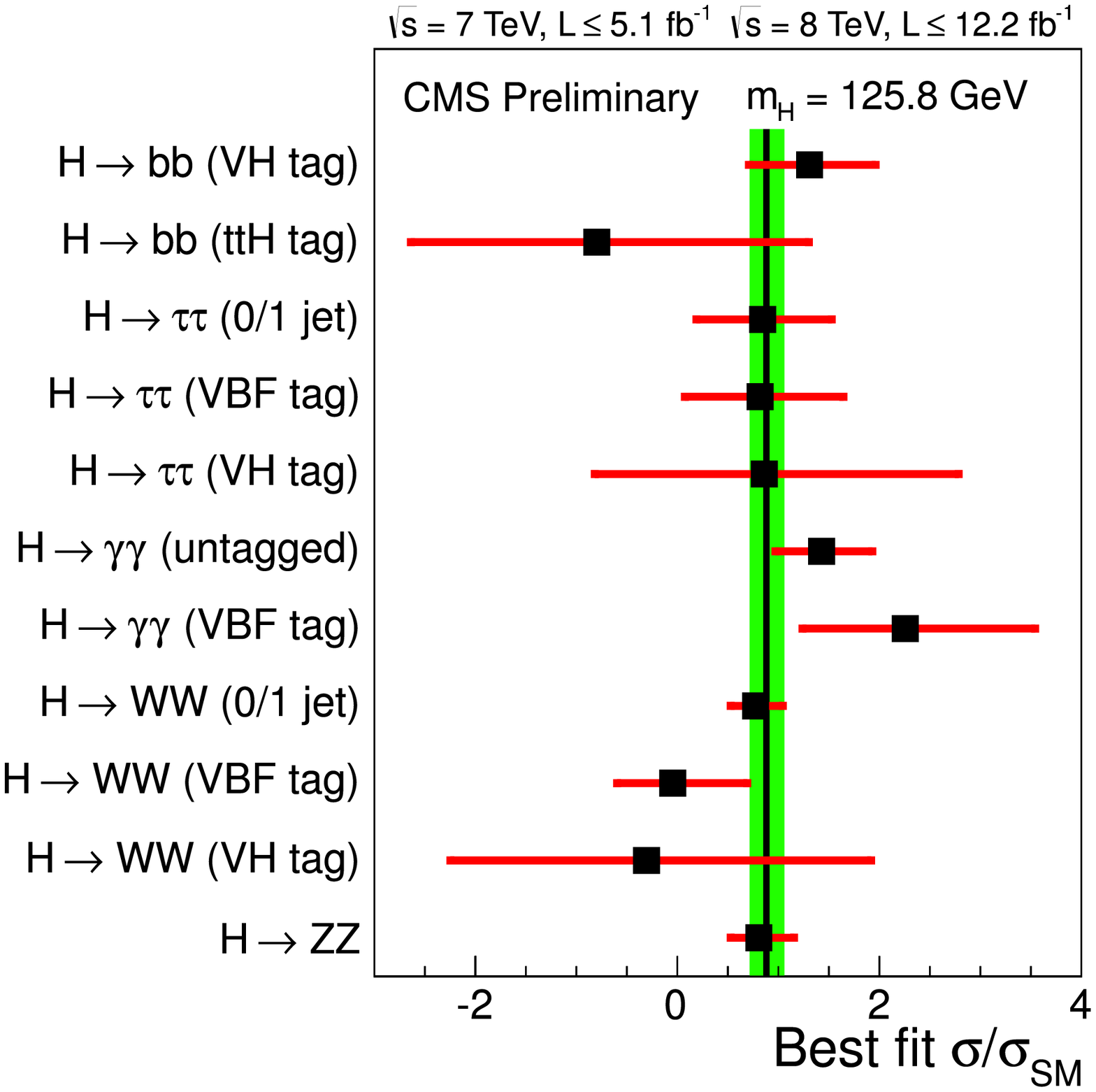}
\hfill
\raisebox{-0.2cm}{\includegraphics[width=.48\linewidth]{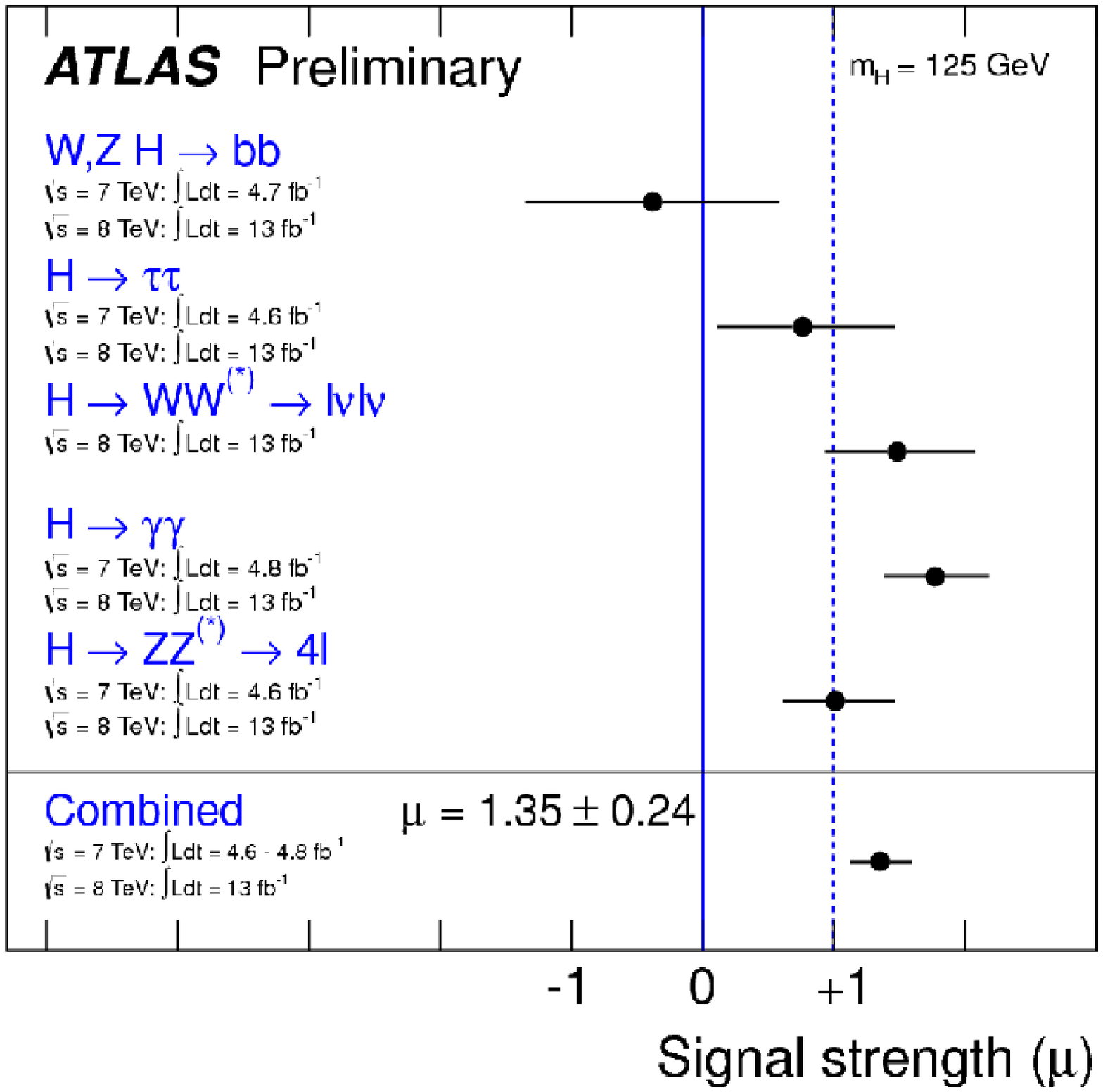}}}
\caption{Signal strength in the several channels sensitive to a Higgs boson with a mass
close to 125~GeV in CMS~(on the left) and ATLAS~(on the right). SM prediction should be
centered at 1, which is compatible with the measured values and with the combined average.}
\label{fig:signalstrength}
\end{figure}

The first set of results is the comparison of the signal strength for the several channels that
have been investigated. The results are summarized in the plots of Fig.~\ref{fig:signalstrength}. As mentioned
in the previous section, values are not completely matching the expectations from the SM, but
they are not significantly discrepant. More data will be needed to reduce the uncertainty and
investigate possible anomalies in the production and decay mechanisms. Explicit disentangling
of the couplings show they are fully compatible with the SM expectations, as in~\cite{bib:cms:h:couplings}.

After the production mechanism has been checked, the first obvious property to measure is the
mass of the found resonance. Dedicated studies has been performed at the two collaborations
using the most sensitive channels. In the case of CMS, the last study has been based on the 4-lepton
sample and provides a mass value
of $m(H)=126.2 \pm 0.6\text{(stat)} \pm 0.2\text{(syst)}$~GeV~\cite{bib:cms:h:mass}.
In the same analysis, studies of the spin and the parity leads to the conclusion that
the data clearly favours a pure scalar versus a pseudoscalar. Additionally, data is not precise
enough to distinguish between spin-0 and spin-2 particles in this channel.

In the case of ATLAS, the results presented in~\cite{bib:atlas:h:mass}
show some tension between the masses
extracted from the 4-lepton and the diphoton channels. In the first case a value of
$m(H)=123.5 \pm 0.9\text{(stat)} \pm 0.3\text{(syst)}$~GeV is obtained. For the second,
the value is $m(H)=126.6 \pm 0.3\text{(stat)} \pm 0.7\text{(syst)}$~GeV, in better agreement
with the measured value at CMS using the 4-lepton channel. This discrepancy will require some
further investigation and perhaps data to be understood. It should be added to the issue that
the signal strength values as measured by ATLAS tend to be higher than the SM expectations.

In addition to the mass measurement, studies of the spin and parity has also been performed
by ATLAS~\cite{bib:atlas:h:properties}. They are similar to those by CMS, but more complete
since information is also extracted from the $H\rightarrow\gamma\gamma$ analysis. This has allow to
add more sensitivity to the distinction between sipn-0 and spin-2 particles.

\subsection{Other searches for SM-like Higgs and within models of new physics}

Even though a boson that is a good candidate to be the Higgs as predicted by the SM has been found, other
analyses looking for SM-like Higgs bosons are still of interest. The main motivation is that they
may be sensitive to scalar resonances with a mass larger than that of the boson, or smaller but with
lower production cross sections.

Most of these searches are following very closely the searches for the SM Higgs at the correspondent
masses, since they inherit from analyses performed before the boson was observed. They are naturally
diverging from the optimal search for the SM Higgs, in order to look for similar particles, but not
with exactly those properties of the SM Higgs. Many searches has been performed by ATLAS~\cite{bib:atlas:h:others}
and CMS~\cite{bib:cms:h:others} and have computed limits for possible presence of particles that
are SM-Higgs alike, since no hint for a resonant scalar has been seen.

Furthermore, several BSM theories include the modification of the Higgs-sector,
which implies that other Higgs particles may be present in Nature, even with the
presence of the SM one. The suggested discrepancies in the Higgs properties add further motivations
for this kind of models. Note we discussed them here even if searches for BSM physics
are included in sections~\ref{sec:exotics} and~\ref{sec:susy}.

As usual in searches for new physics, supersymmetric models are the most attractive to be considered.
In the case of Higgses, Supersymmetry~(SUSY) requires the presence of at least five Higgses, one basically like that
predicted in the SM and others that are relevant due to their properties: charged Higgses and Higgses
with enhanced couplings to bottom quarks and $\tau$ leptons. This later case motivated the search
for a Higgs decaying into $\tau^+\tau^-$ interpreted in SUSY models. Lack of any observed signal
bring the experiments to use the results~\cite{bib:atlas:htautaususy,bib:cms:htautaususy} to set
constraints in the SUSY parameter space.

In addition, searches for charged Higgs have been performed in order to look for their presence in
decays of the top quark. CMS has focused on the $\tau$ channel~\cite{bib:cms:chargedhiggs}, looking 
for an anomalous presence of $\tau$-based decays with respect to other leptonic channels. Limits
were set for several models due to the good agreement of the data with the
W-only-decay hypothesis. In the case of ATLAS, one of the investigated channel was
$H^\pm\rightarrow cs$~\cite{bib:atlas:chargedhiggs},
in which the presence of a dijet resonance not peaking at the mass of the W boson will be identified
as a signal. In addition, we expect a lower yield due to the competing channel that is purely hadronic~(assuming
that the charged Higgs decay preferably into that channel). Data does not confirm these
expected anomalies, so additional limits are set for this kind of model.

Aside for the basic SUSY models, other extensions of the SM incorporate modifications of the Higgs sector
and therefore they have been searched for. There are many possibilities here, and several classes
of Higgses show up. However, we should emphasize that some of them yield topologies
that may have been missed due to kinematic selection, as it is the case of Higgses with
low masses (as the dimuon resonance search in~\cite{bib:cms:dimuonhiggs}) which may be produced just as
boosted objects due to their own couplings. Other possible exotic particle in the Higgs sector
is the presence of doubly-charged particles whose searches, as the one in~\cite{bib:atlas:doublychargedhiggs}, 
have not reported any visible discrepancy with respect to the expected SM backgrounds.

In conclusion, no significant
hint of alternative or extended Higgs sectors has been found to complement the boson
observed at a mass around 125~GeV. However, this does not imply that the physics beyond the SM
is out of reach, since the Higgs sector is well known for providing very elusive particles. For this
reason, searches of new particles have been performed independently of the discovery of
the possible Higgs, as discussed in the following sections.

\section{Searches for new physics}
\label{sec:exotics}

As it has been discussed before, the LHC is intended as a machine to bring information about
new physics beyond the SM. The possibility that the Higgs boson has been found does not only
confirm the validity of the SM, but also its limitations that should be investigated to find
even more correct answers about the structure of the Universe at the smallest distances.

Finding these answers at the LHC requires a huge effort in order to cover the many
possibilities, and therefore corners of the parameter space. This makes the search topic
a very broad field of investigation. In this report we just summarize the most
interesting searches of all those developed at the LHC.

Within the searches for BSM physics, the models involving SUSY are strongly
motivated due to their good theoretical performance to solve the SM limitations. Specifically
the doubling of the particle spectrum, in order to have a supersymmetric partner to each
SM particle, allows a very reach phenomenology that translates into many analyses
investigating several types of final state topologies. Those are discussed in section~\ref{sec:susy}.

On the other hand, there are well-defined alternatives to supersymmetric models that also
provide possible explanations to the issues of the SM as the full description of the Universe.
In the following subsections we focus on summarizing the searches for these alternative models.

\subsection{Searches for unknown high-mass resonances}

\begin{figure}
\centerline{\includegraphics[width=0.51\linewidth]{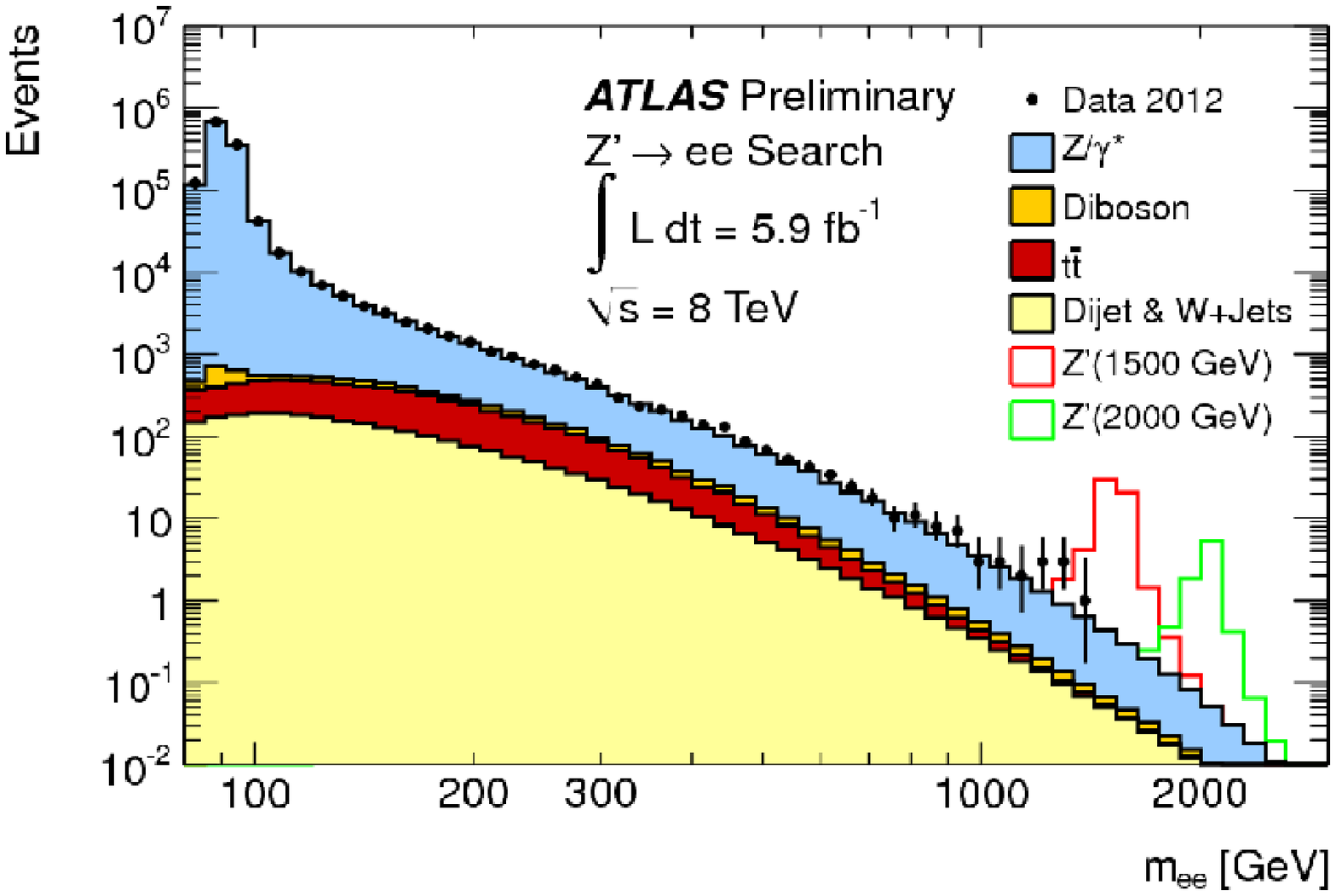}
\hfill
\includegraphics[width=0.51\linewidth]{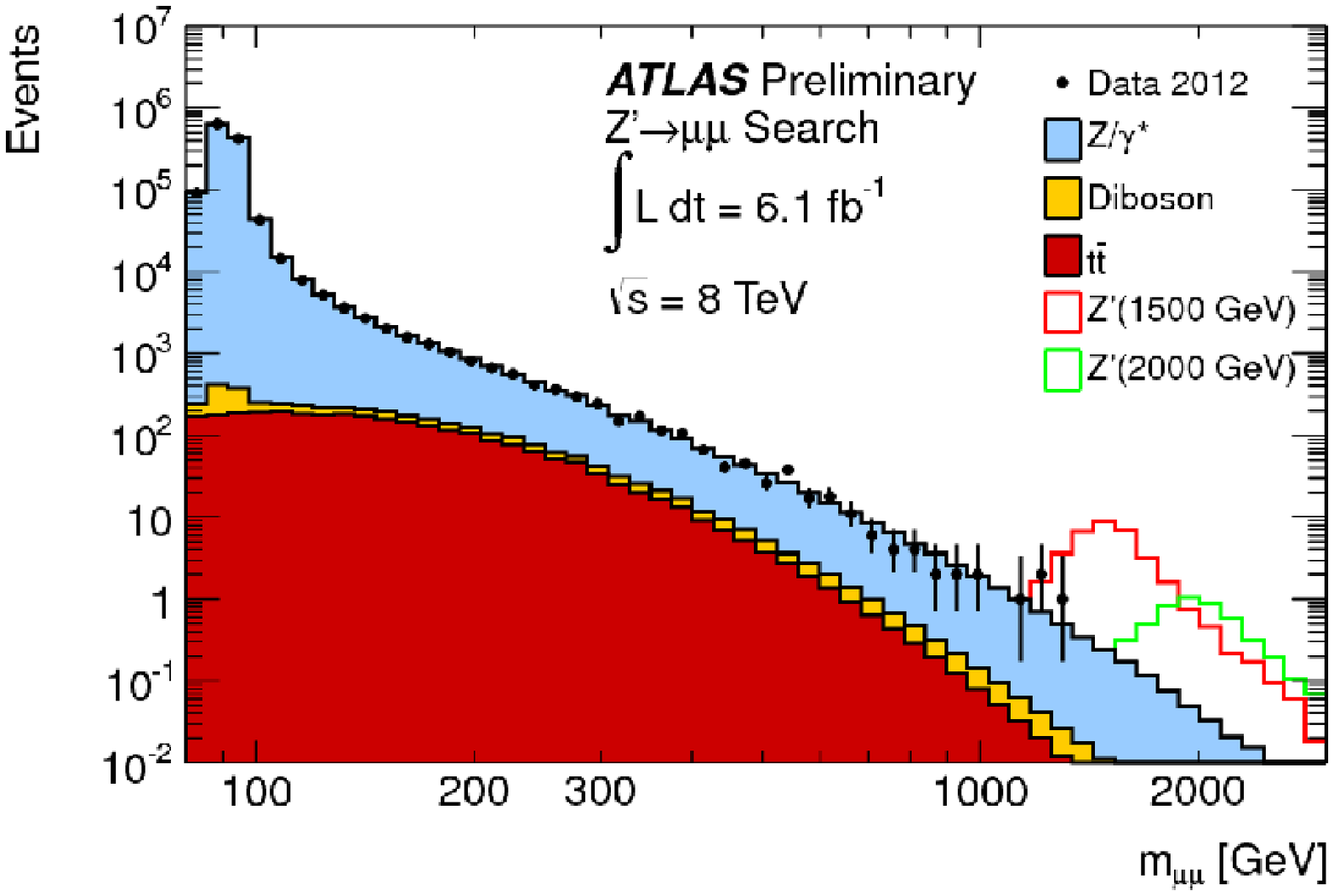}}
\caption{Invariant mass distributions of electron-positron~(left) and dimuon~(right)
in dilepton events collected by ATLAS. Data~(dots) are compared with the SM
predictions~(solid histograms) and some expectations for BSM resonances~(lines).}
\label{fig:dileptonres}
\end{figure}

When looking for new physics, the more direct approach is to look for particles
that are not included in the SM spectrum. For that, the search for resonances
decaying into detectable and well-known particles is the simplest approach. Some
of these resonances are naturally predicted in extensions of the SM, specially with the
addition of new interactions. Figure~\ref{fig:dileptonres} show the 
invariant mass of dileptons as measured by ATLAS when looking for a massive resonance
that would appear as a peak in those distributions. The comparison of the data with the 
SM expectation show
very good agreement and results~\cite{bib:atlas:zprime}~(and~\cite{bib:cms:zprime} for CMS)
are used to set limits on the production cross section for resonances, and lower mass
limits on possible
Z-like particles in the order of 2.5-3~TeV.

Similar to the lepton search, the production of dijet resonances has also been considered,
as in the CMS result documented in~\cite{bib:cms:dijetresonan}, in which special
treatment has been performed in order to separate between resonances decaying into
gluons or into quarks (or a mix). 
Also in this case, a good agreement
has been observed, but the main issue is how to handle the huge background at the lower
invariant mass, that forces to reject events even at the trigger level. 

This has been
the testing analysis of a new technique, called \emph{data scouting}, which allows to
collect interesting events passing around the trigger limits. The idea is to collect
events at a higher rate but storing only the final reconstructed objects, which
allows the reduction of the data content per event. This permitted CMS to trigger and perform
studies for lower invariant masses with competitive results~\cite{bib:cms:datascout}
even with a reduced datasample of 0.13~fb$^{-1}$. 

When looking for
resonances, the presence of neutrinos is not a limitation, and the
search is also extended to the use of the \emph{transverse mass} of a lepton and the
$E_T^{\text{miss}}$, defined as

\begin{equation*}
      M_T = \sqrt{2\cdot p_{T,\ell}\cdot E_T^{\text{miss}}\cdot \left(1-\cos\Delta\phi_{\ell,\nu}\right)}\,\,\;
\end{equation*}

\noindent
to investigate the presence of new resonances decaying into a charged lepton and a neutrino.
In the case of a resonance, this variable shows a Jacobian peak that is on top of a smooth background.
The current results, as those in~\cite{bib:cms:wprime},
do not show any hint of such type of structure, and limits on production of W-like particles has been
set.

However, when we talk about limits on very massive W-like particle, a possible decay channel is into
a top and a bottom quark, which is not allowed for the W.
This was investigated by ATLAS~\cite{bib:atlas:wprime} and found no sign of a resonance decaying into
those quarks, and independently of the number of the identified b-jets. It should be noted that the searches
of this kind of resonance have become very powerful at the LHC due to the available energy for
producing high-mass resonances decaying to the most massive particles in the SM spectrum. This is
also confirmed in the study of resonances decaying to weak bosons, which are predicted to
appear in several BSM theories. A result by ATLAS has taken advantage of the trilepton final
state to look for resonances decaying into WZ~\cite{bib:atlas:wzresonance}, providing a very
competitive result, although usually this kind of search is performed with semileptonic or
fully hadronic channels to make use of the larger branching ratio.

In fact, the energy at the LHC is so large and the possibility for producing resonance so large, that
very massive object could appear and the decay products will be boosted, which may lead to dijet
(e.g. from W) merged into one reconstructed jet. This has been turned into a benefit to enhance signal,
by using merged jets to tag the presence of hadronically decayed bosons. The result of the
analysis by
CMS~\cite{bib:cms:dibosonboosted} shows the good performance of the boosted-jet tools. Unfortunately
no sign of new physics was found. 

A similar analysis by ATLAS looking for a resonance decaying into ZZ in the
semileptonic channel~\cite{bib:atlas:zzresonance} also exploits the merged jet topology to increase
acceptance to very massive resonances and set a much constraining limit than that accessible by the obvious
dijet topology.

Among the searches for resonances indicating BSM physics, one common topic is the studies of
possible excited states of fundamental particles, which could be related to new physics (e.g.
contact interactions or internal substructure). This is the case for the search of excited
muon states decaying into a muon and a photon as the one by ATLAS~\cite{bib:atlas:excitedmuon}
looking for the Drell-Yan production of a muon and an excited muon. The results are in good
agreement with SM predictions for the most discriminant variable: the invariant mass of
the two muons and the photon, which allows to set stringent limits in the possible scale for
such a excited state to exist.

In addition to the searches for resonant states in
the two-body decays, the high masses accesibles at the LHC allows the searches for
more complicated topologies, with more objects in the final state. One example is the search
for boosted resonances decaying into three jets. The search performed by CMS~\cite{bib:cms:threejets}
assumes pair-production of these objects, and therefore the idea is to study three-jet
ensambles whose transverse momentum is large but the corresponding mass may show a peak structure
related to a decaying resonance. The requirement of large transverse momentum allows the
reduction of the combinatorial background, for which the mass and the transverse momentum 
will show a correlation. Although the result of the analysis does not show hints of
any possible resonance, the used technique can be used in other searches in the future. In
the current case, limits are set on the existence of resonances.

Another alternative that is open at the LHC is the cascade decay with initial massive objects
sequentally decaying into states. A very syummetric case considered at CMS consists
on the pair production of objects (e.g. technicolour particles) decaying into pairs of particles
(e.g. other lighter state in the technicolour spectrum) which decay into dijet. This
process will lead to an 8-jet topology in which there are resonant peaks in four dijet masses,
two 4-jet masses and perhaps in the 8-jet mass in case the original pair-production occurs
from the decay of a single-produced particle. All this information is combined into an artificial
Neural-Network to enhance signal-like topologies. The results~\cite{bib:cms:eightjets} show
that there is no peak structure on top of the combinatorial background coming from usually-produced
8-jet events and limits has been set for models motivating this kind of 
signature.

\subsection{Searches for leptoquarks}

One special case of pair-produced resonances that are motivated by unification models is 
\emph{leptoquarks},
particles having both lepton and baryon numbers. They are detected via their
decay into a lepton and a quark, which gives a resonant peak in the invariant mass (in
the case of charged leptons) or significant excess in $E_T^{\text{miss}}$-related variables (in the
case of neutrinos). 

Since these particles carry colour, they are pair-produced with a large cross-section, giving
rise to clean signatures due to the leptons in the decay. Furthermore,
they also have a rich phenomenology, since these
particles could be of different classes (scalar, vector) and also appear in different generations,
although they are usually not mixing fermions of different families.

The basic analyses, mostly oriented to the first two generations are easily identified
by the kind of lepton, which determines the generation we are focusing. Searches by
ATLAS~\cite{bib:atlas:leptoquarks} show good agreement with the SM expectations for the
$eejj$ and $\mu\mu jj$ final states. These results are used to set limits that are going
beyond previous searches of these particles. 

Since the first generations are not providing hints of leptoquarks, even in the channels with
neutrinos, searches have also been focused on the third generation, where $\tau$ leptons and
bottom quarks are expected. Specifically the search by CMS~\cite{bib:cms:leptoquarks}
with the use of b-jets exploits
the sensitivity given by the scalar sum of the transverse momenta of the decay products. The results
are in good agreement with the SM expectations, and they are used to set limits on the
leptoquark production, but also on the production of scalar tops within
R-parity~($R_P$)-Violating SUSY models~(see details in section~\ref{ssec:rpvsusy}),
giving an explicit proof that searches of new physics
are usually sensitive to several classes of models bringing similar final states, an in
similar areas of the phase space.

\subsection{Extradimensions and graviton searches}

The extensions of the SM do not only consider the extension of the particle spectrum or
the interaction sector. Several models introduce the modification of the structure of
the Universe by incorporating additional dimensions, that would be microscopic and whose
existence may explain the large scale difference between the electroweak interaction
and gravitation.
The idea is that the new dimensions will be forbiden to the SM particles and effects, while
gravity expands in all the available dimensions. The signatures will be striking with
the production of gravitons (producing large $E_T^{\text{miss}}$ since they escape
detection) and SM particles, leading to single-photon~(monophoton) or single-jet~(monojet)
topologies, 

These have been looked for by the collaborations. As an example, ATLAS has looked
for events with a photon with large transverse momentum that is
accompanied with large $E_T^{\text{miss}}$,
which is the most significant variable to identify the presence of new
physics~\cite{bib:atlas:monophoton}. Good agreement is observed with respect to
the SM expectations for this signature, dominated by undetected weak bosons (neutrino decays)
in association with a photon. Also some background contribution is present due to detector 
effects generating artificial kinematics looking like the signal.

Furthermore, ATLAS and CMS have also looked for the monojet
topology~\cite{bib:atlas:monojet,bib:cms:monojet}.
Although the main motivation for this signature
is the production of gravitons produced in association with quarks, there
has been an increase use of this kind of search for studying the production of
invisible particles~(as generic Dark Matter candidates) in a model-independent way, being
the jet balancing the $E_T^{\text{miss}}$ produced by initial-state radiation. This
keeps a small fraction of the total signal, but allows to look for hard-to-detect
particles that may be copiously produced at the LHC collisions. It should be remarked that
this makes a strong case when compared to the more clean monophoton signature: results are more sensitive
to other classes of models.

\begin{figure}
\begin{minipage}{0.48\linewidth}
\vspace*{-0.3cm}
\centering{\hspace*{-0.7cm}\includegraphics[width=1.2\linewidth]{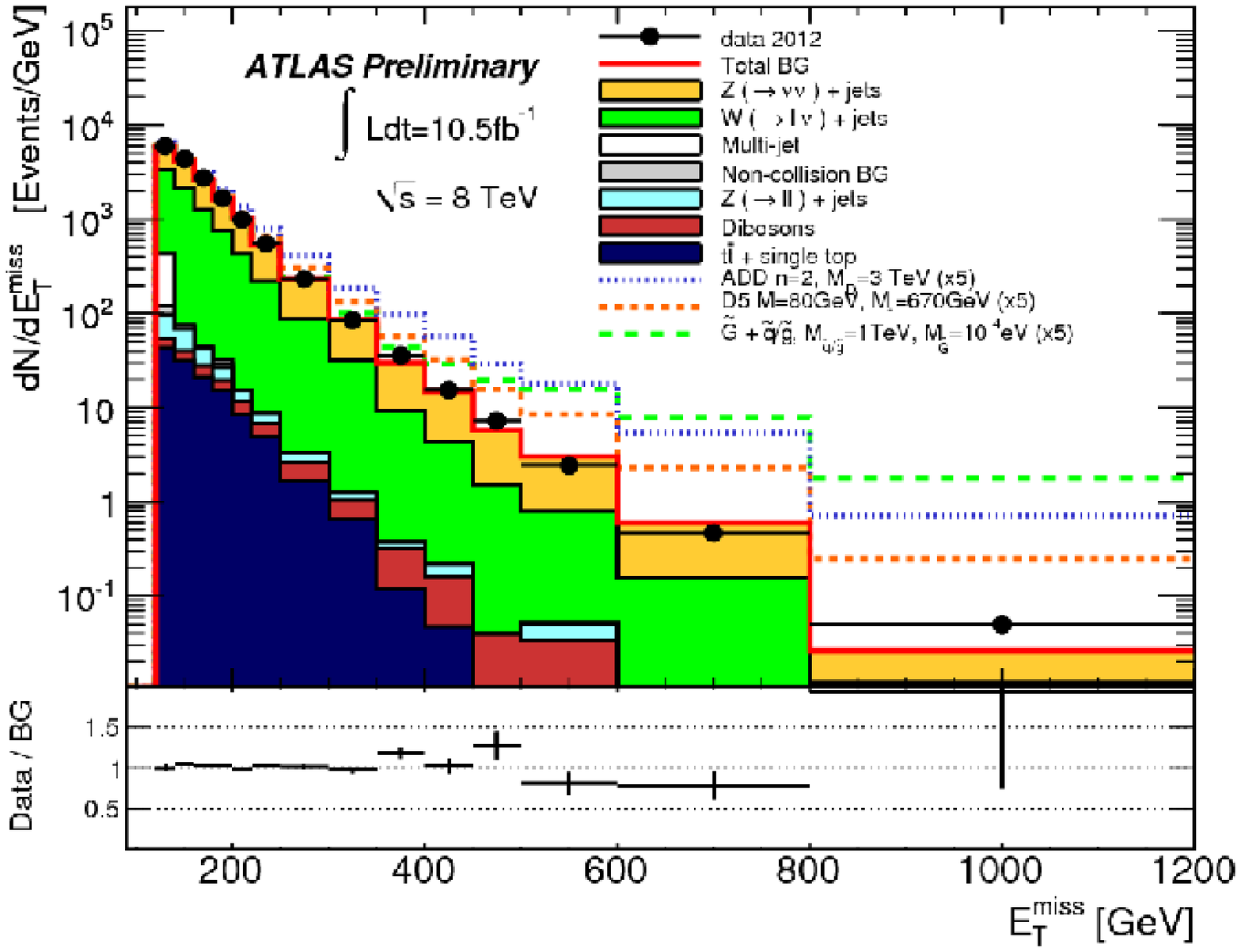}}
\caption{Distribution of the $E_T^{\text{miss}}$ variable in events with a single
jet with large transverse momentum. Data~(dots) are compared to the background expectations~(filled
histograms) and possible signals for BSM physics~(coloured lines).}
\label{fig:monojet}
\end{minipage}
\hfill
\begin{minipage}{0.48\linewidth}
\centering\includegraphics[width=.95\linewidth]{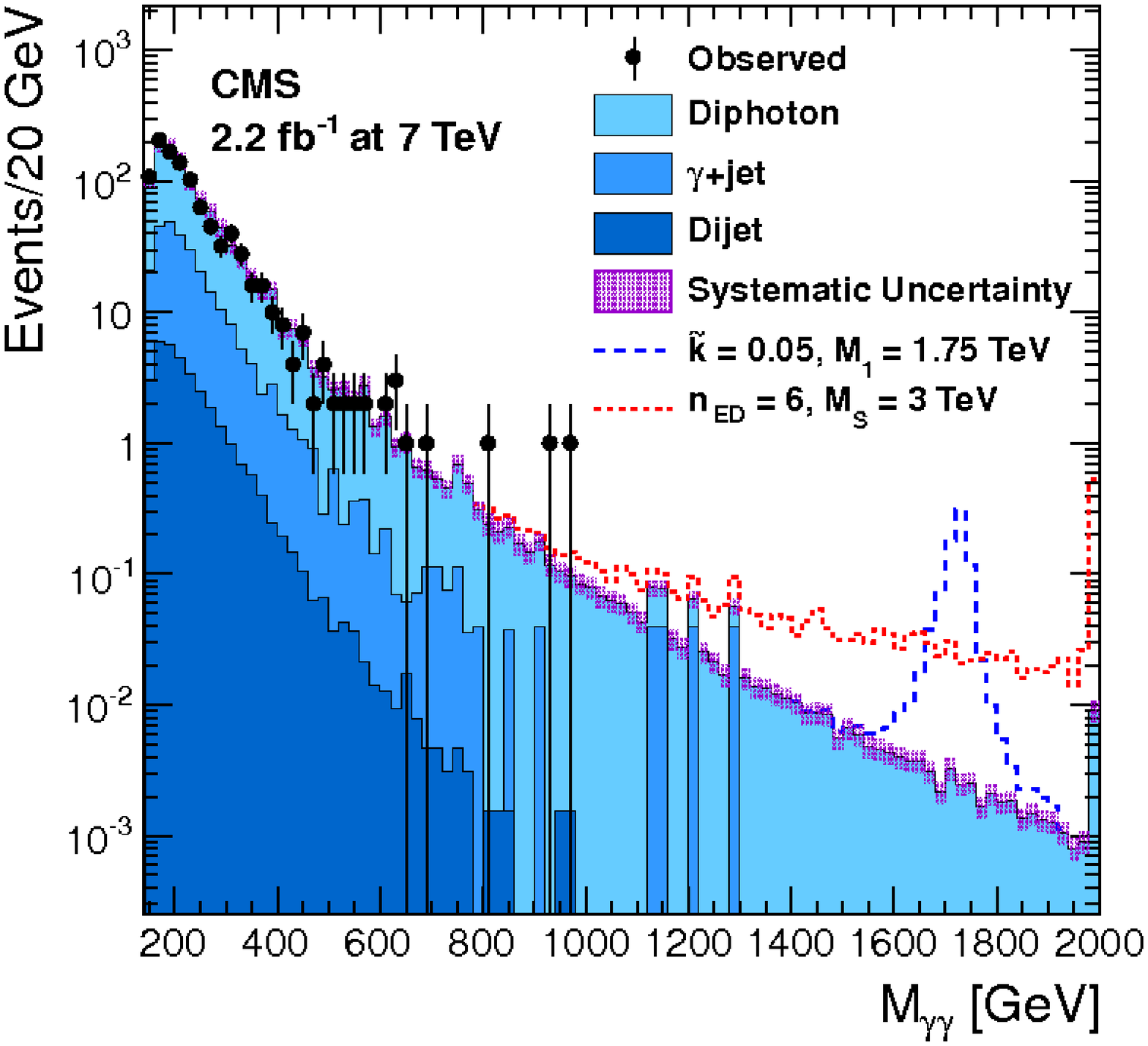}
\caption{Distribution of the diphoton invariant mass in events with two photons with large transverse
momentum. Data~(dots) are compared to the background expectations~(filled
histograms) and possible signals for BSM physics~(coloured lines) containing a Randall-Sundrum
resonance or a generic extradimention model.}
\label{fig:diphoton}
\end{minipage}
\end{figure}

The results of the monojet searches has also found good agreeement with the SM predictions.
Figure~\ref{fig:monojet} shows the $E_T^{\text{miss}}$ distribution of
the ATLAS analysis~\cite{bib:atlas:monojet}, that has also been used to set limits in
the production of gravitino from the decays of squarks and gluinos.

Another possiblity related to extra-dimensions and accessible production of graviton is that
particles may appear as Kaluza-Klein towers which sequentally decay into less massive objects.
Specifically, gravitons may appear as diphoton resonances, which is an easy-to-identify signature,
but it suffers from large backgrounds. Anyway, they have been investigated by the LHC experiments,
as the analysis in~\cite{bib:cms:kkdiphoton}, and no hint of such a resonance has been found
on top of the diphoton high-mass spectrum, as shown in Fig.~\ref{fig:diphoton}, which also includes
the expectation from a resonance as those predicted by Randall-Sundrum models and the expected
effect due to a more generic model including additional dimensions.

\subsection{New physics in the top sector and new generations}

As discussed before, the top quark is usually suggested as the primary candidate
to open the path towards new physics. Its large mass and coupling to the Higgs,
which are the basic quantities related to the loose ends of the SM, make
this quark a very attractive place to search for discrepancies with respect to
the SM expectations,
Since the first step to fix the hierarchy problem is to have a partner
canceling the top-induced corrections to the Higgs mass, such a partner should be
at reach of the LHC independently of its nature. Although the most obvious choice is
a SUSY partner~(see section~\ref{ssec:naturalsusy}), alternative options have been made,
including the possibility of the existence of a very massive 4$^{\text{th}}$ generation.

One option considered by ATLAS~\cite{bib:atlas:top53} is the search for the pair-production
of a top partner having an electric charge of 5/3 (of that of the positron). Appearing in
several models, the decay into a top quark and a W boson allows to have good acceptance
with same-sign dileptons and also to use the hardness of the event (scalar 
sum of transverse momenta of final objects) as discriminating variable. No significant
discrepancy has been observed with respect to the low expected SM background.

As a general rule, the existence of additional generations (containing canonical or exotic particles)
that would contain coloured particles more masive than the SM ones leads to very busy
final states in terms of multiplicity and of energy. This is used in the optimization looking
for this kind of topologies, being very common the requirement of hard events or with
rare combination of objects~(same-sign leptons, leptons and b-jets in high multiplicities and
similar requirements). The performed analyses searching for a 4$^{\text{th}}$ generation, as those
in~\cite{bib:atlas:4gen,bib:cms:4gen}. All searches have brought the conclusion that there
are no hints for the existence of a 4$^{\text{th}}$ generation (in the reachable masses) nor of
any new physics that may look like massive particles regarding busy final-state topologies.

\subsection{Searches for very exotic signatures}

The lack of success to find hint of straightforward
BSM physics has open the possibility that Nature is not as
predictable as we might think and the new physics may appear in some even more exotic signatures
than those considered for the theoretically-motivated BSM final states. This has led to 
the study of final states
that could have escaped the more traditional selection or based on models less
related to the confirmed SM predictions, which bring to new classes of final states.

One option that has been considered is the production of microscopic black holes at the LHC
collisions. Some generic properties of them from quantum gravity provide general
rules of final-state expectations: high multiplicities and democratic treatment of objects. The search
for this kind of events~\cite{bib:cms:blackholes} was performed by exploting that
the scalar sum of transverse energies for the SM background presents a shape that is independent
of the multiplicity. Therefore the lower multiplicity events are used to get the shape that
is compared to the data with high multiplicities. Good agreement has been observed and limits
in a model-independent approach are set.

Other rare topology that is not commonly considered is the presence of long-lived particles
that may escape even the trigger selection. Some of these particles appear in several models
as quasi-stable particles. In this context, searches for charged massive particles (CHAMPs)
by CMS~\cite{bib:cms:champs} or more dedicated searches like the stable chargino using track-disappearance
by ATLAS~\cite{bib:atlas:stablechi} are good examples of the possibilities beyond the usual
approaches and how the detectors <re used with non-standard event reconstructions to look
for unexpected classes of particles. In this, we should also mention the search for magnetic
monopoles~(as that by ATLAS in~\cite{bib:atlas:monopoles}), whose existence is very strongly
motivated due to the electric charge quantization and as part of the electromegnetic
unification. The need for specific reconstruction of the events (since these particles are not
electric charges, and behave very differently inside magnetic fields) add some complication to the
analysis, but still the results are very competitive when compared to direct searches because of
the possibility to produce them with high cross section at the LHC. In any case no hint
for production of monopoles has been observed and further data will help to increase the sensitivity,
specially with the addition of the dedicated experiment for this (MoEDAL~\cite{bib:moedal:webpage}).
 
In conclusion, after the first datasamples provided by the LHC collisions have been analyzed, no
discrepancy with the SM prediction has been found that could be considered as a significant
hint of new physics or particles beyond the SM spectrum. The future running of the LHC at
a higher energy and higher luminosities, discussed in section~\ref{sec:future}, should
provide more information on the possible BSM physics.

\section{Searches for supersymmetry}
\label{sec:susy}

In SUSY models the particle spectrum is at least doubled~\cite{bib:susy}, bringing a lot of possible processes
that could distort the measured values with respect to the SM expectations. Depending on
the considered process, the final state to be investigated is different, providing a rich
phenomenology.

However, since at the LHC the initial state is based on partons, the dominant
production mechanism is usually the production of coloured superpartners. In usual models they are
produced in pairs since R-parity ($R_P$,
a quantity being 1 for particles and -1 for superpartners) is
conserved. In addition, the conservation of $R_P$ implies that the lightest SUSY
particle~(LSP) is stable and a Dark Matter candidate.

These basic properties allow to make general analysis in searches for SUSY which focus
on specific parts of the spectrum. In addition, this also brought a new way of interpreting the
results which are based on ``simplified models'' which provide well-determined processes
for the given final states. This has simplified the interpretation of the results in terms of
the possible theoretical models. On the other hand, the more traditional, ``full model'', approach are
still advantageous to interpret results from different analysis and experiments within a common
framework.

Independently of the model the most basic search for SUSY is to look for jets and $E_T^{\text{miss}}$.
The latter being a hint of the stable LSP, and the jets appearing as the decay products of coloured
superpartners, which are the ones associated
with larger production cross sections: squarks and gluinos. These analyses are
just dependent on the reconstruction of the $E_T^{\text{miss}}$ and they try to quantify its
presence with variables that are less sensitive to misreconstruction. In addition
several categories are investigated in order to be sensitive to different kind of SUSY processes. The
categories are usually identified by the hardness of the event~(with $E_T^{\text{miss}}$ or momenta of
jets), the multiplicity of jets, or the multiplicity of b-jets.

The analyses by the collaborations, as those in~\cite{bib:atlas:jetmet,bib:cms:jetmet}, do not show
any significant discrepancy with respect to the expected backgrounds. Results are used to set limits
in several types of models, and are typically excluding the presence of squarks (of
the first generations) and gluinos below 1-1.5~TeV.

In the case of massive squarks, it is feasible to produce gauginos that are lighter but still hard
to produce directly from the proton collisions. These gauginos may decay in leptons with
large transverse momenta which simplify the identification of the events at the trigger
and reconstruction levels. Both collaborations have searched for SUSY events in final
states with leptons, jets and significant
$E_T^{\text{miss}}$~\cite{bib:atlas:ljetmet,bib:cms:ljetmet} and the results shows good agreement
with the SM expectation. Results have been used to set limits on the production of SUSY particles
that produce leptons in the final state. It should be noted that the studies with leptons include
the $\tau$ lepton~(as in~\cite{bib:cms:taussusy}) since they provide increased sensitivity to
the case of Higgsino-like gauginos.

When one consider leptons in the final state, the presence of multileptons may be a good hint of SUSY
due to the reduced SM backgrounds. Specially when there are at least three leptons and significant
$E_T^{\text{miss}}$, which is the golden final state detecting the production of a pair of
chargino and neutralinos decaying leptonically or even production of scalar leptons.
The background of these kind of studies~\cite{bib:stlas:multilepton,bib:cms:multilepton} is dominated
by diboson (or multiboson) production in which leptons are the decay products of the massive
weak bosons. 

Again, the presence of $\tau$ leptons is fundamental is some areas of the parameter
space since the gauginos may not be as ``flavour symmetric'' as the corresponding SM bosons. In any
case, no significant excess has been observed and the results are used to set limits
on the production of gauginos. It should be noted that this kind of final state is sensitive to
a different area of the SUSY parameter space, so they are complementary to the search
of events in which coloured superpartners are producted and sequentially decay into SM particles.

\subsection{Gauge-mediated Supersymmetry breaking}

After the simplest topologies have been investigated and report negative results
regarding the existence of SUSY, other models providing significant differences
in the final states need to be considered. A qualitative change is set by models in
which SUSY is broken in a hidden sector and communicated via gauge interaction~\cite{bib:gmsb},
since the LSP is the gravitino and the phenomenology depends on the next-to-lightest 
SUSY particles~(NLSP) because most of the decays go preferably via that particle.

In the cases where such particle is a scalar lepton, usually the scalar $\tau$, the
final state contains leptons that are easy to identify. Searches by both
collaborations~\cite{bib:atlas:taugmsb,bib:cms:taugmsb} show good agreement with
expectations in several types of final states.

Other case that is very relevant is when the NLSP is a neutralino, decaying into a gauge
boson (usually a photon) and the gravitino. This is also a relatively simple
final state, since the presence of photons helps to make the event selection much cleaner.
The analysis searching for diphoton and $E_T^{\text{miss}}$ by CMS~\cite{bib:cms:metdiphoton}
observed a good agreement between the observed data and the expected SM backgrounds,
as displayed in Fig.~\ref{fig:susydigamma}, where the $E_T^{\text{miss}}$ distribution in
events with two photons is shown, including some possible signals to explicitly shown the
sensitivity to a signal in this variable.

Even if the considered final state in models with gauge-mediated SUSY breaking was able to avoid
limits set for MSSM-inspired searches, the results are not showing any significant
discrepancy that could be attributed to the production of SUSY particles. 

\subsection{Natural SUSY and third generation squarks}
\label{ssec:naturalsusy}

After the studies of the more obvious SUSY final states, the obtained limits are
moving the SUSY scale to high values so it starts to approach the decoupling with
respect to the electroweak scale. Since the motivation for SUSY is to fix problems
at this latter scale, new concepts are required to keep the connections between the
two scales and, at the same time, avoid the current limits from more
inclusive final states.

In this sense the two obvious things is first to keep the neutralino~(or equivalent) as the LSP
in order to have a Dark Matter candidate that is stable and weakly coupled. Secondly, we need
the lightest scalar top to be light enough to keep the divergences in the Higgs mass as
smaller as possible. This means $m(\tilde{t})\lesssim 400$~GeV.
This expression also requires a gluino not far from 2~TeV to avoid a strong
correction on the scalar top mass. With these requirements, all other SUSY particles
may have any value, since their influence is much smaller. Therefore current limits
on general searches are avoided.

However, this ``natural'' SUSY becomes only completely natural when other superpartners
are associated to the needed ones. For this reason it is not uncommon to have also light
scalar bottom quarks or scalar $\tau$~(as mentioned above). Additionally, the LSP
could be a family of degenerated gauginos of several classes. It should be noted that
in spite of the reduced number of superparticles involved, the possible final
states are very complex due to the involvement of the third generation of fermions.

For example, with the described spectrum, it is feasible to have gluino-pair production
as the process with higher cross section. These gluinos decay into quarks and $E_T^{\text{miss}}$.
In the case the scalar bottom is available, the gluinos may give rise to final states
containing $E_T^{\text{miss}}$ and four bottom quarks, that may be identified as b-jets.
This topology is very clean due to the reduced backgrounds and therefore sensitivity
may be enhanced by the b-jet requirements, allowing some additional room with respect to
the more inclusive limits, where the limitation was the huge backgrounds. The study
done by ATLAS~\cite{bib:atlas:gluino4bs} shows no hint for anomalous production of
multi-b-jets and significant $E_T^{\text{miss}}$, a selection sensitive to this
final state. Limits in SUSY and other models are set. Regarding the interpretation,
it should be noted that this analysis is also sensitive to the
decay into top quarks, since also four bottom-quarks appear in the final state.

On the other hand, the case of top and scalar top quarks produced via gluino production is
much richer than just the presence of b-jets, due to the large multiplicity
of W bosons. It is possible then to identify the events containing four top quarks
and significant $E_T^{\text{miss}}$ in several approaches and with very challenging
final states for the SM expectations:
analyses in this topic~\cite{bib:atlas:4tops,bib:cms:4tops} are testing the SM predictions
in very specific corners of the phase space, and specifically in regions
that were not tested before. Even there the SM predictions provide a very good
description of the measurements, which translates into further contraints to SUSY
production.
 
\begin{figure}
\begin{minipage}{0.48\linewidth}
\centering\includegraphics[width=1.1\linewidth]{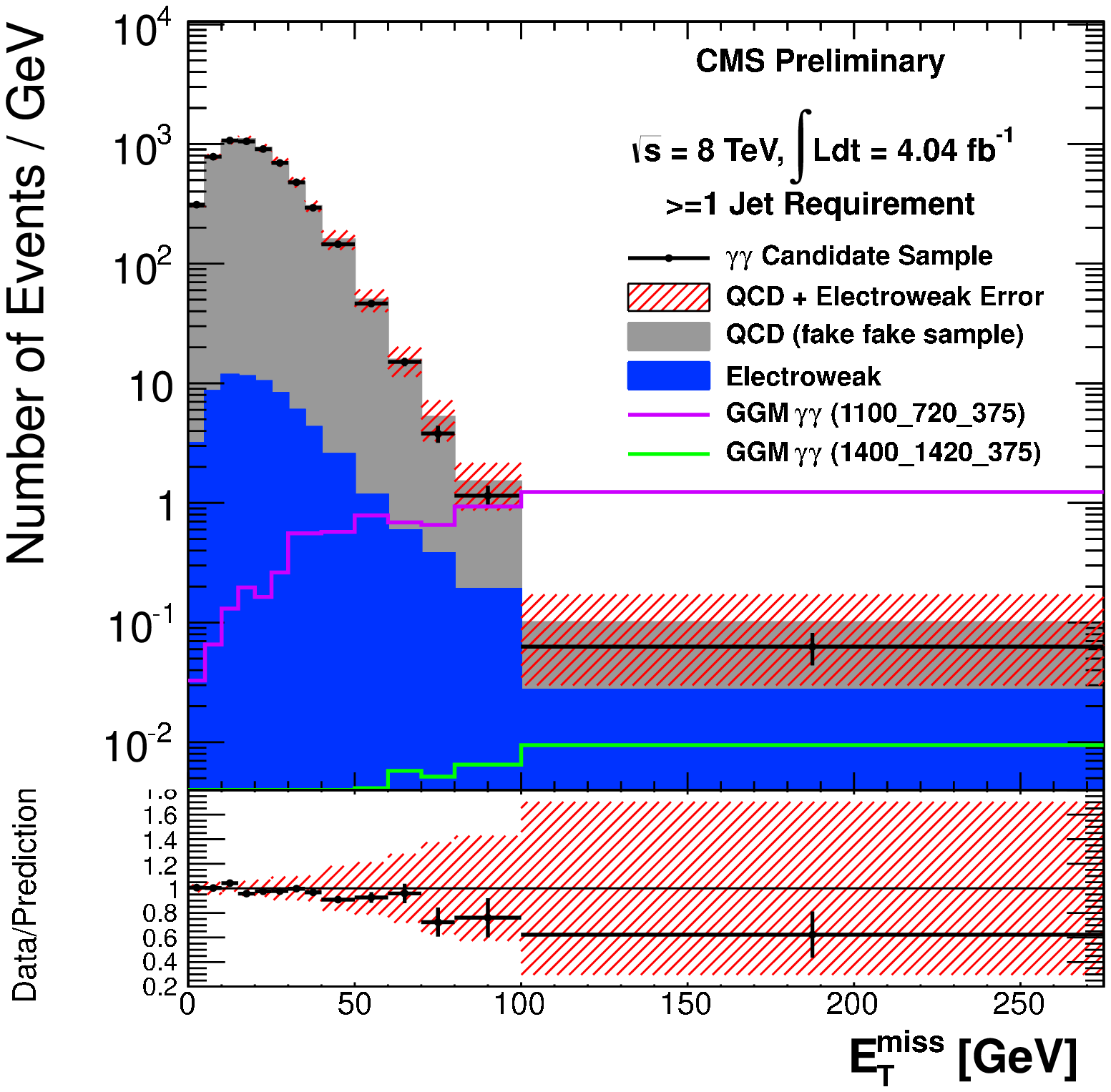}
\caption{Distribution of the $E_T^{\text{miss}}$ in diphoton events as measured by the
CMS collaboration. Data~(dots) are compared to the SM predictions (solid histograms) and
to possible models of new physics~(lines).}
\label{fig:susydigamma}
\end{minipage}
\hfill
\begin{minipage}{0.48\linewidth}
\vspace*{1.6cm}
\centering{\hspace*{-0.5cm}\includegraphics[width=1.1\linewidth]{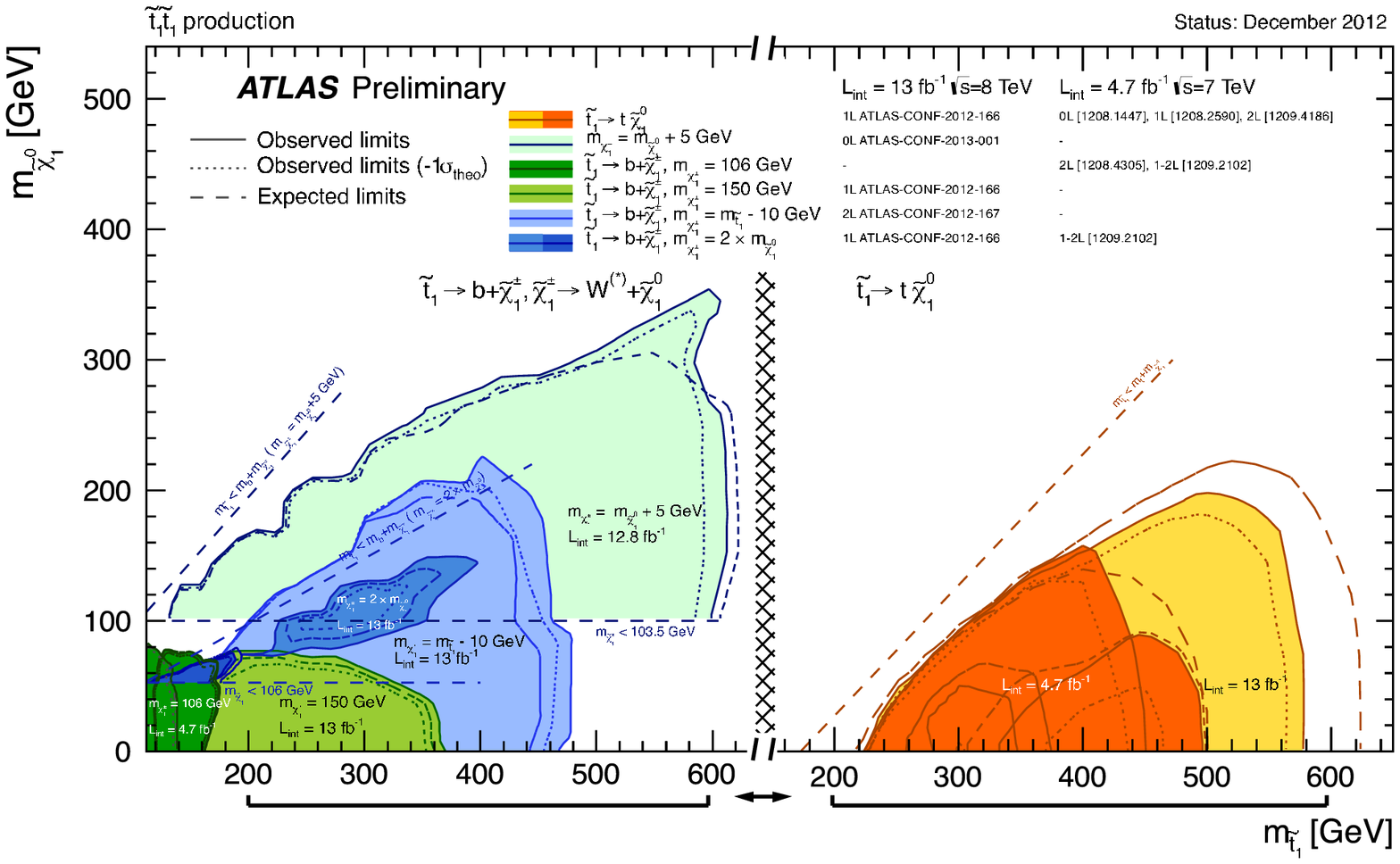}}
\caption{Summary of the limits for scalar tops from the available analyses by ATLAS,
drawn in the neutralino~(LSP)-stop plane according to the assumptions of all the channels.}
\label{fig:susystoplimits}
\end{minipage}
\end{figure}

Even if the use of gluino-mediated production allows the use of striking signatures, it is
more attractive the direct production of squarks of the third generation which are those strongly
motivated to be relatively light, according to ``naturalness''. Therefore experiments performed searches
of scalar bottom quarks as that in~\cite{bib:atlas:sbottom} in which the identification of b-jets is
fundamental to reduce the SM background. In addition, searches for direct production 
of scalar top quarks~\cite{bib:atlas:stop1,bib:atlas:stop2,bib:atlas:stop3,bib:cms:stop}
still provide enough complexity in the final state to allow several classes of searches.
This is seen in summary plots as that displayed in Fig.~\ref{fig:susystoplimits}, containing the
exclusion areas from several searches of direct production of scalar top.quarks.

As the summary plot shows, the several assumptions on the decay and kinematics of the final states
allows to exclude large areas of the parameter space. But in summary, the lack of observation
of hints for scalar top quarks just bring the scale for SUSY~(in this case given by the mass of
the scalar top) to higher values, similarly of the results in more inclusive searches. Threfore,
it seems that SUSY may not show up in the most obvious way to fix the issues of
the SM and particle physics.

\subsection{Searches for $R_P$-Violating SUSY}
\label{ssec:rpvsusy}

Although usually it is assumed that $R_P$ is conserved because it directly provides
a Dark Matter candidate, it is obvious that there is no reason a priori why that quantity
needs to be conserved.
By relaxing the conservation condition it is possible to avoid many of the most stringent limits,
since they are usually obtained with the requirement on $E_T^{\text{miss}}$, which is inspired
by the assumption of conserving $R_P$. In addition, the phenomenology becomes much richer
due to the possibilities in the spectrum and in the possible interactions. For example with
the presence of unusual resonances in the final
state (like $\tilde{\nu}_{\tau}\rightarrow e\mu$).

One general characteristic of the $R_P$-Violating signatures is that since all the superpartners decay into
SM particles, the final state usually is related with high multiplicity of objects, and involving
many different types of them. This also brings the fact that basically every final state is
available in $R_P$-Violating SUSY due to the rich phenomenology.

As reference analyses, we should mention the multilepton searches, as that by
ATLAS~\cite{bib:atlas:4lepton} looking for anomalous production of events
containing 4 or more leptons having either large $E_T^{\text{miss}}$ or large energy
activity~(quantified via the concept of effective mass). Reasonable agreement with the
small SM expectations has been observed.

Other typical search in the context of $R_P$-Violating models is the search for resonances
decaying into two leptons of different type, as the one performed by ATLAS in~\cite{bib:atlas:rpvsusy},
wich considers the case of $e\mu$, $e\tau$ and $\mu\tau$ resonances. No hint of such states
was found and limits were set in the relevant models. It should be noted that the open possibilities
in this set of models have the clear disadvantage that the application of the limits is very
reduced in comparison with the parameter space.

A last analysis that needs to be discussed is the search for events containing multileptons
and identified b-jets, that has been investigated by CMS~\cite{bib:cms:rpvsusy}. The interest
of this search is not only about possible presence of new physics, but also since it is sensitive to
very rare SM processes, whose observation is as interesting as the search for BSM physics. This includes
some of the associated production of top quark and weak bosons mentioned in
section~\ref{ssec:topquarkproperties}. Although
no hint of new physics has beeen observed, the analysis already probes the sensitivity to
the rare SM processes that should be investigated in future datasamples collected at the LHC.

\section{Future of the LHC experiments and physics}
\label{sec:future}

After the running ended in March 2013, the LHC accelerator is currently in a shutdown
period which is needed for maintenance and repair work which will allow the running
at the highest energy and luminosity conditions. This shutdown will last until
2015 and it is also used by the experiments for additional improvements and work.

The plan after the shutdown is to run for a few years at nominal energy (probably 13~TeV)
and collect a sample of 100~fb$^{-1}$ . Afterwards a new shutdown is expected to bring
the luminosity to the design value and run for a few more years~(2019-2022) to collect
additional 350~fb$^{-1}$ at a center of mass energy of 14~TeV.

Afterwards a third shutdown will bring the machine to a Phase-2 upgrade that may allow to collect
additional 3000~fb$^{-1}$ along the next decade. All these data will allow accurate
studies for particles and interactions observed during the first runs of the collider. An
alternative will be to upgrade the LHC so it may be able to reach higher energies and
set a new frontier on the investigated energy scales.

In addition to the improvements by the accelerator, the experiments are getting ready to
upgrade their components in order to exploit the possibilities the several stages
of the LHC will provide. ATLAS and CMS will need to face new challenges in terms
of collection rate, luminosity and radiation and are therefore working on improvements
for the DAQ and trigger selection, upgrades of the internal parts of the detectors and
replacements of the parts that may be limiting factors in the incoming phases.

In the case of ALICE, the main goal is to have the best possible detector for the
run after the second shutdown, in order to get all the reachable information about the
heavy ion program of the LHC, hopefully understanding the Quark-Gluon Plasma with
unprecendent accuracy and being able to provide enough information for the theoretical
characterization of its properties. It is not completely clear yet whether ALICE will
be present in the LHC running beyond 2022,

The case of the LHCb is special due to the reduced need for luminosity. The plan is to
collect 5~fb$^{-1}$ after the current shutdown and then collect 50~fb$^{-1}$ during the
main part of the main run of the current LHC. As in the case of ALICE, it is not clear
whether LHCb will be present in future improvements of the LHC projects, either
in terms of luminosity or of new energy regimes.

To summarize, the LHC is planning the future runnings with improved performance
in order to provide large amount of data
that will yield to important measurements
during the several stages of the accelerator. The expected program and
the results from the experiments are awaited from the particle-physics community
to confirm and improve the results already obtained at the LHC and described in
previous sections.

However, it should be
remarked that even the current datasample are still providing important and relevant
results, as reported on the web pages of the experiments~\cite{bib:webpages}.

\section{Overview and conclusions}

The LHC experiments have finished a very sucessful \emph{Run I} with very important
milestones and discoveries in all the topics planned for the program. Confirmations of
the SM expectations, measurements of heavy-flavour and top quark physics and results
related to heavy-ion collisions have clearly overrule most of the previous achievements
due to the new energy frontier, the good performance of the accelerator and the detectors
and also to the high quality of the studies.

In the part dedicated to searches for new particles, which is
the main goal of the LHC, the current results already made the first big discovery
by finding of
a new boson having a mass of 125~GeV. For other possible particles expected in extensions
of the SM, new limits have been set, highly increasing the constraints for BSM physics.

The properties of the new boson has been measured in the current datasample and they seem to
confirm that this boson may be the long-awaited Higgs boson expected in the standard model, the last
missing piece of this theory. Further studies are on-going, and others waiting for further running
of the LHC, in order to increase the precision of the measurements and confirm this extrem.

Expectations for the future running in 2015 at 13~TeV are getting higher with the
increase in reach for possible new particles and also the improved precision of the measurements
with the larger data samples expected. Specifically, precision measurements of the properties
of the new boson and of other observations that have been accessible at the LHC keep the
focus on the LHC results as the more-likely 
door to the new discoveries in the second half of this decade.


\subsection*{Acknowledgements}

I would like to thank the organizers for giving me the opportunity to take part at the school
and for making it possible with their support and help,
Furthermore both the organizers and the participants
should be recognized for the nice atmosphere and enjoyable time
created around the School.

\end{document}